\documentclass[twocolumn,graphics,floatfix,a4paper,prb]{revtex4}
\usepackage{graphicx, epsfig,bm}
\usepackage{amsmath, amssymb,braket,bbm,turnstile}
\usepackage{tikz}
\usepackage{color}
\usepackage{fancyhdr}
\usepackage{marvosym}

\newcommand{\Fig}[1]{Fig.~\ref{#1}}

\newcommand{\Eq}[1]{Eq.(\ref{#1})}

\newcommand{\bit}{\begin{itemize} \setlength{\itemsep}{0ex} \setlength{\topsep}{0ex} } 
\newcommand{\eit}{\end{itemize}}
\newcommand{\be}{\begin{equation}}
\newcommand{\ee}{\end{equation}}
\newcommand{\bea}{\begin{eqnarray}}
\newcommand{\eea}{\end{eqnarray}}
\newcommand{\ba}{\begin{align}}
\newcommand{\ea}{\end{align}}
\newcommand{\SKIP}[1]{}

\def \ra{\rightarrow}

\def \la{\leftarrow}

\def \eps{\epsilon}
\def \1{\mathbbm{1}}

\begin{document}
\title{Continuous matrix product states for non-relativistic quantum fields: \\
a lattice algorithm for inhomogeneous systems}
\author{Martin Ganahl}
\email[corresponding author: ]{martin.ganahl@gmail.com}
\affiliation{Perimeter Institute for Theoretical Physics, 31 Caroline Street North, Waterloo, ON N2L 2Y5, Canada}
\author{Guifre Vidal}
\affiliation{Perimeter Institute for Theoretical Physics, 31 Caroline Street North, Waterloo, ON N2L 2Y5, Canada}

\begin{abstract}
  By combining the continuous matrix product state (cMPS) representation for quantum fields in the
  continuum with standard optimization techniques for matrix product states (MPS) on the lattice,
  we obtain an approximation $|\Psi\rangle$, directly in the continuum, of the ground state of
  non-relativistic quantum field theories. This construction works both for translation invariant
  systems and in the more challenging context of inhomogeneous systems, as we demonstrate for an
  interacting bosonic field in a periodic potential. Given the continuum Hamiltonian $H$, we
  consider a sequence of discretized Hamiltonians $\{H(\epsilon_{\alpha})\}_{\alpha=1,2,\cdots,p}$ on
  increasingly finer lattices with lattice spacing $\epsilon_1 > \epsilon_2 > \cdots > \epsilon_p$.
  We first use energy minimization to optimize an MPS approximation $|\Psi(\epsilon_1)\rangle$
  for the ground state of $H(\epsilon_1)$. Given the MPS $|\Psi(\epsilon_{\alpha})\rangle$
  optimized for the ground state of $H(\epsilon_{\alpha})$, we use it to initialize the
  energy minimization for Hamiltonian $H(\epsilon_{\alpha+1})$, resulting in the optimized MPS
  $|\Psi(\epsilon_{\alpha+1})\rangle$. By iteration we produce an optimized MPS
  $|\Psi(\epsilon_{p})\rangle$ for the ground state of $H(\epsilon_p)$, from which we finally
  extract the cMPS approximation $|\Psi\rangle$ for the ground state of $H$. Two key ingredients of
  our proposal are: (i) a procedure to discretize $H$ into a lattice model where each site
  contains a two-dimensional vector space (spanned by vacuum $|0\rangle$ and one boson $|1\rangle$
  states), and (ii) a procedure to map MPS representations from a coarser lattice to a
  finer lattice.
\end{abstract}

\maketitle
\section{Introduction}
Tensor networks for quantum lattice systems 
have their origins in the proposal twenty-five years ago of 
\textit{matrix product states} (MPS) \cite{fannes_finitely_1992,
  schollwock_density-matrix_2011,verstraete_matrix_2009} 
(introduced as finitely 
correlated states \cite{fannes_finitely_1992}) 
and the advent of the density matrix renormalization group 
(DMRG) \cite{white_density_1992}. The tensor network formalism was later on generalized using 
concepts and tools from quantum information theory, leading to the proposal e.g. of the 
multi-scale entanglement renormalization ansatz (MERA)\cite{vidal_entanglement_2007} and 
projected entangled pair states (PEPS) \cite{verstraete_renormalization_2004}, 
and is nowadays the basis of powerful variational 
approaches to approximate the ground-state of local lattice Hamiltonians both in one and 
two spatial dimensions \cite{schollwock_density-matrix_2011,verstraete_matrix_2009}. 
This formalism can also be used to study quantum field theories (QFTs), after a suitable 
lattice discretization \cite{stoudenmire_one-dimensional_2012,milsted_matrix_2013,
  wagner_guaranteed_2013,wagner_kohn-sham_2014,baker_one-dimensional_2015,baker_erratum:_2016,
  stoudenmire_sliced_2017,baker_chemical_2017}. 
More recently, important proposals were made to extend tensor networks from the lattice to 
the continuum, so that they can be applied to quantum fields directly --that is, 
without resorting to a lattice discretization. 
Prominent examples are the continuous MPS (cMPS)
\cite{verstraete_continuous_2010,haegeman_applying_2010} and the continuous MERA (cMERA) 
\cite{haegeman_entanglement_2013,miyaji_continuous_2015,hu_spacetime_2017,franco-rubio_entanglement_2017,wen_holographic_2016,molina-vilaplana_information_2015,cotler_cmera_2016}.

Ever since the original proposal of cMPS for non-relativistic QFTs in one spatial dimension 
by Verstraete and Cirac \cite{verstraete_continuous_2010}, several optimization algorithms 
for ground-states of translation invariant systems have been put 
forward\cite{haegeman_applying_2010,draxler_particles_2013,quijandria_continuous_2014,
  chung_matrix_2014,quijandria_continuous-matrix-product-state_2015,haegeman_quantum_2015,
  rincon_lieb-liniger_2015,chung_matrix_2015,
  draxler_atomtronics_2016,ganahl_continuous_2017,ganahl_continuous_2017-1} and applied to 
a number of systems. These include non-relativistic QFTs of interacting bosons 
\cite{draxler_particles_2013,quijandria_continuous_2014,
  quijandria_continuous-matrix-product-state_2015,haegeman_quantum_2015,
  rincon_lieb-liniger_2015,ganahl_continuous_2017,ganahl_continuous_2017-1} 
and fermions \cite{chung_matrix_2014,chung_matrix_2015}, 
and even a relativistic QFT of fermions (after modifying the kinetic 
term at high energies) \cite{haegeman_applying_2010}. 
However, in inhomogeneous (that is, non-translation 
invariant) systems, the cMPS formalism has proven notoriously challenging and, besides 
important first steps \cite{haegeman_quantum_2015}, optimization algorithms have remained 
largely unaddressed. Only very recently, one of the authors has proposed a new, spline-based 
parametrization of inhomogeneous cMPS \cite{ganahl_continuous_2017-1} and has shown 
how a ground-state optimization for such states can be implemented.

\subsection{Hybrid lattice/continuum strategy}
In the present manuscript we propose a hybrid lattice/continuum strategy which combines
lattice MPS optimization techniques with the cMPS representation. We use
this strategy to produce a cMPS approximation $\ket{\Psi}$ to the ground-state of 
the continuum Hamiltonian $H$ of an inhomogeneous QFT. 
Our central insight is that given a sufficiently fine lattice with spacing $\eps$, 
and a discretized version $H(\eps)$ of a continuum Hamiltonian $H$, 
an MPS approximation $\ket{\Psi(\epsilon)}$ to the ground 
state of $H(\epsilon)$ already has a hidden cMPS structure.
We make use of this hidden cMPS in two different ways:
(1) We exploit the cMPS representation for continuum 
systems to relate two MPS representations on two lattices with different lattice 
spacing $\epsilon$ and $\epsilon'$. Indeed, using rescaling properties of the cMPS representation, 
we can translate an optimized MPS $\ket{\Psi(\epsilon)}$ for $H(\eps)$ into an
MPS $\ket{\Psi'(\epsilon')}$ on a lattice with lattice spacing $\epsilon' \not = \epsilon$. 
The state $\ket{\Psi'(\epsilon')}$ is of interest because it is already remarkably 
close to the ground-state of the Hamiltonian 
$H(\epsilon')$ on this second lattice, and thus provides an excellent starting point for an MPS 
energy minimization algorithm on that lattice. 
(2) Given the optimized MPS $\ket{\Psi(\eps)}$ for the ground-state of $H(\eps)$
for sufficiently small $\eps$, we can extract its hidden cMPS structure and use it to directly 
build a cMPS approximation $\ket{\Psi}$ for the continuum Hamiltonian $H$.
The resulting cMPS approximation $\ket{\Psi}$ for the ground-state of the QFT Hamiltonian $H$ is 
seen to already be quite accurate. However, it is by no means an optimal cMPS approximation. 
Indeed, $\ket{\Psi}$ can be used as initialization of the recently introduced, spline-based, 
cMPS approach\cite{ganahl_continuous_2017-1}, 
which would then produce a better cMPS approximation. 

A natural question arises: if we already have an inhomogeneous cMPS algorithm directly in the 
continuum \cite{ganahl_continuous_2017-1}, why should we then use the proposed hybrid MPS/cMPS 
algorithm at all? 
The answer is that cMPS algorithms are generally much more delicate than lattice MPS algorithms 
and work best starting from a sufficiently pre-converged cMPS. Indeed, as it is well documented 
\cite{ganahl_continuous_2017}, 
even in the much better behaved case of translation invariant systems (both 
with euclidean time evolution \cite{haegeman_time-dependent_2011} 
and energy minimization \cite{ganahl_continuous_2017} 
algorithms), a properly pre-converged cMPS is key to preventing fatal optimization instabilities 
and to very significantly improve convergence time. 

We have mentioned above that the cMPS representation plays an important role not just in the 
continuum, but also in lattices approximating the continuum, since it provides a procedure to map 
wave-functions between lattices with different spacing $\epsilon$ and $\epsilon'$. 
As a matter of fact, the cMPS representation has inspired a new scheme for 
discretizing a bosonic continuum Hamiltonian $H$. 
A standard discretization of $H$ would lead to a boson Hubbard model where each site contains an 
infinite-dimensional complex vector space (corresponding to a bosonic degree of freedom). 
This vector space would be then truncated down to a $d$-dimensional complex vector 
space $\mathbb{C}_{d}$,
with its $d$ levels representing the vacuum (or no-boson) state $\ket{0}$ and $n$-boson 
states $\ket{n}$ for $n=1,2, \cdots, d-1$. 
The acceptable value of $d$ on a given site might be hard to predict a priori and depends 
on the lattice spacing: at fixed particle density, the number of particles per site
is proportional to $\epsilon$ and thus a coarser lattice requires a larger $d$ than a 
finer lattice.
The discretization scheme we propose in this manuscript
produces instead a lattice where each site is 
described by a two-level complex vector space $\mathbb{C}_2$, spanned by the vacuum 
state $\ket{0}$ and the one-boson state $\ket{1}$. In this case $H(\epsilon)$ can be thought 
of as a hard-core boson Hamiltonian with modified kinetic term. This has clear advantages 
over the standard discretization scheme. On the one hand, the same vector space 
dimension $d=2$ independent of the lattice spacing $\epsilon$ avoids having to introduce 
a different MPS format (with tensors of different sizes) in changing from one lattice to 
another. On the other hand, the cost of MPS manipulations grows with the dimension of the 
local vector space, and therefore using the smallest possible dimension $2$ leads to 
lower computational times.

\subsection{Previous work on multigrid DMRG}

Our algorithm belongs to the context of multigrid MPS approaches, as pioneered by M. Dolfi, 
B. Bauer, M. Troyer, and Z. Ristivojevic\cite{dolfi_multigrid_2012}. 
In that work, a bosonic field on a periodic potential was also studied by first discretizing 
it onto a sequence of lattice Hamiltonians, and then sequentially finding an MPS approximation 
to the ground-state of each Hamiltonian, starting with the coarsest lattice and finishing 
with the finest one, and using the optimized MPS on one lattice to build an initial MPS for 
the optimization on the next, finer lattice. As the authors there 
demonstrated\cite{dolfi_multigrid_2012} , 
a multi-grid MPS approach successfully circumvents the problem one faces when 
trying to directly optimize an MPS on the finest lattice, namely that the optimization 
gets stuck in local minima with an incorrect spatial profile of particle density. 
Our approach thus shares important similarities with this proposal.
However, it also contains several significant improvements, based on exploiting the cMPS 
representation hidden in a lattice MPS, which we outline next. 

On the one hand, while both approaches have as input a continuum Hamiltonian $H$, the multi-grid 
approach of Ref. \cite{dolfi_multigrid_2012} 
outputs an approximate ground-state wave-function on some fine 
lattice, whereas our hybrid algorithm outputs an approximate ground-state wave-function 
directly back in the continuum. This is not just a more natural format for the output 
(recall that the lattice was introduced only as a computational device), but one which may lead 
to easier comparison with other continuum approaches, such as perturbative methods
\cite{cazalilla_bosonizing_2004}, Bethe Ansatz \cite{lieb_exact_1963-1,lieb_exact_1963} or
the Gross-Pitaevskii approach 
\cite{pitaevskii_1961,gross_structure_1961,gross_hydrodynamics_1963}.

A second major difference is in how the transition from a coarser lattice to a finer lattice 
takes place in the two approaches. Dolfi et al.\cite{dolfi_multigrid_2012} use
an ad hoc MPS splitting procedure that is essentially independent of the 
actual wave-function under consideration. 
Instead, by extracting and exploiting the approximate cMPS structure hidden in a lattice 
MPS, we will see that our approach produces a better pre-converged initial state on the 
finer lattice.
In addition, the hidden cMPS description allows for much more flexibility in how the 
coarser and finer lattice relate. While the method proposed by Dolfi et al. 
\cite{dolfi_multigrid_2012} is restricted to mapping one site to $n$ sites, 
here we can map $m$ sites to $n$ 
sites for any pair of integers $m,n$ -- and one can even consider transitions between two 
irregular lattice discretizations (where both coarser and finer lattices have a lattice spacing 
that depends on position). Such flexibility is relevant when studying inhomogeneous systems 
with e.g. a position-dependent potential $V(x)$ that changes faster in some places than others, 
since it allows us to locally adjust the lattice spacing to the needs of the problem.

Finally, a third important difference is that Dolfi et al.\cite{dolfi_multigrid_2012} used 
a standard scheme to discretize the continuum bosonic Hamiltonian $H$ into a lattice 
Hamiltonian on a lattice where each site had a complex vector space truncated to $d=3$ 
levels, whereas we propose and use a discretization scheme that produces $d=2$-level sites, 
and thus potentially smaller computational times.

\subsection{Outline}

The rest of the manuscript is organized into the following sections:

In Sect. II we review necessary background material, including standard 
discretization of a continuous Hamiltonian $H$ into a lattice, and the cMPS and MPS
variational states. In the remaining sections we present our proposed algorithm, together
with a demonstration of its performance. More specifically, 
in Sect. III we explain how to extract a hidden, approximate cMPS representation from an MPS, and 
how to discretize a continuous, bosonic Hamiltonian $H$ into a two-level lattice model.
In Sect. IV we describe the algorithm that takes as input a continuum Hamiltonian $H$ and 
produces a cMPS approximation $\ket{\Psi}$ for its ground-state,
with technical details described in the Appendix.
In Sect. V we present results for a concrete system, and in Sect. VI we compare our approach
to the multi-grid method proposed by Dolfi et al.\cite{dolfi_multigrid_2012}.

\section{Background material}

In this section we review the type of QFT Hamiltonian $H$ whose ground-state we would like to investigate, its discretization onto a sequence of lattice Hamiltonians $H(\eps_{\alpha})$, as well as 
the continuous MPS to approximate ground-states directly in the continuum and the matrix product state (MPS) to approximate ground-states of the lattice.

\subsection{Hamiltonian in the continuum}

For the sake of concreteness, in this manuscript we focus on a specific inhomogeneous Hamiltonian 
for a non-relativistic bosonic field of the form
\begin{eqnarray}\label{eq:Ham}
H &=& \frac{1}{2m} \int dx\, \partial_x\psi^\dagger(x) \partial_x\psi(x) + \int dx\, 
\mu(x)\psi^\dagger(x) \psi(x)\nonumber \\
 &+& g \int dx\,\psi^\dagger(x) \psi^\dagger(x) \psi(x) \psi(x),
\end{eqnarray}
which describes the dynamics of a single species of a bosonic field on the real line. 
The creation operator $\psi(x)$ fulfills the canonical bosonic field commutation relations
\begin{equation}
\left[\psi(x), \psi^{\dagger}(y) \right] = \delta(x-y).
\end{equation}
Above, $\frac{1}{2m} \int dx\, \partial_x\psi^\dagger(x) \partial_x\psi(x)$ 
is the non-relativistic kinetic term, with $m$ the mass of a boson, 
$g \int dx\,\psi^\dagger(x) \psi^\dagger(x) \psi(x) \psi(x)$ is a quartic local interaction 
term with interaction strength $g$ and $\int dx\, \mu(x)\psi^\dagger(x) \psi(x)$ 
is a chemical potential term. In this work we will focus on the case of a periodic
chemical potential of the form
\be\label{eq:pot}
\mu(x)=\mu_0+V_0(\cos(\frac{2\pi x}{L})-1)^2
\ee
parametrized by offset $\mu_0$, amplitude $V_0$ and periodicity $L$.
More generally, one could consider a similar Hamiltonian involving multi-species of bosons, 
single species or multiple species of fermionic fields, and even a mixture of bosonic and 
fermionic fields.

\subsection{Hamiltonian on the lattice}

Hamiltonian $H$ in \Eq{eq:Ham} can be discretized following a standard procedure.
First, the real line $\mathbb{R}$ is replaced by a lattice $\mathcal{L}(\epsilon)$ of equidistant 
points $\{ x_i \equiv \epsilon i \}_{i\in \mathbb{Z}}$, 
with lattice spacing $\epsilon \equiv x_{i+1}-x_i$. At 
each position $x_i$, one then replaces 
the bosonic annihilation operator $\psi(x)$ in the continuum 
with a bosonic annihilation operator $\psi(x_i)$ defined as
\begin{equation}
\psi(x) \rightarrow \psi(x_i) \equiv \frac{1}{\sqrt{\epsilon}}c_i.
\end{equation}
$c_i$ is a bosonic lattice annihilation operator at site $i$ 
fulfilling the canonical bosonic lattice commutation relations
\begin{equation}
\left[c_i, c_j^{\dagger} \right] = \delta_{ij}.
\end{equation}
Using the standard replacements $\int dx \rightarrow \epsilon \sum_i $ and $\partial_x f(x) \rightarrow (f(x+\epsilon)-f(x))/\epsilon$, the non-relativistic kinematic term becomes
\begin{eqnarray}
&&\frac{1}{2m} \int dx\, \partial_x\psi^\dagger(x) \partial_x\psi(x) ~~~\rightarrow\\
&&~~~~~~~ -\frac{1}{2m\epsilon^2} \sum_i \left(c_i^{\dagger} c_{i+1} + h.c. \right) + \frac{1}{m\epsilon^2} \sum_i c_i^{\dagger} c_i ,~~~
\end{eqnarray}
whereas the chemical potential term transforms into
\begin{eqnarray}
\int dx\, \mu(x)\psi^\dagger(x) \psi(x) ~~~ \rightarrow ~~~ \sum_i \mu_i c_i^{\dagger} c_i,
\end{eqnarray}
with $\mu_i \equiv \mu(x_i)$, and the interaction term reads
\begin{eqnarray}
g \int dx\,\psi^\dagger(x) \psi^\dagger(x) \psi(x) \psi(x) ~~\rightarrow~~ \frac{g}{\epsilon} \sum_{i} c_i^{\dagger}c_{i}^{\dagger} c_i c_{i}.~
\end{eqnarray}

We therefore end up with the lattice Hamiltonian
\begin{eqnarray}
&& H(\epsilon) \equiv -\frac{1}{2m\epsilon^2} \sum_i \left(c_i^{\dagger} c_{i+1} + h.c. \right) \nonumber \\
&&~~+ \sum_i \left(\mu_i + \frac{1}{m\epsilon^2}\right) c_i^{\dagger} c_i + \frac{g}{\epsilon} \sum_{i} c_i^{\dagger}c_{i}^{\dagger} c_i c_{i}.~~~~~ \label{eq:Ham_eps}
\end{eqnarray}

We emphasize that later on we will propose a more convenient, 
alternative discretization scheme which results in modified kinetic and interaction terms, 
see \Eq{eq:Ham_eps2}, acting on a lattice $\mathcal{L}(\epsilon)$ where each site is described 
by a two-dimensional complex vector space spanned by $\ket{0}$ and $\ket{1}$, that is $d=2$, for any value of $\epsilon$.

Hamiltonian $H(\epsilon)$ acts on a lattice where each site $i$ is described by an infinite-dimensional complex vector space spanned by states $\{\ket{n_i}\}$, for $n_i=0,1,2,\cdots$, 
where the state $\ket{n_i}$ fulfills $c_i^{\dagger}c_i \ket{n_i} = n_i \ket{n_i}$ and thus corresponds to having $n_i$ bosonic particles on that site. In practical calculations one must truncate this basis to a finite number $d$ of states, so that $n_i=0,1,\cdots, d-1$. In a ground-state calculation, this approximation may be well justified (see below).

\subsection{Scaling of expectation values with $\eps$}
Let us assume that the ground-state expectation value of the 
particle density $\langle n(x)\rangle \equiv \langle \psi^{\dagger}(x)\psi(x)\rangle$, 
the quartic term $\langle \psi^{\dagger}(x)\psi^{\dagger}(x)\psi(x)\psi(x)\rangle$, and
and the kinetic term $\langle \partial_x\psi^{\dagger}(x)\partial_x\psi(x) \rangle$ are finite 
and that the corresponding lattice expectation values converge smoothly to the continuum. 
This implies that, for small lattice spacing $\epsilon$ and to leading order in $\epsilon$, we have
\begin{eqnarray}
&&\braket{c_i^{\dagger}c_i} \approx\eps\braket{\psi^{\dagger}(x)\psi(x)}=O(\eps),\label{eq:expect1}\\
&&\braket{c_i^{\dagger}c_i^{\dagger}c_i c_i}\approx\epsilon^2 \braket{\psi^{\dagger}(x)\psi^{\dagger}(x)\psi(x)\psi(x)}=O(\epsilon^2),~~~~~~~ \label{eq:expect2}\\
&&\langle \left(c_{i+1}- c_i\right)^{\dagger} \left(c_{i+1}- c_i\right)\rangle\approx\nonumber  \\
&&~~~~~~~~~~~~~~~~~~~\epsilon^3 \langle \partial_x\psi^{\dagger}(x)\partial_x\psi(x) \rangle = O(\epsilon^3) \label{eq:expect3}.
\end{eqnarray}

In particular, if $P_{n_i} \equiv \braket{\ket{n_i}\bra{n_i}}$ is the probability 
that site $i$ is found in state $\ket{n_i}$ when the lattice is in the ground-state 
of $H(\epsilon)$, with $\sum_{n_i} P_{n_i}=1$ and $P_{n_i} \geq 0$, then we 
have $\langle c_i^{\dagger}c_i\rangle = \sum_{n_i} n_iP_{n_i}$ 
,$\langle c_i^{\dagger}c_i^{\dagger}c_i c_i \rangle = \sum_{n_i} n_i(n_i-1)P_{n_i}$ and 
Eqs. (\ref{eq:expect1})-(\ref{eq:expect2}) imply
\begin{eqnarray}
O(\epsilon) &=& \sum_{n_i=0}^{\infty} n_i P_{n_i} \geq \sum_{n_i=1}^{\infty} P_{n_i},\label{eq:Oeps}\\
O(\epsilon^2) &=& \sum_{n_i=0}^{\infty} n_i(n_i-1)P_{n_i} \geq \sum_{n_i=2}^{\infty} P_{n_i}.\label{eq:Oeps2}
\end{eqnarray}
\Eq{eq:Oeps} tells us that the probability $P_1$ of finding one or more boson on one site is 
at most $O(\epsilon)$. 
Eq.(\ref{eq:Oeps2}) implies that the probability $P_2$ of finding two or more bosons at one site is at 
most $O(\epsilon^2)$.
If we were to truncate the Hilbert space dimension to $d=2$, then we would be throwing away 
contributions of order $\eps^2$.
However, this approximation would be incompatible with 
the interaction term in the Hamiltonian \Eq{eq:Ham}, which requires at least two bosons 
on one site, that is $d=3$. We conclude that a valid truncation of the local Hilbert space
requires at least $d=3$.

\subsection{Continuous matrix product states}

The continuous matrix product state (cMPS) \cite{verstraete_continuous_2010,haegeman_calculus_2013} 
is a variational ansatz for ground-states of non-relativistic QFT 
Hamiltonians such as $H$ in \Eq{eq:Ham}. On the real line it has the form
\begin{equation}
  \ket{\Psi}=v_l^{\dagger}\mathcal{P}e^{\int_{-\infty}^{\infty}dx\, Q(x)\otimes \1+R(x)\otimes\psi^{\dagger}(x)}v_r|0\rangle,
  \label{eq:cmps}
\end{equation}
where for each value $x \in \mathbb{R}$, $Q(x),R(x) \in \mathbbm{C}^{D\times D}$ are $D\times D$ 
matrices of complex coefficients,  $\mathcal{P}e$ denotes a path ordered exponential, $\ket{0}$ 
is the physical vacuum, that is $\psi(x)\ket{0}=0 ~~\forall x\in \mathbb{R}$, and $v_l,v_r$ 
are arbitrary boundary vectors at $x=\pm \infty$.

The matrix functions $Q(x)$ and $R(x)$ contain the variational parameters of the ansatz. The expectation value of local observables can be expressed in terms of these matrices. For instance, we have
\begin{align}
&\langle \psi^{\dagger}(x)\psi(x)\rangle = \bra{l(x)} R(x)\otimes \bar{R}(x) \ket{r(x)},\label{eq:Obs1}\\
&\langle \psi^{\dagger}(x)\psi^{\dagger}(x)\psi(x)\psi(x)\rangle =\bra{l(x)} R^2(x)\otimes \bar{R}^2(x) \ket{r(x)},~~~ \label{eq:Obs2}\\
&\langle \partial_x\psi^{\dagger}(x)\partial_x\psi(x) \rangle = \bra{l(x)} \left( \left[Q(x),R(x)\right] + \frac{dR(x)}{dx}\right) ~~~~~~~~~\nonumber \\
&~~~~~~~~~~~~~~\otimes \left(\left[\bar{Q}(x),\bar{R}(x)\right] + \frac{d\bar{R}(x)}{dx}\right) \ket{r(x)},\label{eq:Obs3}
\end{align}
where $\bra{l(x)}$ and $\ket{r(x)}$ are $D^2$-dimensional vectors that depend on the cMPS description at smaller and larger values of $x$, respectively, and fulfill $\braket{l(x)|r(x)}= \braket{\Psi|\Psi}=1$. In particular, $\bra{l(x)}$ and $\ket{r(x)}$ can be chosen to have only finite entries and therefore finite matrices $Q(x)$ and $R(x)$ and $dR(x)/dx$ will guarantee a finite value for Eqs. (\ref{eq:Obs1})-(\ref{eq:Obs3}).

\subsection{Lattice matrix product states}

The MPS is a variational ansatz for ground-states of lattice Hamiltonians such as $H(\epsilon)$ in \Eq{eq:Ham_eps}. It has the form
\begin{equation}
\ket{\Psi} = \sum_{\cdots, n_i, n_{i+1}, \cdots} \cdots A^{[i]}_{n_i}A^{[i+1]}_{n_{i+1}} \cdots \ket{\cdots  n_{i} n_{i+1}\cdots} \label{eq:mps}
\end{equation}
where $i$ labels lattice sites, and
$n_i=0,1\cdots,d-1$ labels states $\ket{n_i}$ of the occupation number basis
on that site. $A_{n_i}^{[i]}  \in \mathbbm{C}^{D\times D}$ is a $D\times D$ matrix of complex 
coefficients. We often refer to the set of $d$ matrices $\{A_{0}^{[i]},  A_{1}^{[i]}, \cdots, A_{d-1}^{[i]}\}$ on site $i$ as a three-index MPS tensor and denote it as $A^{[i]}$.

The MPS tensors $A^{[i]}$ for all sites $i\in \mathbb{Z}$ contain the variational parameters of this ansatz.

\subsection{Periodic potential}

In this work we consider that the chemical potential $\mu(x)$ in the continuum Hamiltonian $H$ in 
\Eq{eq:Ham} is a periodic function,
\begin{equation}
\mu(x+L) = \mu(x).
\end{equation}
We will then assume that the ground-state of $H$ has the same periodicity, which we will incorporate into the cMPS through the periodicity conditions
\begin{equation}
Q(x+L) = Q(x),~~~R(x+L) = R(x).
\end{equation}

By choosing a lattice spacing $\epsilon$ such that $L/\epsilon \equiv N \in \mathbb{N}$, the 
discretized Hamiltonian $H(\epsilon)$ in \Eq{eq:Ham_eps} is periodic under shifts by $N$ 
sites, with the chemical potential $\mu_i$ fulfilling $\mu_{i+N} = \mu_i$. In that case, if the 
ground state is also periodic under shifts of $N$ sites, we can restrict out attention to an 
MPS with a unit cell consisting of $N$ MPS 
tensors $\{ A^{[i]}, A^{[i+1]}, \cdots, A^{[i+N-1]} \}$, with
\begin{equation}
A^{[i+N]} = A^{[i]}.
\end{equation}
Several results presented in the next sections do not rely on having a periodic potential. 

\subsection{cMPS as the continuous limit of an MPS}

The cMPS wave function \Eq{eq:cmps} can be understood \cite{verstraete_continuous_2010} as the continuum limit of a continuous family of MPS $A^{[i]}(\epsilon)$ parametrized by the lattice 
spacing $\epsilon$ of the lattice $\mathcal{L}(\epsilon)$ and defined as
\begin{align}\label{eq:disccmpsA0}
  A_{0}^{[i]}(\epsilon)&\equiv\1+\eps Q(x_i) &n_i=0, \\
  A_{n_i}^{[i] }(\epsilon)&\equiv\sqrt{\frac{\eps^{n_i}}{n_i!}} [R(x_i)]^{n_i} &n_i\geq 1,\label{eq:disccmpsAn}
\end{align}
where $Q(x_i)$ and $R(x_i)$ are the cMPS matrices $Q(x)$ and $R(x)$ evaluated at $x=x_i$. Indeed, in the limit $\epsilon \rightarrow 0$ one can recover the cMPS decomposition in \Eq{eq:cmps} from the MPS decomposition in \Eq{eq:mps} \cite{verstraete_continuous_2010}.

The particular form of the MPS tensors $A^{[i]}$ in \Eq{eq:disccmpsA0}-\Eq{eq:disccmpsAn}
ensures that expectation values of local operators such as $\langle c_i^{\dagger}c_i\rangle$, $\langle c_i^{\dagger}c_i^{\dagger}c_i c_i \rangle$, and $\langle (c_{i+1}- c_i)^{\dagger} \left(c_{i+1}- c_i\right)\rangle$ are continuous as a function of the lattice spacing $\epsilon$ and scale as in Eqs. (\ref{eq:expect1})-(\ref{eq:expect3}) for small values of $\epsilon$.

\subsection{Gauge freedom}

The cMPS representation contains a gauge freedom, meaning that there are redundant parameters. 
If $Q(x),R(x)$ is a cMPS representing a state $\ket{\Psi}$, then the pair $\tilde{Q}(x)$ and $\tilde{R}(x)$ given by
\begin{eqnarray}
\tilde{Q}(x) &=& g(x)Q(x)g(x)^{-1} - \frac{dg(x)}{dx}g(x)^{-1}\\
\tilde{R}(x) &=& g(x)R(x)g(x)^{-1}
\end{eqnarray}
is another cMPS that represents the same state $\ket{\Psi}$. Here $g(x)$ is some invertible $D\times D$ matrix. We can use this freedom to impose certain conditions on the cMPS matrices. In this work we will mostly work on the left canonical gauge, characterized by
\begin{equation}
Q^{\dagger}(x)+ Q(x)+ R^{\dagger}(x)R(x) = 0,
\end{equation}
for which the $D^2$-dimensional left vector $\bra{l(x)}$ and right vector $\ket{r(x)}$ 
become  the identity operator and a diagonal density matrix, after they are reorganized as $D\times D$ matrices.

Analogously, on the lattice the MPS representation $A^{[i]}$ and $\tilde{A}^{[i]}$ represent the same state $\ket{\Psi}$ if they are related by
\begin{equation} \label{eq:gaugeA}
\tilde{A}^{[i]}=g^{[i-1]}A^{[i]}(g^{[i]})^{-1},
\end{equation}
where, for all $i\in \mathbb{Z}$, the $D\times D$ matrix $g^{[i]}$ is invertible. 
The left canonical gauge reads
\begin{equation} \label{eq:leftgauge}
\sum_{n_i} (A^{[i]}_{n_i})^{\dagger}A^{[i]}_{n_i} = \1,~~~\sum_{n_i} A^{[i]}_{n_i}\rho ~(A^{[i]}_{n_i})^{\dagger} = \rho,
\end{equation}
where $\rho$ is a diagonal density matrix containing the squares of the Schmidt values 
$\lambda_{\alpha}$, i.e.
\be
[\rho]_{\alpha\beta}=\lambda^2_{\alpha}\delta_{\alpha\beta}\nonumber
\ee

\section{Exploiting the cMPS representation on the lattice}

In this section we explain how to extract a cMPS representation from an MPS that approximates the ground-state on a fine lattice. We then propose a modified discretization scheme that produces discrete Hamiltonians $H(\eps)$ on a lattice made of two-level sites. Finally, we explain how to use the hidden cMPS representation to map an MPS on one lattice to an MPS on another lattice.

\subsection{Extraction of a cMPS from a lattice MPS}
\label{sec:extraction}

Given a lattice MPS $\ket{\Psi(\epsilon)}$ that approximates the ground-state of $H(\epsilon)$, 
we extract cMPS matrices $Q(x)$ and $R(x)$ evaluated on the points 
$x_i \in \mathcal{L(\epsilon)}$ through
\begin{eqnarray}
Q(x_i) &\equiv& \frac{A^{[i]}_0-\1}{\epsilon}, \label{eq:QA0}\\
R(x_i) &\equiv& \frac{A^{[i]}_1}{\sqrt{\epsilon}}, \label{eq:RA1}
\end{eqnarray}
and then extend these matrices $Q(x_i)$ and $R(x_i)$ from $\mathcal{L}(\epsilon)$ to the real 
line by interpolation (see below). 
\\
It is important to recognize that $Q(x_i)$ depends non-trivially on the choice of gauge in the MPS. Indeed, if $\tilde{A}^{[i]}$ relates to $A^{[i]}$ through \Eq{eq:gaugeA}, then it is easy to check that for $\tilde{Q}(x_i) \equiv \left(\tilde{A}^{[i]}_0-\1\right)/\epsilon$ we get
\begin{eqnarray}
\tilde{Q}(x_i) &=& \frac{1}{\epsilon}\left(g^{[i-1]}A^{[i]}(g^{[i]})^{-1} - \1\right)\\  
&\not =& \frac{1}{\epsilon} g^{[i-1]}\left(A^{[i]} - \1\right)(g^{[i]})^{-1}\\ 
&=&  g^{[i-1]}Q(x_i)(g^{[i]})^{-1}
\end{eqnarray}
which means that the final cMPS state $\ket{\Psi}$ extracted from the MPS state $\ket{\Psi(\epsilon)}$ is affected by the specific gauge choice we work with. In particular, a random choice of gauge may result in matrices $Q(x_i)$ and $R(x_i)$ that change abruptly with $x_i$. This 
typically leads to a cMPS that is not as useful for our current purposes. 
We have heuristically determined that it is best to work with a diagonal gauge, such as e.g. the 
left canonical gauge of \Eq{eq:leftgauge}, which we use from now on
(see Appendix \ref{app:shift} for a more detailed explanation). In particular, 
interpolation of the matrices $Q(x_i), R(x_i)$ is carried out in this diagonal gauge. 
To extend the extracted state to the continuum, we use a standard basis-spline 
interpolation on the matrix entries $[Q(x_i)]_{\alpha\beta}$, $[R(x_i)]_{\alpha\beta}$. 
For brevity, let $M\in\{Q,R\}$ label the two different matrices in the following.
Each interpolation takes as input a set of data 
points $(x_i,[M(x_i)]_{\alpha\beta})$. A continuous and smooth function $f^{M}_{\alpha\beta}(x)$ 
is then constructed such that $f^{M}_{\alpha\beta}(x_i)\equiv[M(x_i)]_{\alpha\beta}$. 
The function $f^{M}_{\alpha\beta}(x)$ is chosen as an 
expansion in so-called order $k$ basis-spline polynomials $B_i^k(x)$, i.e.
\be
f^{M}_{\alpha\beta}(x)=\sum_i B_i^k(x)\mathcal{F}^{M,i}_{\alpha\beta}
\ee
with properly chosen expansion coefficients $\mathcal{F}^{M,i}_{\alpha\beta}$\cite{ganahl_continuous_2017-1,bachau_applications_2001}. 
The polynomials $B_i^k(x)$ have finite support on a small region inside the interval 
$[0,L]$, with $L$ the periodicity of the Hamiltonian $H$.
We use order $k=5$ polynomials in this manuscript, and use periodic spline interpolation.
The cMPS matrix functions are then given by
\begin{align}
  Q_{\alpha\beta}(x)=f^Q_{\alpha\beta}(x)\nonumber\\
  R_{\alpha\beta}(x)=f^R_{\alpha\beta}(x)\nonumber
\end{align}

The cMPS matrices $Q(x)$ and $R(x)$ extracted from an MPS are useful for a number of purposes. For instance, when working on a lattice $\mathcal{L}(\epsilon)$ with sufficiently small lattice spacing $\epsilon$, the extracted cMPS may already provide a good approximation to the ground-state of the continuum Hamiltonian $H$. 

\subsection{Mapping an MPS from $\mathcal{L}(\epsilon)$ to $\mathcal{L}(\epsilon')$}\label{sec:mapping}

On the other hand, we can use the extracted cMPS matrices $Q(x)$ and $R(x)$ and  
Eqs. (\ref{eq:disccmpsA0})-(\ref{eq:disccmpsAn}) to define a new state $\ket{\Psi'(\epsilon')}$ on another lattice $\mathcal{L}(\epsilon')$ with lattice spacing $\epsilon' \not = \epsilon$, such that site $i$ in $\mathcal{L}(\epsilon)$ and in $\mathcal{L}(\epsilon')$ is on positions $x_i$ and $x'_i$, respectively, with
\begin{eqnarray}
x_i \equiv \epsilon i,~~~~x'_i \equiv  \epsilon' i.
\end{eqnarray}
Indeed, we can define MPS tensors for $\ket{\Psi'(\epsilon')}$ according to
\begin{align}\label{eq:prime1}
  A'^{[i]}_{0}&\equiv\1+\eps' Q(x'_i) &n_i=0, \\
  A'^{[i]}_{n_i}&\equiv\sqrt{\frac{\eps'^{n_i}}{n_i!}} [R(x'_i)]^{n_i} &n_i\geq 1,\label{eq:prime2}
\end{align}
so that, overall, we map $\ket{\Psi(\epsilon)}$ into $\ket{\Psi'(\epsilon')}$ through three steps
\begin{eqnarray}
A^{[i]} ~~&&\stackrel{\rm Eqs. (\ref{eq:QA0}),(\ref{eq:RA1})} {\longrightarrow} ~~Q(x_i),R(x_i) \\
&&  \stackrel{\rm interpolation} {\longrightarrow} ~~Q(x),R(x) \\
&&\stackrel{\rm Eqs. (\ref{eq:prime1}),(\ref{eq:prime2})}
{\longrightarrow} ~~A'^{[i]}.
\end{eqnarray}
\\
\subsection{Modified discretization}

Recall that on the lattice we can truncate the infinite-dimensional vector space on each site 
down to $d$ dimensions, where $d$ depends on the lattice spacing $\epsilon$. Recall also that the 
cost of an MPS computation grows with $d$, and therefore it would be convenient to work with the 
smallest non-trivial local space, which corresponds to $d=2$. It would therefore be of interest 
if we could discretize the continuum Hamiltonian $H$ into a lattice Hamiltonian $H(\epsilon)$ 
acting on a lattice where each site has a vector space of dimension $d=2$. It turns out that by 
thinking in terms of the cMPS decomposition, this is possible.

To make a local Hilbert space dimension of $d=2$ compatible with
a bosonic theory, the discretized kinetic and interaction energy operators have to be modified.
Let us first discuss the necessary modifications for the kinetic energy operator.
For illustration purposes, let us for a moment consider
a {\it translation invariant} MPS $\ket{\Psi}$
with tensors $A_{n_i}^{[i]}(\eps)$ of the form
\begin{align}
  A^{[i]}_{0}(\eps)&=\1+\eps Q &n_i=0\nonumber\;\\
  A^{[i]}_{n_i}(\eps)&=\sqrt{\frac{\eps^{n_i}}{n_i!}} [R]^{n_i} &n_i\geq1\nonumber.
\end{align}
However, the following results are equally valid for an inhomogeneous state.

Let us recall that the action of the annihilation operator $c_i$ on the state $\ket{\Psi}$ is
given by
\begin{align}
    c_{i}\ket{\Psi}=\cdots A^{[i-1]}_{n_{i-1}}
    \Big[\sqrt{\eps}R\ket{0}{+}\mathcal{O}(\eps^{3/2})\Big]A^{[i+1]}_{n_{i+1}}\cdots\nonumber.
 \end{align}
If we compute the action of the discretized derivative operator
$\frac{1}{\eps^{3/2}}(\1_ic_{i+1}-c_i\1_{i+1})$ on this state, we obtain, to
lowest orders in $\eps$, an expression of the form
\begin{widetext}
  \begin{align}
    \frac{1}{\eps^{3/2}}(\1_ic_{i+1}-&c_i\1_{i+1})\ket{\Psi}=\nonumber\\
    \cdots&A^{[i-1]}_{n_{i-1}}
      \frac{1}{\eps^{3/2}}
      \Big[({\1+}\eps Q)\sqrt{\eps}R\ket{00}{+}({\1+}\eps Q)\sqrt{\frac{\eps^2}{2}}\sqrt{2}R^2
        \ket{01}{+}\eps R^2\ket{10}\nonumber\\
        &\qquad\quad{-}\sqrt{\eps}R({\1+}\eps Q)\ket{00}
        {-}\sqrt{\frac{\eps^2}{2}}\sqrt{2}R^2({\1+}\eps Q)\ket{10}{-}\eps R^2\ket{01}{+}
            \mathcal{O}(\eps^{2})\Big]A^{[i+2]}_{n_{i+2}}\cdots\nonumber\\
            =&\cdots A^{[i-1]}_{n_{i-1}}\frac{1}{\eps^{3/2}}
      \Big[\underbrace{\Big(({\1+}\eps Q)\sqrt{\eps}R-\sqrt{\eps}R({\1+}\eps Q)\Big)}_{\eps^{3/2}[Q,R]}\ket{00}+\nonumber\\
      &\qquad\qquad\qquad\qquad\underbrace{\Big(\eps R^2-\sqrt{\frac{\eps^2}{2}}\sqrt{2}R^2\Big)}_{0}\ket{10}+
      \underbrace{\Big(\eps R^2-\sqrt{\frac{\eps^2}{2}}\sqrt{2}R^2\Big)}_{0}\ket{01}+\mathcal{O}(\eps^2)\Big]A^{[i+2]}_{n_{i+2}}\cdots.\label{eq:kinop}
  \end{align}
\end{widetext}
Here $\ket{n_in_{i+1}}\equiv \ket{n_i}\otimes\ket{n_{i+1}}$ are occupation number states.
All terms of order $\eps$ inside the square brackets are seen to cancel, leaving a leading term 
$\propto \eps^{3/2}$ which cancels with the pre-factor $\frac{1}{\eps^{3/2}}$ to produce a finite 
result.
Crucially, the cancellations of order $\eps$ involve basis states $\ket{n_i=2}$ and 
$\ket{n_{i+1}=2}$.
A similar calculation for the case of $d=2$ will result in a non-vanishing contribution
of order $\eps$, which will diverge in the limit $\eps\ra 0$ (unless $R^2=0$). Thus,
a Hilbert space dimension of $d=2$ seems irreconcilable with a bosonic theory. However,
if we modify the term $\frac{1}{\eps^{3/2}}(\1_ic_{i+1}-c_i\1_{i+1})$ to
\begin{align}\label{eq:modisckin}
  \frac{1}{\eps^{3/2}}(\1_ic_{i+1}-c_i\1_{i+1})\ra \frac{1}{\eps^{3/2}}(\mathcal{P}^0_{i}c_{i+1}-c_{i}\mathcal{P}^0_{i+1}),
\end{align}
with $\mathcal{P}^0_{i}$ a projection operator onto the empty state $\ket{0}$ at position $x_i$,
it is easy to verify that all terms of order $\eps$ will vanish. Furthermore, the 
result of the operator
application $\frac{1}{\eps^{3/2}}(\mathcal{P}^0_{i}c_{i+1}-c_{i}\mathcal{P}^0_{i+1})\ket{\Psi}$
is, to lowest order in $\eps$, seen to be identical to \Eq{eq:kinop}. 
Similarly, the operator $c_ic_i$ in the interaction is not compatible
with a local Hilbert space dimension of $d=2$. To overcome this problem, we 
make the replacement
\be
c_ic_i\ra c_{i}c_{i+1}.
\ee
Similar to the kinetic energy operator, the result of 
$c_ic_{i+1}\ket{\Psi}$ is, to lowest order in $\eps$, seen to 
be identical to that of $c_ic_i\ket{\Psi}$.
After these replacements the final
lattice Hamiltonian $H$ assumes the form
\begin{eqnarray}
&& H(\epsilon) \equiv -\frac{1}{2m\epsilon^2} \sum_i \left(c_i^{\dagger} c_{i+1}+ h.c. \right) \nonumber\\
&&~~+ \frac{1}{2m\epsilon^2} \sum_{i} \left(\mathcal{P}_i^0 c^{\dagger}_{i+1}c_{i+1} + c^{\dagger}_{i}c_{i} \mathcal{P}_{i+1}^0\right) \nonumber\\
&&~~ + \sum_i \mu_i  c_i^{\dagger} c_i \nonumber \\
&&~~+ \frac{g}{\epsilon} \sum_{i} c_i^{\dagger}c_{i+1}^{\dagger} c_i c_{i+1}.~~~~~ \label{eq:Ham_eps2},
\end{eqnarray}
and the MPS tensors are given by
\begin{align}\label{eq:disccmps_final}
  A^{[i]}_{0}(\eps)&=\1+\eps Q\\
  A^{[i]}_{1}(\eps)&=\sqrt{\eps} R \nonumber.
\end{align}

\begin{figure}
  \includegraphics[width=1\columnwidth]{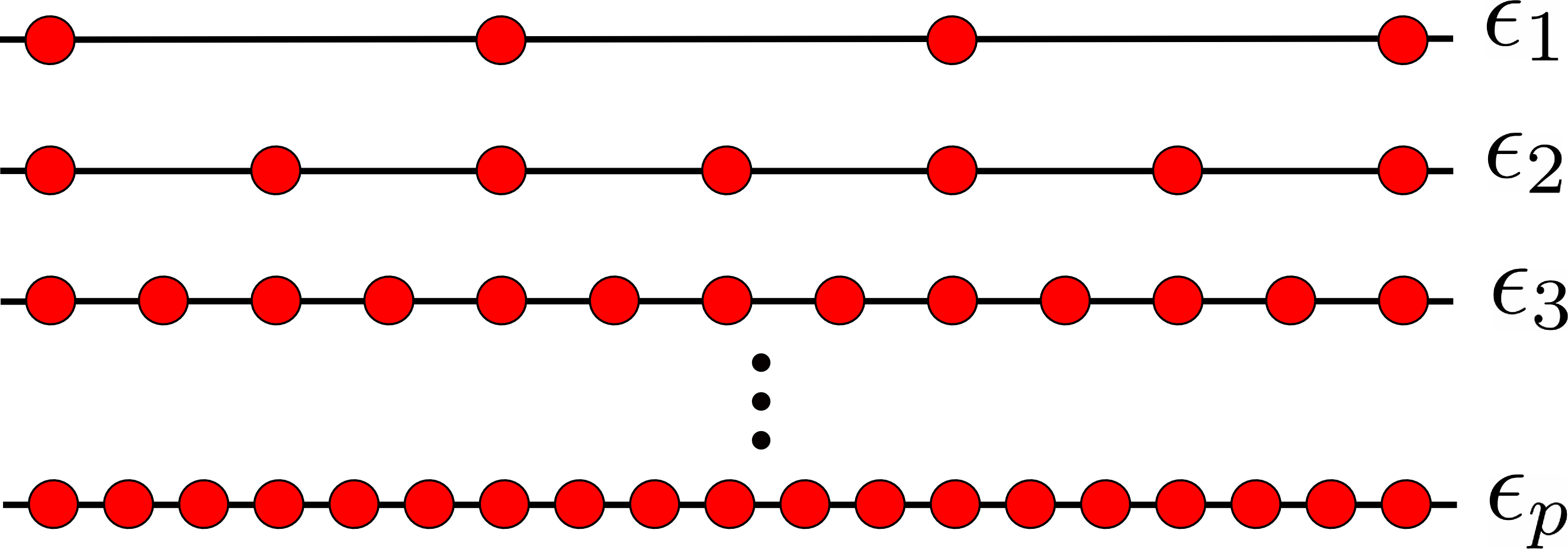}
  \caption{{\bf Consecutive fine-graining of a lattice.} For each lattice 
    spacing $\eps_{\alpha}$ we obtain a Hamiltonian $H(\eps_{\alpha})$ from \Eq{eq:Ham_eps2}}
  \label{fig:finegrain}
\end{figure}

\begin{figure}
  \includegraphics[width=1\columnwidth]{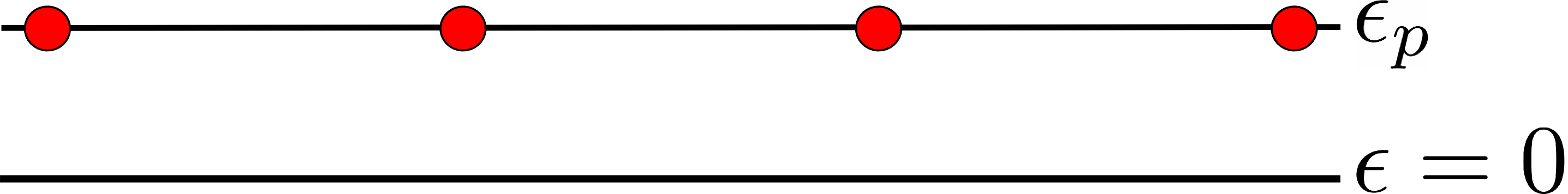}
  \caption{{\bf Obtaining a cMPS from a lattice MPS.} We extract the 
    cMPS matrices $\{Q(x_i),R(x_i)\}$ from the ground-state $\ket{\Psi(\eps_p)}$ of a lattice
    optimization and use interpolation techniques 
    to obtain continuous matrices $Q(x), R(x)$ (see Sec.\ref{sec:extraction}).}
  \label{fig:finegrain_2}
\end{figure}
\begin{figure*}
  \includegraphics[width=0.28\paperwidth]{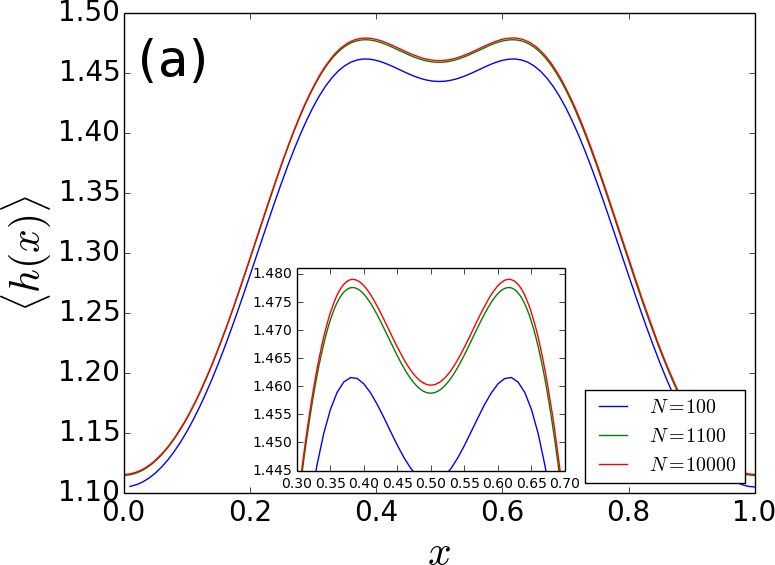} 
  \includegraphics[width=0.28\paperwidth]{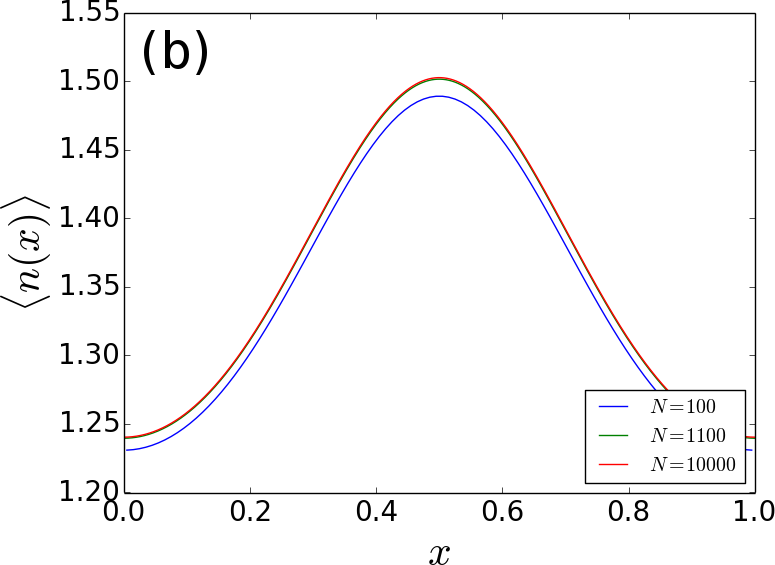} 
  \includegraphics[width=0.28\paperwidth]{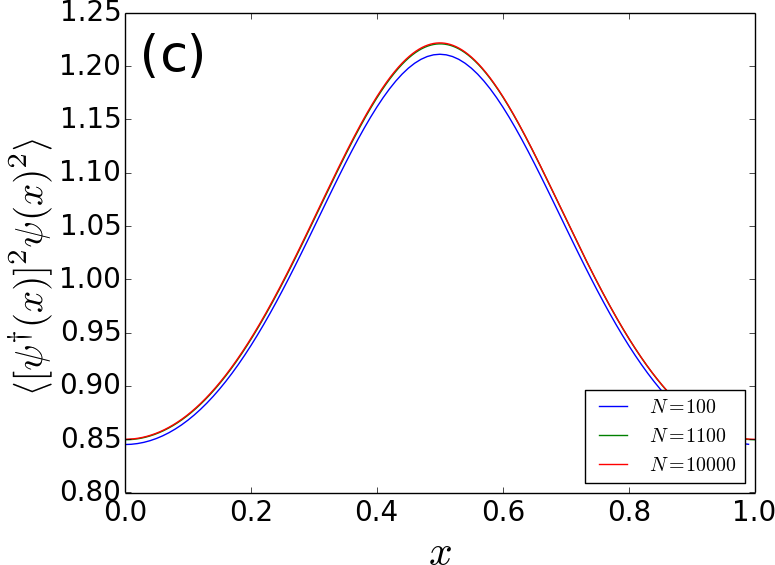} 
  \caption{{\bf Ground-state observables for different lattices.}
(a) Energy density $\braket{h(x)}\equiv\braket{\frac{1}{2m}\partial_x\psi^{\dagger}(x)\partial_x\psi(x)+g\psi^{\dagger}(x)\psi^{\dagger}(x)\psi(x)\psi(x)}$ of the ground-state of \Eq{eq:Ham} for $D=16,\mu_0=-0.5,V_0=-1.0,g=1.0$, for different number of lattice points per unit-cell. (b) Particle density $\braket{n(x)}$
    and (c) interaction term $\braket{[\psi^{\dagger}(x)]^2\psi^2(x)}$ for the same ground-states as in (a).}\label{fig:diffgrids}
\end{figure*} 

\begin{figure}
  \includegraphics[width=1.0\columnwidth]{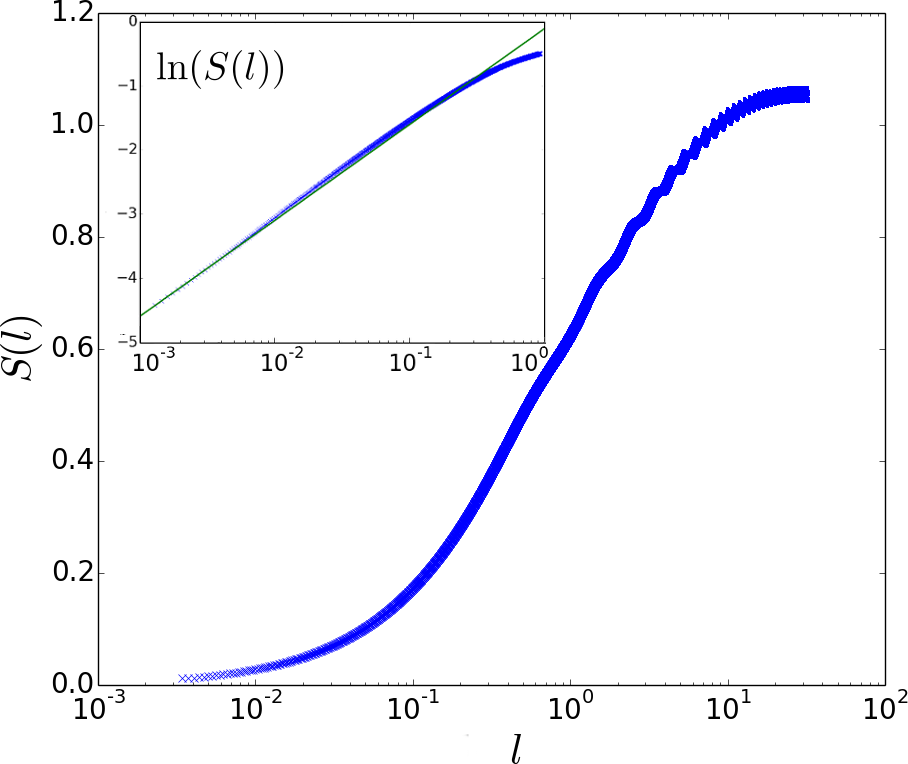} 
  \caption{{\bf Entanglement entropy of a bulk region of length $\bm{l}$.} 
    Main figure: entanglement entropy $S(l)$ in the ground-state of
    \Eq{eq:Ham_eps2} of a small region of length $l$ as a function 
    of $l$, for $D=16,\mu_0=-0.5,V_0=-1.0,g=1.0,L=1.0$, 
    and $N=10^4$ lattice points per unit-cell.
    Oscillations are due to the periodicity of the state.
    Inset:  $\ln(S(l))$ vs. $\log(l)$. For the smallest $l$ visible, we see a power law 
    increase in $S(l)$ which levels off for larger $l>10^{-1}$. The point
    where the curvature in the figure changes sign is approximately at $l \approx 0.05$. 
    The green line is a fit to the first 15 data-points. }
  \label{fig:entropy}
\end{figure}

\begin{figure}
  \includegraphics[width=1.0\columnwidth]{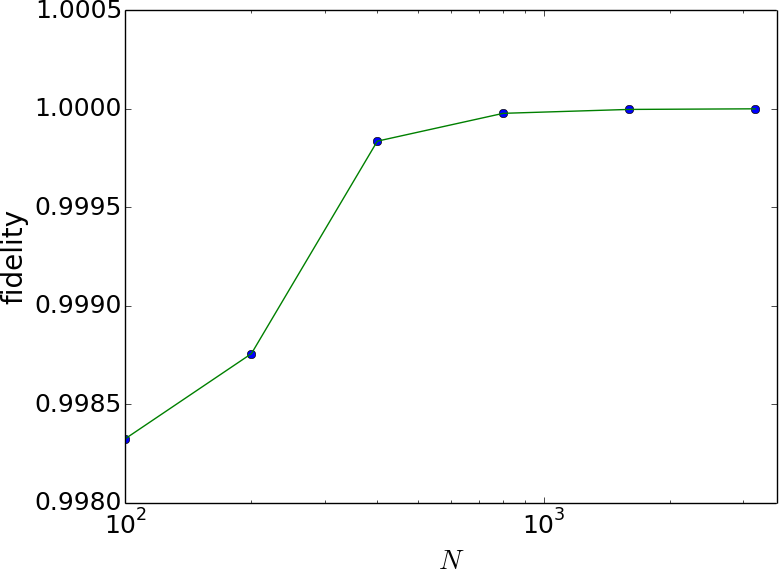} 
  \caption{{\bf Quality of interpolation.} 
    Fidelity per unit-cell of ground-states on $N=100,200,400,800,1600$ and 3200 
    grid points per unit-cell after interpolation to $N_{final}=3200$ with ground-state at 
    $N_{final}=3200$. Parameters are the same as in \Fig{fig:diffgrids}.
    Lines are guides to the eye.}\label{fig:fidelity}
\end{figure}

\begin{figure}
  \includegraphics[width=1.0\columnwidth]{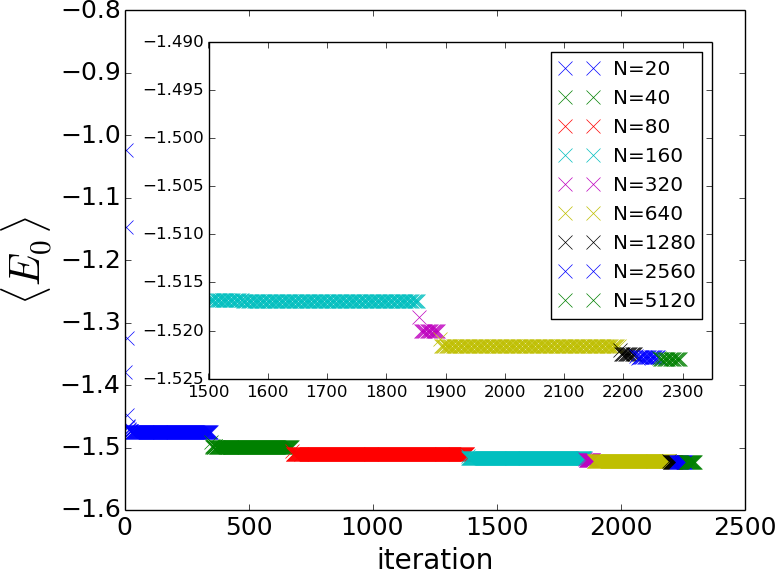} 
  \includegraphics[width=1.0\columnwidth]{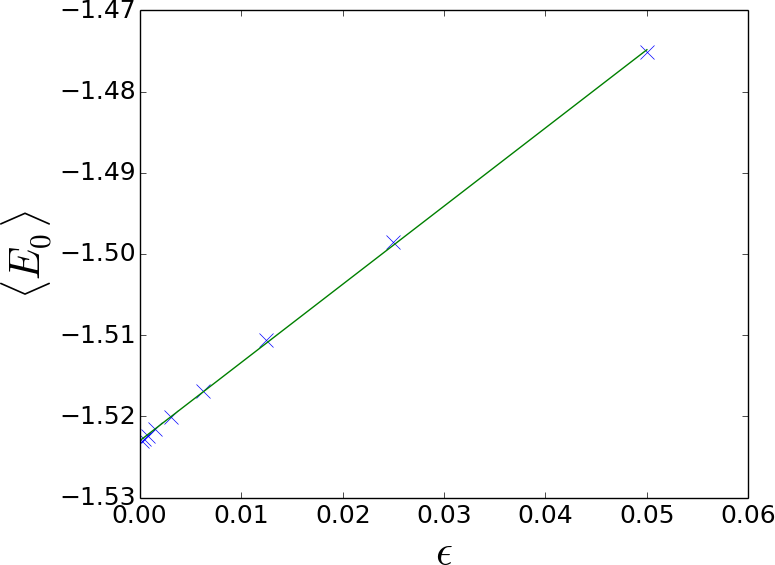} 
  \caption{{\bf Convergence of the ground-state energy.} 
    Upper panel: ground-state energy per unit-cell, 
    as a function of DMRG sweeps for 
    $\mu_0=-0.5, V_0=-1.0,g=1.0,D=32,L=1.0$. The different colours correspond to different 
    fine-graining. At each lattice spacing, ground-state energies are converged to
    $10^{-8}$. Blue crosses at small iteration numbers are for $N=20$ 
    lattice sites;  following crosses are obtained by taking the MPS from the previous 
    discretization and using a spline interpolation on the matrices $Q(x_i),R(x_i)$ to obtain 
    a finer discretization, and continue optimization for the fine-grained lattice. 
    Lower panel: converged ground-state energy per unit-cell 
    for different grid spacings $\eps$.}\label{fig:diffgrids2}
\end{figure}

\section{A hybrid MPS/cMPS algorithm}
In this section we describe an algorithm that takes a continuous, inhomogeneous Hamiltonian $H$ 
of a non-relativistic QFT, and produces a cMPS approximation for its ground-state.
\subsection{Sequence of lattice Hamiltonians}
To obtain a cMPS approximation $\ket{\Psi}$ for the ground-state of the continuum Hamiltonian 
$H$, we first construct a sequence of lattice Hamiltonians $H(\eps_{\alpha}),\alpha = 1\dots p$,
with $\eps_1>\eps_2\dots >\eps_p$. This is schematically represented in \Fig{fig:finegrain}.
We then sequentially compute an MPS approximation $\ket{\Psi(\eps_{\alpha})}$ to the ground-state
of Hamiltonian $H(\eps_{\alpha})$. 
We initialize the ground-state optimization of $H(\eps_{\alpha})$ using the extension
of $\ket{\Psi(\eps_{\alpha-1})}$ to the lattice $\mathcal{L}(\eps_{\alpha})$, as described above.
$\ket{\Psi(\eps_1)}$ is initialized with
random, constant matrices $Q,R$. From the converged state $\ket{\Psi(\eps_p)}$ 
we then construct a cMPS approximation $\ket{\Psi}$ to the ground-state of the
continuum Hamiltonian $H$, see \Fig{fig:finegrain_2}.

\subsection{Optimization of MPS with a finite unit-cell}
For each Hamiltonian $H(\eps_{\alpha})$ we obtain the 
ground-state from am infinite DMRG (iDMRG) method, 
as introduced by McCulloch \cite{white_density_1992,mcculloch_infinite_2008}, 
extended to the case of a large unit-cell with $N\gg 2$ sites. 
In Appendix \ref{app:opt} we present in detail the implementation for the cases
considered in this manuscript, and explain possible variants of the algorithm, 
including a gradient-based optimization method for translation invariant lattices.
Once the ground-state $\ket{\Psi{(\eps_{\alpha})}}$ of $H(\eps_{\alpha})$
has been found, we extract the cMPS matrices $\{Q(x_i), R(x_i)\}$ and interpolate them
to obtain an initial state $\ket{\Psi'(\eps_{\alpha+1})}$ for the optimization
of the next Hamiltonian $H(\eps_{\alpha+1})$ 
(see Secs. \ref{sec:extraction} and \ref{sec:mapping}).
Finally, after having obtained the ground-state $\ket{\Psi{(\eps_{p})}}$ of $H(\eps_p)$, 
we use an interpolation to extract a cMPS approximation $Q(x), R(x)$ to the 
continuous ground-state, see Sec. \ref{sec:extraction}.

\section{Example}
In this section we present results for ground-states observables of \Eq{eq:Ham_eps2} 
with a periodic background potential \Eq{eq:pot} with periodicity $L=1$.
In particular, we show results for 
lattices $\mathcal{L}(\eps=\frac{L}{N})$ 
with an increasing number $N$ of discretization points per 
unit-cell. For all calculations, we set $m=0.5, g=1.0, \mu_0=-0.5$ 
and $V_0=-1.0$. 
On each grid $\mathcal{L}(\frac{L}{N})$
we run the optimization until the energy per unit-cell is converged to $10^{-8}$. 
In \Fig{fig:diffgrids} (a), (b) and (c) we compare converged energy densities
$\braket{h(x)}\equiv\braket{\frac{1}{2m}\partial_x\psi^{\dagger}(x)\partial_x\psi(x)+g\psi^{\dagger}(x)\psi^{\dagger}(x)\psi(x)\psi(x)}$, particle densities $\braket{n(x)}$ and 
interaction term $\braket{[\psi^{\dagger}(x)]^2\psi^2(x)}$
from optimization on three different grids $N=100$ (blue line),
$N=1100$ (green line) and $N=10000$ (red line), obtained by successive interpolation from 
$N=100$ to $N=1100$, and from $N=1100$ to $N=10000$.
We observe a clear convergence with increasing number of discretization points 
(the inset in \Fig{fig:diffgrids} (a) zooms into the central region of the figure).

For large lattice spacing $\eps\geq 0.05$, we find that 
the interpolation method tends to perform worse than for small $\eps$. 
Qualitatively, this follows from a dependence of the lattice matrices $Q(x_i,\eps),R(x_i,\eps)$ 
on the discretization $\eps$.
For large $\eps \geq 0.05$, we find that an interpolation to e.g. $\eps'=\eps/2$ and 
successive normalization of the MPS introduces changes in $Q(x_i,\eps),R(x_i,\eps)$ 
which are comparable to $Q(x_i,\eps),R(x_i,\eps)$. Thus, interpolation 
changes drastically the state.
However, once the discretization is fine enough such that 
$\eps$ is well below a transition scale $\xi$ (see below),
corrections to $Q(x_i,\eps),R(x_i,\eps)$ 
from interpolation become much smaller than $Q(x_i,\eps),R(x_i,\eps)$, 
and interpolation in this case gives accurate ansatz states for finer grids.
This can quantitatively be seen from analyzing for example 
the behaviour of the entanglement entropy 
$S(l)=-{\rm tr}\rho(l)\log(\rho(l))$ of a bulk region of length $l$, where
$\rho(l)$ is the reduced density matrix of this region.
In \Fig{fig:entropy} we show $S(l)$ for a ground-state on a very fine grid
with $N=10^4$ discretization points. 
Oscillations are due to the periodicity of the state.
The inset shows $S(l)$ on a log-log scale. $S(l)$
exhibits a transition from a power law increase 
for $l<0.05$ to a weaker increase for $l>0.05$. 
This transition length scale $\xi\approx 0.05$ coincides with the region of the breakdown of the 
interpolation method, and depends on the location of the region $l$ within the unit-cell
(see Appendix \ref{app:ent} ). The presence of a space dependent effective cutoff $\xi=\xi(x)$ 
suggests that it can be used to create optimal discretization grids for 
MPS by adjusting the grid distances $\eps(x)$ to $\xi(x)$. 
Note that this effective cutoff can be similarly seen in other 
physical observables like e.g. the superconducting correlation function 
$\braket{\psi^{\dagger}(l)\psi(0)}$ (not shown).

An appealing feature of the presented method is the fact that it gives access to a 
parametrization of the ground-state of an interacting quantum
field theory in terms of a set of smooth functions $Q(x)$ and $R(x)$.
This opens up new possible applications of lattice MPS methods. 
As described above, one application is the 
extension by interpolation of the ground 
state at one discretization to a new 
state at an arbitrarily finer discretization. Even though in general
this new state will not be the ground-state of the finer lattice, it will be a good 
approximation to it, which can be used to initialize an optimization
on the finer lattice. 
In particular, interpolation becomes an increasingly soft perturbation 
as one fine-grains the lattice. This is illustrated in 
\Fig{fig:fidelity}, where we plot the fidelity per unit-cell of states 
obtained by interpolating the ground-states $\ket{\Psi(\eps_{\alpha})}$ 
of lattices $\mathcal{L}(\eps_{\alpha})$ with 
$N_{\alpha}=\{100,200,400,800,1600\}, \eps_{\alpha}=\frac{L}{N_{\alpha}}$
to a very fine lattice $\mathcal{L}(\eps_{final})$ 
with $N_{final}=3200, \eps_{final}=\frac{L}{N_{final}}$.
Each ground-states $\ket{\Psi(\eps_{\alpha})}$ is obtained using a standard iDMRG
optimization, and then extended to a
state $\ket{\Psi'_{\alpha}(\eps_{final})}$ on $\mathcal{L}(\eps_{final})$.
We then calculate the fidelity per unit-cell of the overlap
$\braket{\Psi'_{\alpha}(\eps_{final})|\Psi(\eps_{final})}$
As can be seen from the figure, the fidelity rapidly converges towards 1.0,
indicating that the interpolated, non-optimized state $\ket{\Psi'_{\alpha}(\eps_{final})}$
is very close to the true ground-state of the final lattice
(see also Appendix \ref{app:cont} for further discussion on interpolation).

As we argue above, the speed of convergence of an optimization on a given 
lattice $\mathcal{L}(\eps')$ can be drastically improved by a proper choice
of initialization. To this end, we use the proposed interpolation method to
extend the MPS $\ket{\Psi(\eps)}$ from a previous optimization to 
the current lattice  $\mathcal{L}(\eps')$. The resulting state 
$\ket{\Psi'(\eps')}$ is then used as initial state to the optimization on
 $\mathcal{L}(\eps')$.
The merit of this approach is demonstrated in the upper panel of \Fig{fig:diffgrids2}, 
where we show the ground-state energy per unit-cell as a function of the iteration number 
in the optimization (each iteration corresponds to a DMRG sweep over the unit-cell). 
After the energy is converged within $10^{-8}$ we take the 
state, prolong it to a finer grid and continue optimization.
Different colors in the figure correspond to different number of lattice sites $N$ per unit-cell. 
The inset zooms onto the right part of the figure.
The lower panel in \Fig{fig:diffgrids2} 
shows the ground-state energy per unit-cell as a function of discretization parameter $\eps$, 
with a clear linear behaviour in $\eps$.

\begin{figure}
  \includegraphics[width=1.0\columnwidth]{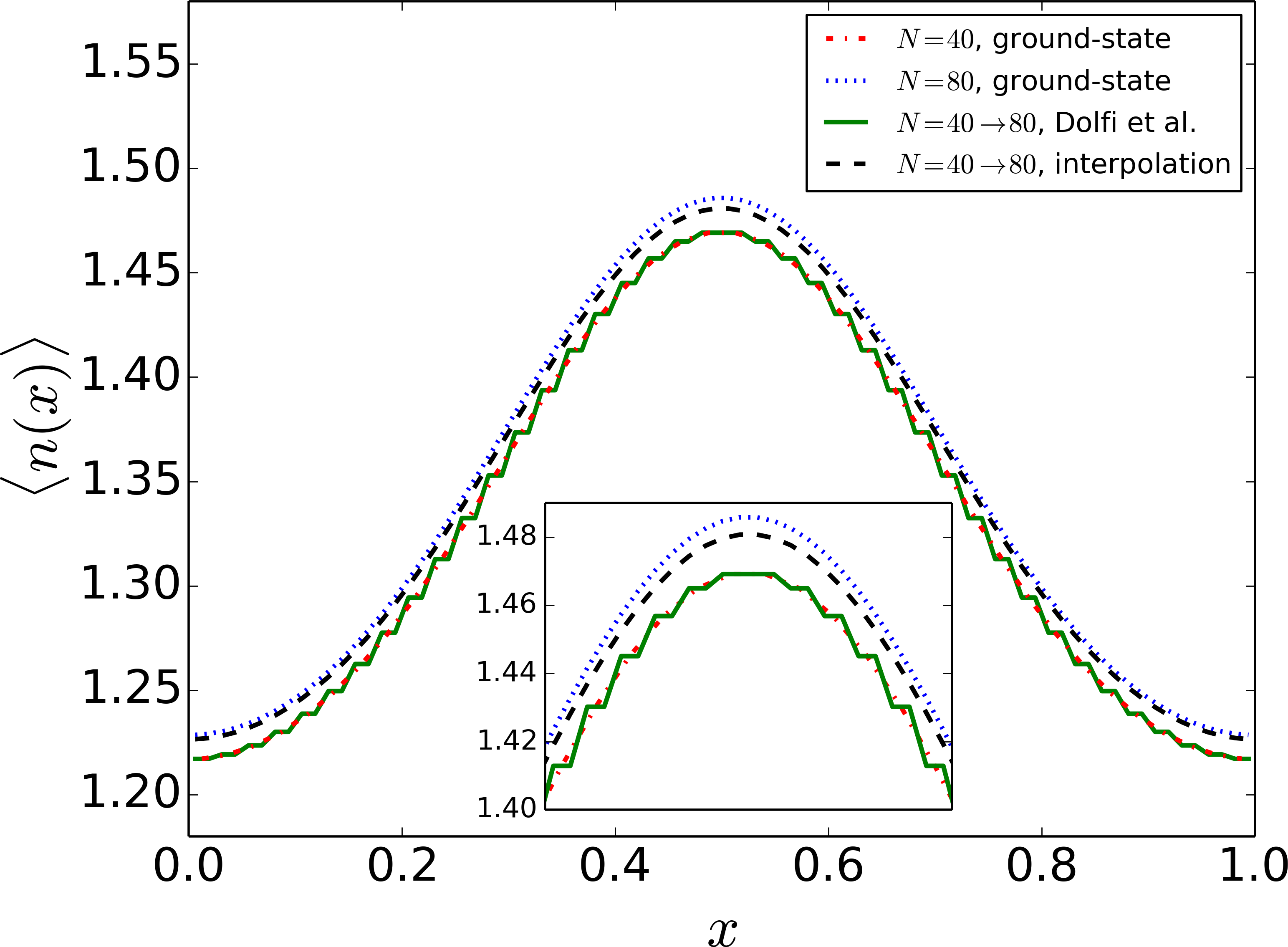}
  \caption{{\bf Particle density $\bm{\braket{n(x)}}$ for different interpolation schemes.} 
    Starting from the ground-state of 
    lattice with $N=40$ per unit-cell (red dash-dotted line), 
    we use prolongation as proposed by Dolfi et al. (green solid line) 
    and our proposed interpolation method (black dashed line)
    to obtain a state on a finer lattice with $N=80$ sites per unit-cell. For comparison
    we also show $\braket{n(x)}$ for the optimized ground-state on $N=80$ sites per unit-cell.
    The inset shows a magnification of the central peak.}
  \label{fig:dolfi}
\end{figure}

\section{Discussion}
As we have already mentioned in the introduction, our proposed method shares important similarities
with the multigrid DMRG method proposed by Dolfi et al. \cite{dolfi_multigrid_2012}. In this
section we give a quantitative comparison between the multigrid approach and our 
proposed interpolation method. We start with a quick summary of the multigrid approach.

To transform an MPS from a lattice with $N$ sites to a lattice with $N'=K\,N$ 
sites, with $K\in \mathbbm{N}$, 
the multigrid approach splits up each tensor $A^{[i]}_{n_i}$ into a product of $K$ new 
tensors 
$\tilde A^{[i'_1]}_{\tilde n_1}\dots \tilde A^{[i'_K]}_{\tilde n_K}$. 
The splitting is performed by first contracting
$A^{[i]}_{n_i}$ with a tensors $T^{n_i}_{\tilde n_1,\dots, \tilde n_K}$,
\be
T^{n_i}_{\tilde n_1,\dots, \tilde n_K}=\frac{\delta(n_i,\sum_{\alpha=1}^K\tilde n_{\alpha})}{\sqrt{\sum_{n'_1,\dots}\delta(n_i,\sum_{\alpha=1}^{K}n'_{\alpha})}},
\ee
and then using an SVD to decompose the resulting tensor into a product of $K$ tenors
$\tilde A^{[i'_1]}_{\tilde n_1}\dots \tilde A^{[i'_K]}_{\tilde n_K}$. 
For the case of $K=2$, this is graphically 
represented as 
\begin{align}
  \raisebox{-0.8cm}{\includegraphics[width=0.15\columnwidth]{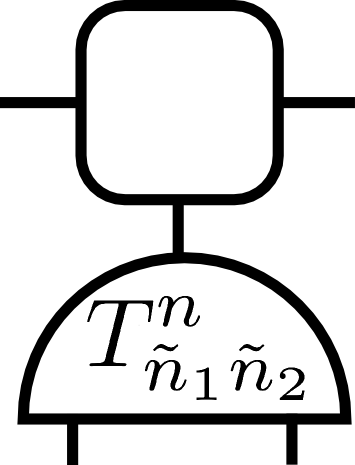}}\stackrel{\textrm{SVD}}{=}
  \raisebox{-0.48cm}{\includegraphics[width=0.25\columnwidth]{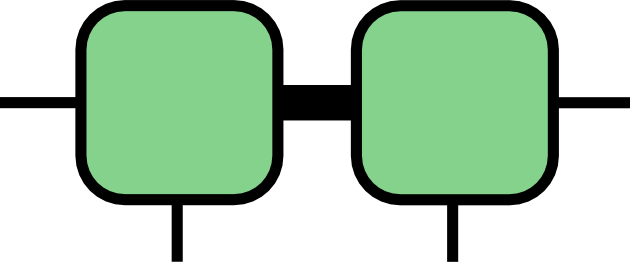}}
\end{align}
(see the Appendix for a short summary of diagrammatic MPS notations).
Note that the bond dimension between the two new tensors has increased, which we have 
indicated by drawing a thicker line between the two.

To compare the multigrid approach with our interpolation method, we start from a
ground-state on a lattice with $N=40$ sites per unit-cell, and 
extend this state to a grid with $N=80$ sites per unit-cell, using the multigrid approach and our
proposed interpolation method. In \Fig{fig:dolfi} we show results for the particle density
$\braket{n(x_i)}$ for the different approaches. 
The red dash-dotted line shows $\braket{n(x_i)}$ for the ground-state on $N=40$ lattice sites
per unit-cell. The green solid and black dashed line are the results from the extension to 
a lattice with $N=80$ sites using a multigrid approach and our interpolation method, respectively.
The blue dotted line is the ground-state for the $N=80$ site lattice. From this result we see
that the multigrid approach already gives a good ansatz state. However, the prolongation 
here produces step-like artifacts. 
The interpolation on the other hand produces a smooth density profile which is seen to 
be already very close to the true ground-state density profile on a lattice with 
$N=80$ sites per unit-cell. The inset shows a zoom onto the central region.
Note that the total particle number is not fixed in our model.

\section{Conclusion}
We have presented a method which unifies the continuous matrix product state 
representation for quantum fields with standard optimization techniques for MPS on the lattice.
Our method is equally applicable to 
both translation invariant and inhomogeneous systems, 
as we demonstrate for an interacting bosonic field in a periodic potential.
Starting from a continuum Hamiltonian $H$, 
we construct a sequence of discretized Hamiltonians $\{H(\epsilon_{\alpha})\}_{\alpha=1,2,\cdots,p}$ 
on increasingly finer lattices $\mathcal{L}(\eps_{\alpha})$ with lattice spacings 
$\epsilon_1 > \epsilon_2 > \cdots > \epsilon_p$.
Our method is initialized by using energy minimization for lattice MPS to optimize an
MPS approximation $\ket{\Psi(\epsilon_{1})}$ to the ground-state of $H(\eps_1)$. 
We then use the MPS $\ket{\Psi(\epsilon_{\alpha})}$ optimized for the ground state of 
$H(\epsilon_{\alpha})$ to initialize the energy minimization 
for Hamiltonian $H(\epsilon_{\alpha+1})$, resulting in the optimized 
MPS $\ket{\Psi(\epsilon_{\alpha+1})}$. To initialize the optimization
we make use of the hidden cMPS structure of the MPS
$\ket{\Psi(\epsilon_{\alpha})}$ to extend it
from the lattice $\mathcal{L}(\eps_{\alpha})$ to the lattice $\mathcal{L}(\eps_{\alpha+1})$.
From the final MPS $\ket{\Psi(\epsilon_{p})}$ for the ground state of $H(\epsilon_p)$, we then
extract the cMPS approximation $\ket{\Psi}$ for the ground state of $H$
directly in the continuum. For the variational energy optimization, we introduce a new
procedure to discretize $H$ into a lattice model where each site 
contains a two-dimensional vector space (spanned by vacuum $\ket{0}$ and one boson $\ket{1}$ 
states).
Our method can be generalized to the case of multiple species of bosons or fermions, or even
mixtures of both.
\\
\\
{\it Acknowledgements}\\
The authors thank J. Rinc\'on, F. Verstraete and D. Draxler for 
insightful discussions. 
The authors also acknowledge support by the Simons Foundation (Many Electron Collaboration). 
Computations were made on the supercomputer Mammouth parall\`ele II from University of 
Sherbrooke, managed by Calcul Qu\'ebec and Compute Canada. 
The operation of this supercomputer is funded by the Canada Foundation for Innovation (CFI), 
the minist\`ere de l'\'Economie, de la science et de l'innovation du Qu\'ebec (MESI) and the 
Fonds de recherche du Qu\'ebec - Nature et technologies (FRQ-NT). 
This research was supported in part by Perimeter Institute for Theoretical Physics. 
Research at Perimeter Institute is supported by the Government of Canada through Industry Canada 
and by the Province of Ontario through the Ministry of Economic Development \& Innovation.


\appendix
\setcounter{equation}{0}
\setcounter{figure}{0}
\setcounter{table}{0}
\renewcommand{\thefigure}{A\arabic{figure}}

\begin{figure*}
  \includegraphics[width=0.28\paperwidth]{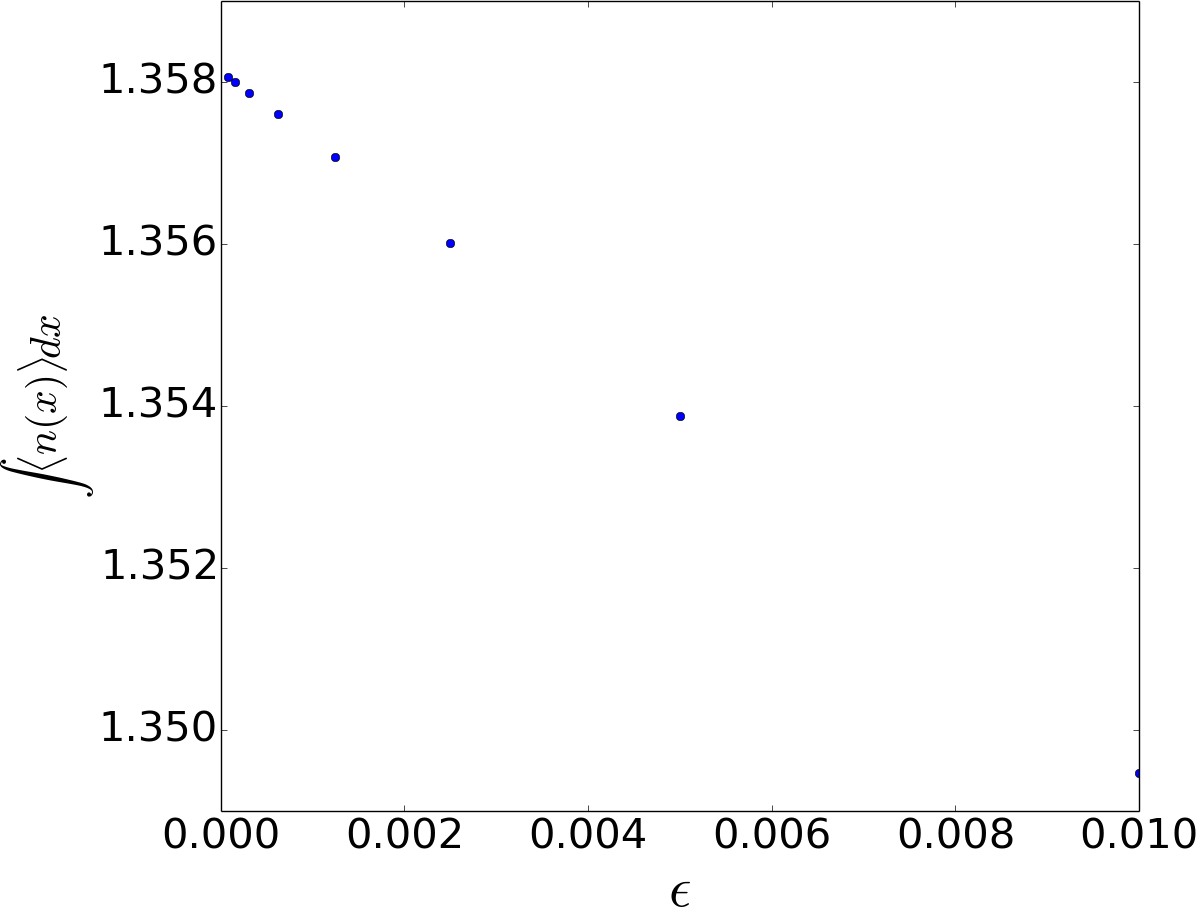}
  \includegraphics[width=0.28\paperwidth]{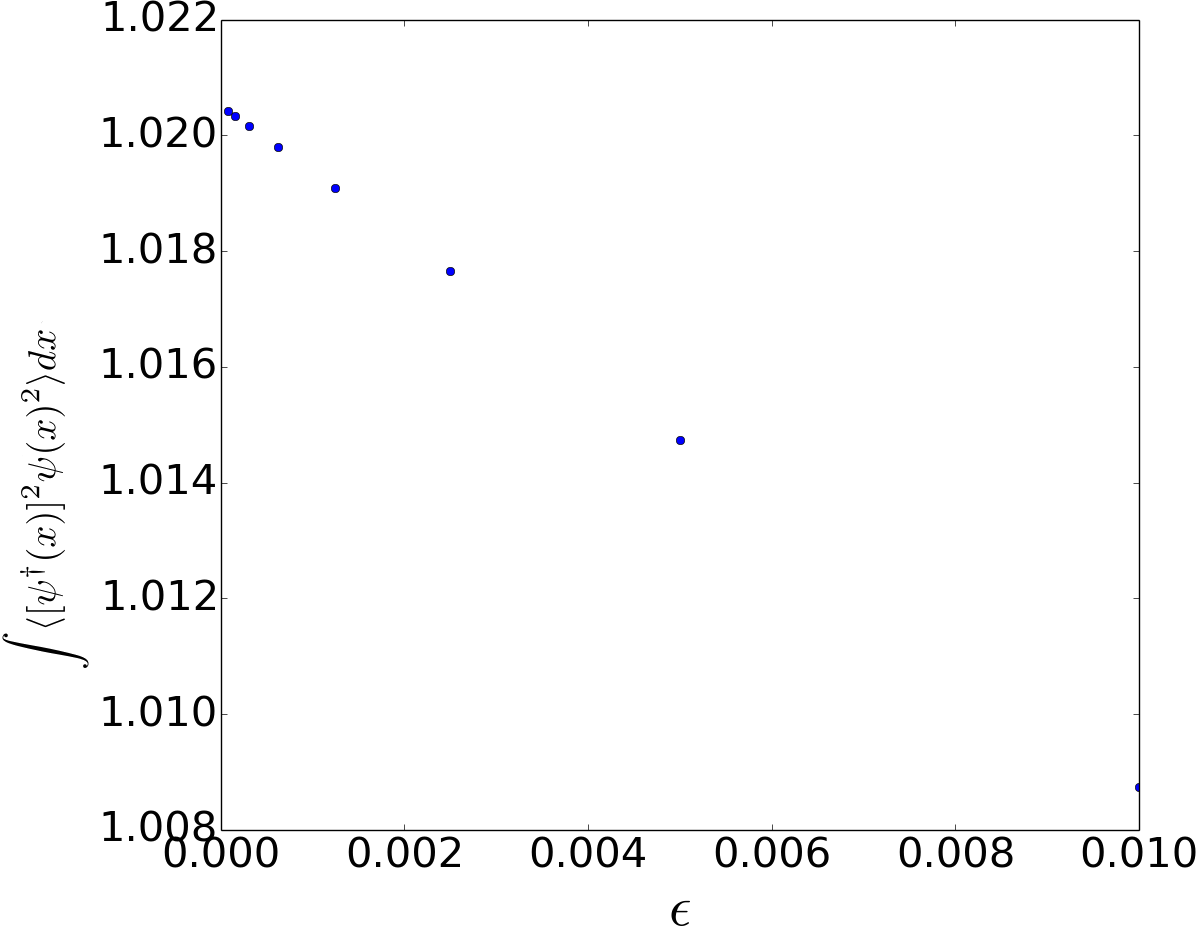}
  \includegraphics[width=0.28\paperwidth]{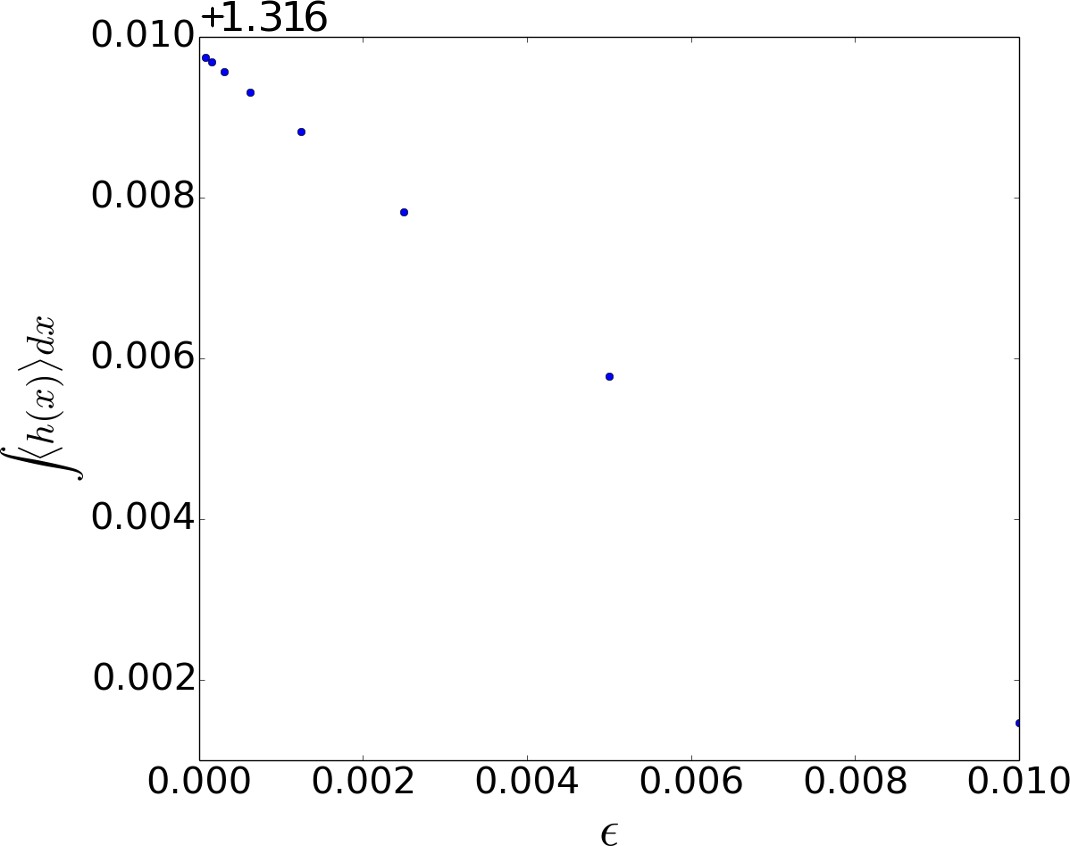}

  \includegraphics[width=0.28\paperwidth]{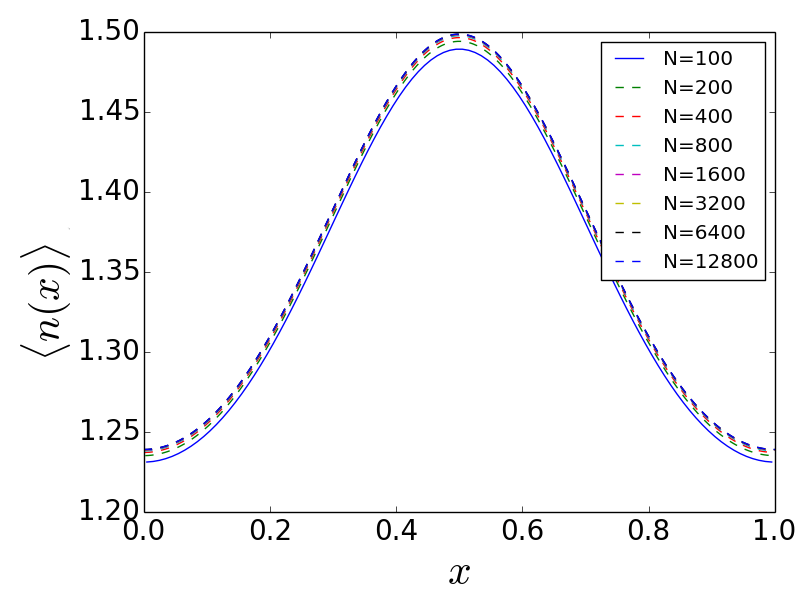}
  \includegraphics[width=0.28\paperwidth]{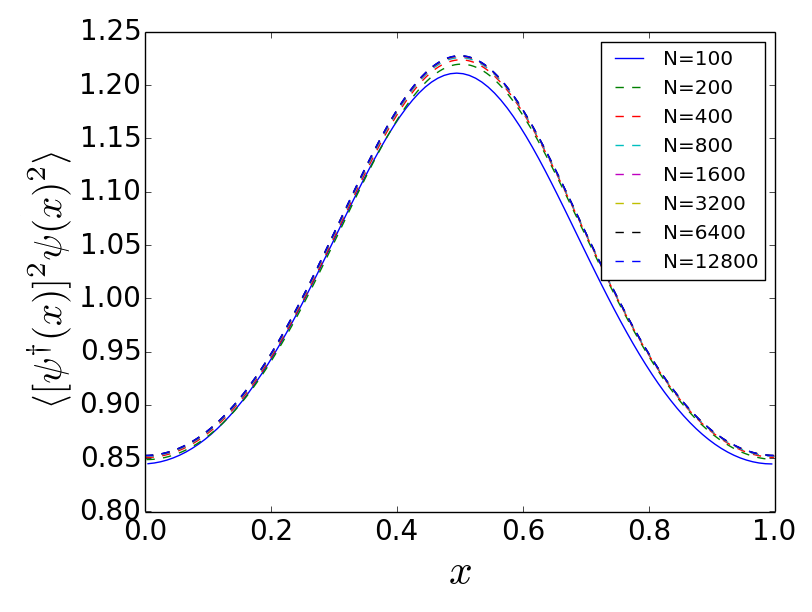}
  \includegraphics[width=0.28\paperwidth]{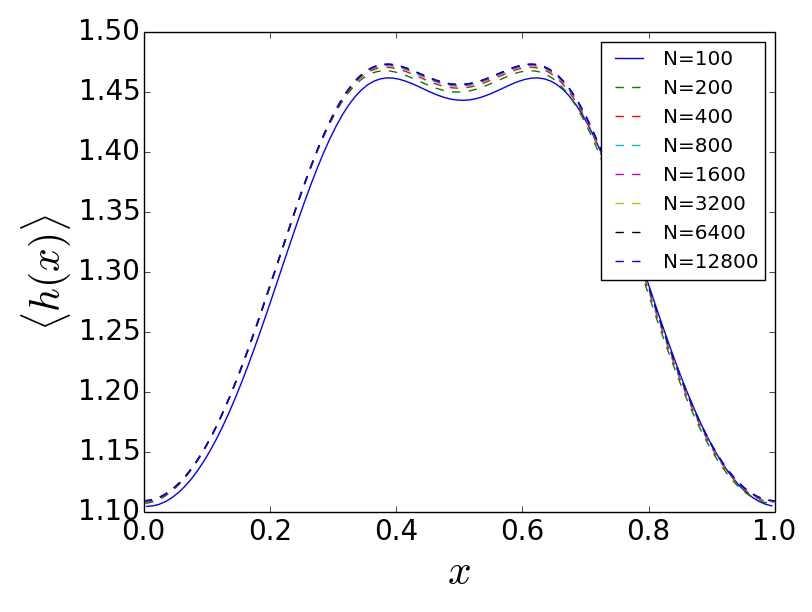}
  \caption{{\bf Convergence of interpolation scheme for different final grid-spacings.}
    We start with a ground-state on a grid of $N=100$ sites per unit-cell and interpolate
    it, using spline interpolation, to different final grids (see legend).
    The upper panel (from left to right) shows the total particle number 
    $\int_0^L\braket{n(x)}dx$, the total 
    interaction energy $\int_0^L \braket{[\psi^{\dagger}(x)]^2\psi^2(x)} dx$
    and total energy $\int_0^L\braket{h(x)}dx$ (see main text)
    as a function of grid spacing $\eps$.
    Each shows linear dependence on $\eps$. 
    Lower panel: spatially resolved $\braket{n(x)},\braket{[\psi^{\dagger}(x)]^2\psi^2(x)}$ 
    and $\braket{h(x)}$ for different number $N$ of 
    final lattice points per unit-cell.}\label{fig:epsconv}
\end{figure*}

\begin{figure}
  \includegraphics[width=1.0\columnwidth]{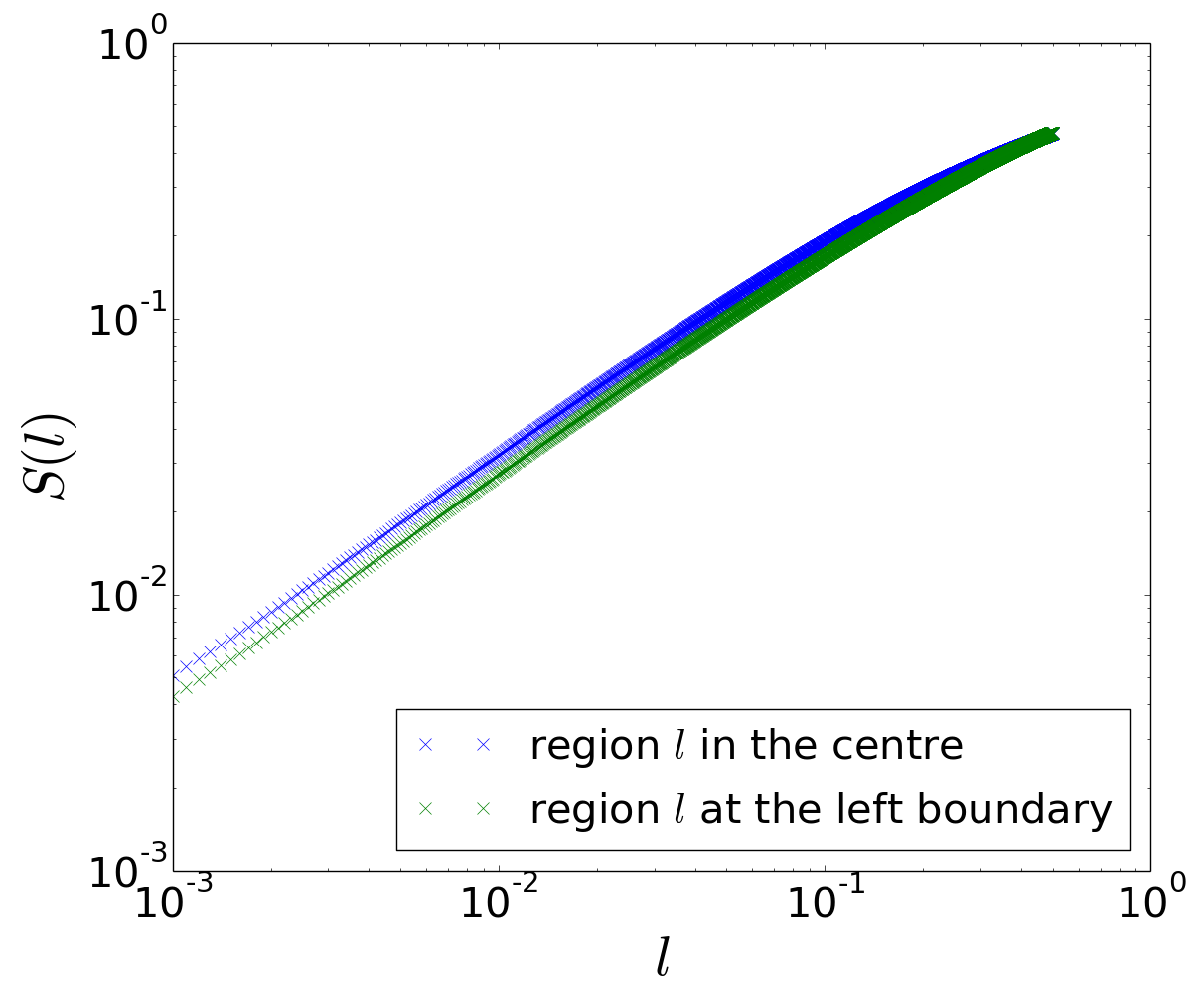}
  \caption{{\bf Entanglement entropy $\bm{S(l)}$ as a function of length $\bm{l}$}.
    Entanglement entropy $S(l)$ of region of length $l$ in the ground-state of
    \Eq{eq:Ham_eps2} for $D=16,\mu_0=-0.5,V_0=-1.0,g=1.0,L=1.0$, 
    and $N=10^4$ lattice points per unit-cell. We show $S(l)$ for two different regions $l$: 
    blue crosses show $S(l)$ as a function of $l$ for $l$ starting at the center of the unit-cell, 
    and green crosses show $S(l)$ for $l$ starting close to the left boundary. 
    For the latter case we see that $S(l)$ departs from the linear behaviour at a larger 
    length scale than for $l$ in the center of the unit-cell.
    This fact could be used devise non-homogeneous discretization schemes for MPS 
    that use a varying grid spacing based on this length scale. Note that the slope
    of both curves is very similar, which could suggest a universal short lengths scale behaviour 
    of the entanglement entropy.}\label{fig:entropy2}
\end{figure}

\section{Shifting of the orthogonality center and regauging}\label{app:shift}
In the following we describe how to shift the orthogonality center of an MPS.
For the discrete lattice case presented in this manuscript, we use 
a QR decomposition on the tensors $A^{[i]}$ to shift the orthogonality center.
However, to keep the cMPS form explicit, that is
keep matrices in the form \Eq{eq:disccmps_final}, 
we use an approach which is slightly different
from the standard procedure. 
The generic form of an MPS after a local optimization step at site $i+1$ is
\begin{align*}
\cdots\left(
  \begin{array}{c}
    \1+\eps Q^l(x_i)\\
    \sqrt{\eps}R^l(x_i)
  \end{array}
\right)\left(
  \begin{array}{c}
    V(x_{i+1})\\
    \sqrt{\eps}R(x_{i+1})
  \end{array}
\right)\left(
  \begin{array}{c}
    \1+\eps Q^r(x_{i+2})\\
    \sqrt{\eps}R^r(x_{i+2})
  \end{array}
\right)\cdots
\end{align*}
where superscripts $l/r$ indicate left or right orthonormal matrices 
\cite{ganahl_continuous_2017-1,haegeman_calculus_2013}.
Pulling out $V(x_{i+1})$ yields
\begin{align*}
\cdots\left(
  \begin{array}{c}
    \1+\eps Q^l(x_i)\\
    \sqrt{\eps}R^l(x_i)
  \end{array}
\right)\left(
  \begin{array}{c}
    \1\\
    \sqrt{\eps}R(x_{i+1})[V(x_{i+1})]^{-1}
  \end{array}
\right)V(x_{i+1})\\
\times\left(
  \begin{array}{c}
    \1+\eps Q^r(x_{i+2})\\
    \sqrt{\eps}R^r(x_{i+2})
  \end{array}
\right)\cdots.
\end{align*}
Using a QR decomposition, the center term in brackets is then orthonormalized:
\begin{align*}
\left(
  \begin{array}{c}\1\\ 
    \sqrt{\eps}R(x_{i+1})[V(x_{i+1})]^{-1}
  \end{array}
\right)
\stackrel{QR}{\ra}
\left(
  \begin{array}{c}\1+\eps Q^{l}(x_{i+1})\\ 
    \sqrt{\eps}R^{l}(x_{i+1})
  \end{array}
\right)C(x_{i+1}).
\end{align*}
The matrix product $C(x_{i+1})V(x_{i+1})$ is then normalized and absorbed into the
matrices at site $i+2$. A similar procedure is applied to shift the center site to the left.
Crucially, we find that the standard QR algorithm in {\it numpy},
equipped with a phase-convention for the diagonal of $R$ (not be confused with
the cMPS matrix), preserves
the cMPS form of the tensors.
Note that due to the gauge freedom of (c)MPS, the matrices $R(x_i)$ and $Q(x_i)$ do
not need to be continuous, even in the limit $\eps\ra 0$. 
For a random initial state, we observe that during the 
local updates, matrices $Q(x_i)$ and $R(x_i)$ develop discontinuities. These jumps, however, 
become smaller and eventually disappear as one approaches convergence, resulting in
smooth matrix functions $R(x_i),Q(x_i)$ within the unit-cell.

However, the orthonormalization procedure described above 
introduces a non-trivial gauge change as one moves from one side of the unit-cell to the other. 
As a result, the matrices $Q(x_i), R(x_i)$ do not trivially connect back to themselves at the end 
of the unit-cell, even if they originally did.
For the purpose of interpolating the matrices a smooth gauge is favourable.
To this end, we first canonize the unit-cell MPS by calculating the left and right
eigen-vectors $l$ and $r$ of the unit-cell transfer operator 
\begin{align}
  T_{UC}=\sum_{\{n_i\}}\big(A_{n_1}^{[1]}\dots A_{n_N}^{[N]})\otimes {\big(\bar A_{n_1}^{[1]}\dots \bar A_{n_N}^{[N]})},
\end{align}
from which we obtain $X\equiv\sqrt{r}$ and $Y\equiv\sqrt{l}$. 
We then compute $U\lambda V=YX$, and absorb 
$V^{\dagger} X^{-1}$ into the leftmost
and $Y^{-1}U\lambda$ into the rightmost MPS matrix. Sweeping from right to left, we then 
successively orthonormalize the matrices using an SVD on the matrices 
$A_{n_i}^{[i]}=U^{[i]}\lambda^{[i]}V_{n_i}^{[i]}$ (indices $[i]$ in square brackets are position labels). 
To fix the gauge freedom of 
the SVD, we fix the diagonal of the right isometry $V_{n_i}^{[i]}$ to be real and positive. 
In a successive sweep from the left to the right boundary we fix the diagonal of the matrix 
$U_{n_i}^{[i]}$ in $A^{[i]}_{n_i}=U^{[i]}_{n_i}\lambda^{[i]} V^{[i]}$ to be real and positive.
The right boundary matrix $V^{[N]}$
will then be a diagonal unitary matrix  which can be distributed over the unit-cell:
\begin{align}
  e^{i\mathcal{H}L}\equiv V^{[N]}=\prod_i^Ne^{i\eps \mathcal{H}}\equiv \prod_i^NG
\end{align}
where $L$ is the length of the unit-cell, $N$ is the number of lattice points, $\eps=L/N$, and 
$\mathcal{H}$ is a hermitian operator. $G$ is close to 
the identity and can be written as $G=\1+\eps \tilde{\mathcal{H}}$, with $\tilde{\mathcal{H}}$ 
an almost hermitian 
operator. $V^{[N]}$ can now be equally distributed over
all matrices by the following two steps: First, at every site $i\in \{1\dots N\}$ transform the 
matrices $Q(x_i), R(x_i)$ according to
\begin{align}
  Q(x_i)\la (G^{\dagger})^{i-1}Q(x_i)G^{i-1}\nonumber\\
  R(x_i)\la (G^{\dagger})^{i-1}R(x_i)G^{i-1}.\label{eq:sym1}
\end{align}
Second, transform each matrix 
\begin{align}
  Q(x_i)\la Q(x_i)+\tilde{\mathcal{H}} +\eps Q(x_i)\tilde{\mathcal{H}}\nonumber \\
  R(x_i)\la  R(x_i) +\eps R(x_i)\tilde{\mathcal{H}}\label{eq:sym2}
\end{align}
The resulting matrices $Q(x_i), R(x_i)$ will connect smoothly back to themselves.
\section{Continuous limit of fine-graining}\label{app:cont}
In this section we explore in more detail how interpolation affects 
different observables. In particular, we will focus on the observables
$\braket{n(x)},\braket{[\psi^{\dagger}(x)]^2\psi^2(x)}$ and $\braket{h(x)}$ (see also main text). 
We first obtain the ground-state of \Eq{eq:Ham_eps2} for $D=16,\mu_0=-0.5,V_0=-1.0,g=1.0$ 
and $N=100$ sites per unit-cell. We then 
interpolate the corresponding matrices $Q(x_i),R(x_i)$ to different finer grids 
with $N=200,400,800,1600,3200,6400$ and 12800 sites per unit-cell, and calculate the 
observables $\braket{n(x)},\braket{[\psi^{\dagger}(x)]^2\psi^2(x)}$ 
and $\braket{h(x)}$ after normalizing the state. 
The results are shown in \Fig{fig:epsconv}. The upper panel
shows $\int_0^L dx\braket{n(x)},\int_0^L dx \braket{[\psi^{\dagger}(x)]^2\psi^2(x)}$ 
and $\int_0^L dx \braket{h(x)}$, 
integrated over the unit-cell, as a function of $\eps$. We observe a clear linear dependence 
on $\eps$ as we decrease the lattice spacing. 
The lower panel shows the spatially resolved quantities for different $N$.

\section{Short range entanglement}\label{app:ent}
In \Fig{fig:entropy} in the main text we show how the entanglement entropy $S(l)$ of a region of size $l$ in the center of the unit-cell (
for parameters $D=16,\mu_0=-0.5,V_0=-1.0,g=1.0$, see \Fig{fig:diffgrids}). As pointed out in the main text we observe a transition from a power law increase for small $l$ 
to a less pronounced increase for larger $l$ at a certain length scale $\xi\approx 0.05$. \Fig{fig:entropy2} shows $S(l)$ for two different locations of the region $l$, namely in the center of the unit-cell
and close to the left boundary of the unit-cell (with $l$ growing to the right). The plot illustrates that $\xi$ varies along the unit-cell, which could be used as basis for determining grid spacings 
for non-homogeneous discretizations as e.g. shown in \Fig{fig:finegrain} in the main text.

\section{Optimization of periodic MPS on a lattice}\label{app:opt}
In this section we give a detailed description of the optimization method used
to obtain the ground-state of a Hamiltonian $H(\eps_{\alpha})$ (see main text).
We employ the infinite DMRG (iDMRG) method as introduced by McCulloch
\cite{mcculloch_infinite_2008}, extended to the case of a large unit-cell with $N\gg 2$ sites.
To make the document self-contained, we give in the following
a detailed explanation of the algorithm.
For the sake of simplicity, we restrict ourselves 
to the case of $N=4$. Results for arbitrary $N$
follow straightforwardly.

Before going into more detail, we first introduce some necessary diagrammatic 
notations for MPS. In the following we consider periodic MPS with $N$ tensors
$\{A^{[1]},\dots,A^{[N]}\}$ per unit-cell. Such a state is pictorially represented as 
\begin{equation}\label{eq:canonical}
  \ket{\Psi}=\cdots\raisebox{-0.3cm}{\includegraphics[width=0.35\columnwidth]{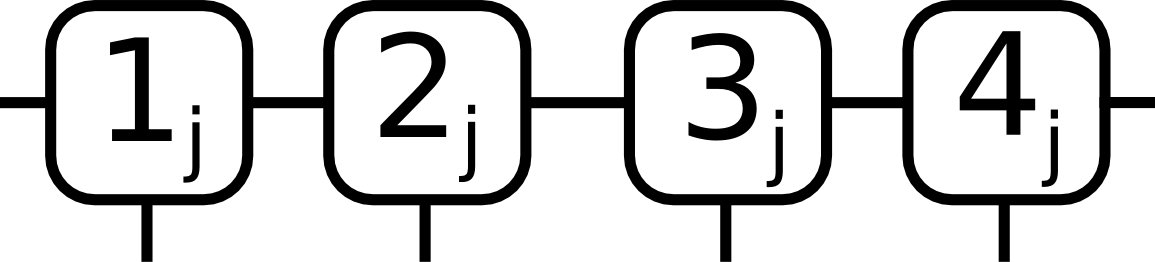}}\cdots,
\end{equation}
where dots indicate an infinite repetition of the unit-cell. Subscripts
$j\in \mathbbm{Z}$ label the unit-cell. For brevity, we will in 
the following omit dots and subscripts $j$ if no confusion can arise.
Using standard regauging techniques 
\cite{orus_infinite_2008} (see Appendix \ref{app:shift}), the state can 
be regauged into left and right orthogonal forms
\begin{align}\label{eq:lr_orthoMPS}
  \ket{\Psi}=\raisebox{-0.3cm}{\includegraphics[width=0.35\columnwidth]{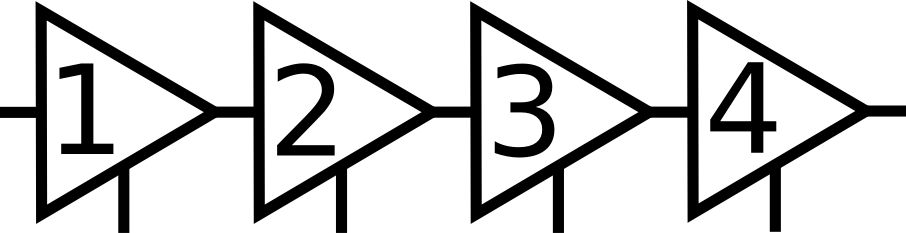}}
  =\raisebox{-0.3cm}{\includegraphics[width=0.35\columnwidth]{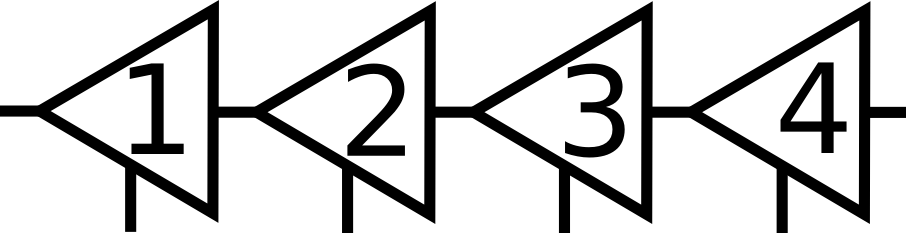}}.
\end{align}
The left and right orthogonal tensors 
\raisebox{-0.25cm}{\includegraphics[width=0.08\columnwidth]{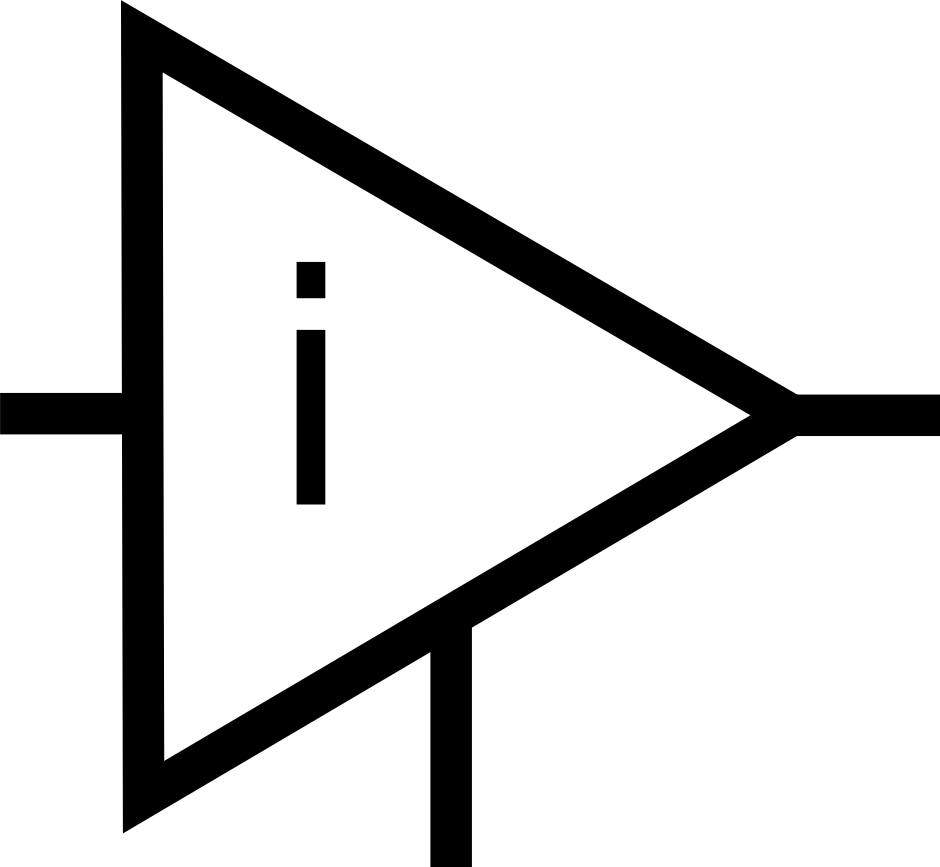}} and 
\raisebox{-0.25cm}{\includegraphics[width=0.08\columnwidth]{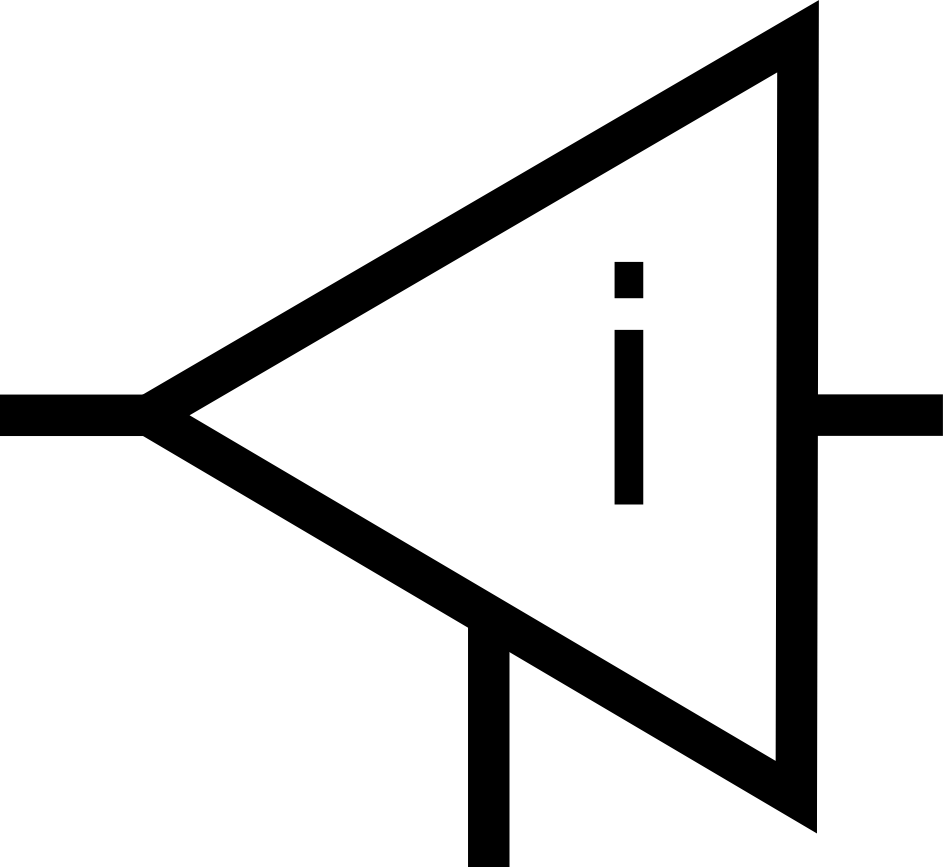}}
obey 
\begin{align}\label{eq:lr_ortho}
  \raisebox{-0.7cm}{\includegraphics[width=0.2\columnwidth]{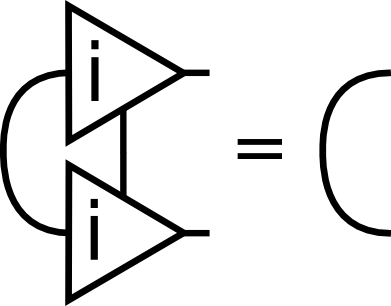}}\\
  \raisebox{-0.7cm}{\includegraphics[width=0.2\columnwidth]{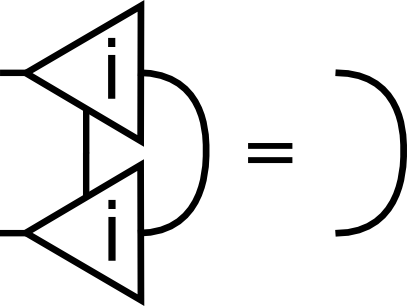}}
\end{align}
(see also \Eq{eq:leftgauge}).

Of particular use is the so-called canonical form \cite{orus_infinite_2008} 
of an MPS, given by the following decomposition:
\begin{equation}\label{eq:canonical}
  \ket{\Psi}=\raisebox{-0.35cm}{\includegraphics[width=0.6\columnwidth]{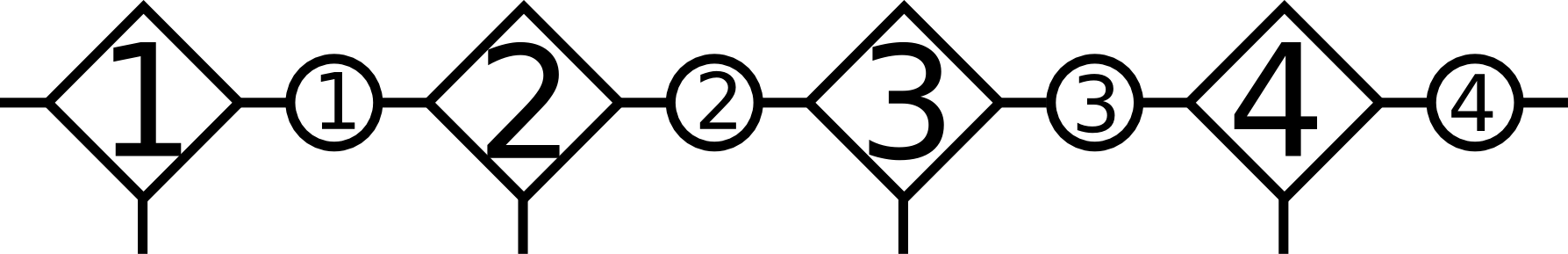}}.
\end{equation}
\raisebox{-0.14cm}{\includegraphics[width=0.11\columnwidth]{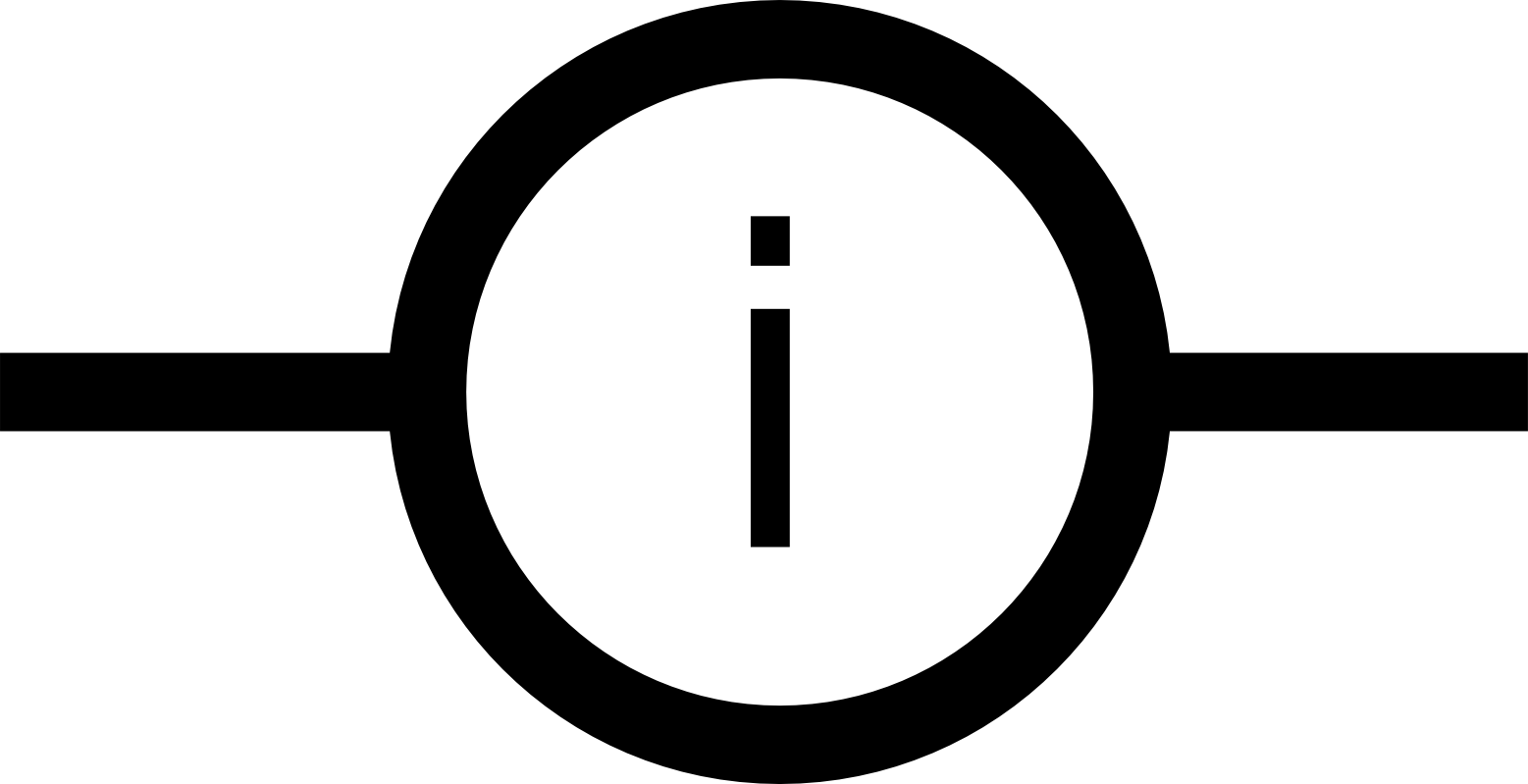}} are 
diagonal bond matrices containing the Schmidt values 
$\lambda_{\alpha}$ on bond $i$, 
and $\Gamma_{\alpha\beta}^{n,[i]}\equiv \raisebox{-0.4cm}{\includegraphics[width=0.12\columnwidth]{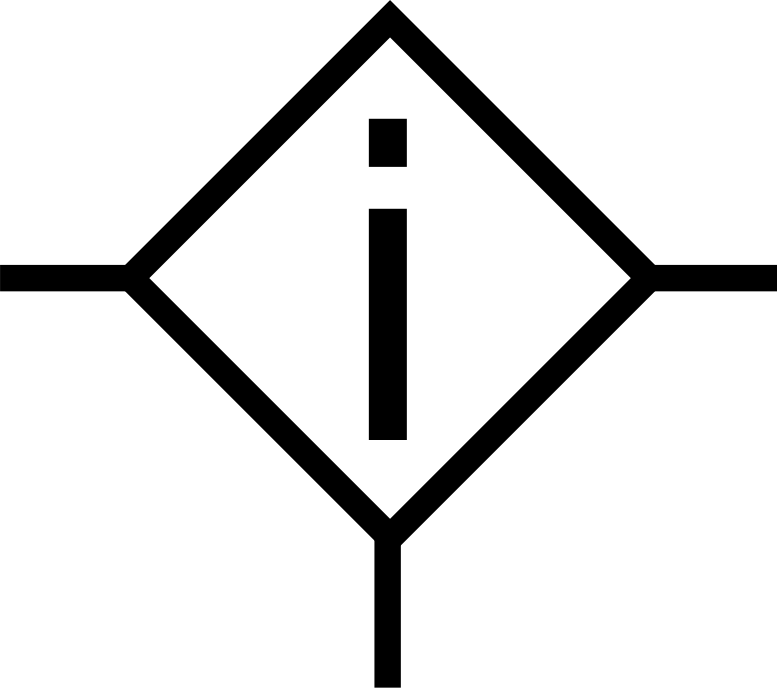}}$ is a $D\times D\times d$ tensor at site $i$. In our convention, the 
index label $i$ of a bond matrix
can be deduced from the index of the left tensor next to it.
For the sake of brevity, we will thus omit it if no confusion can arise.
We will use angle brackets to denote inverses of matrices, i.e. 
\raisebox{-0.11cm}{\includegraphics[width=0.08\columnwidth]{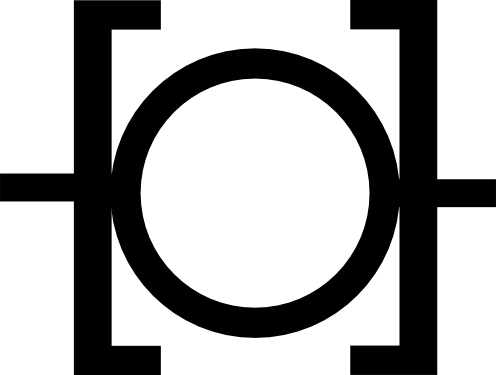}} is the inverse of \raisebox{-0.11cm}{\includegraphics[width=0.08\columnwidth]{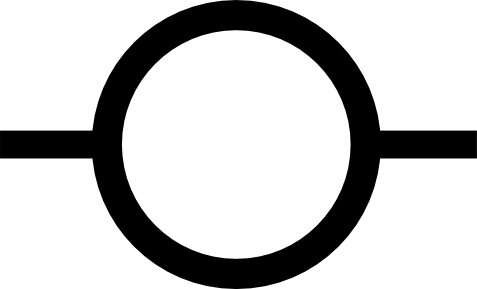}}, with 
$\raisebox{-0.17cm}{\includegraphics[width=0.4\columnwidth]{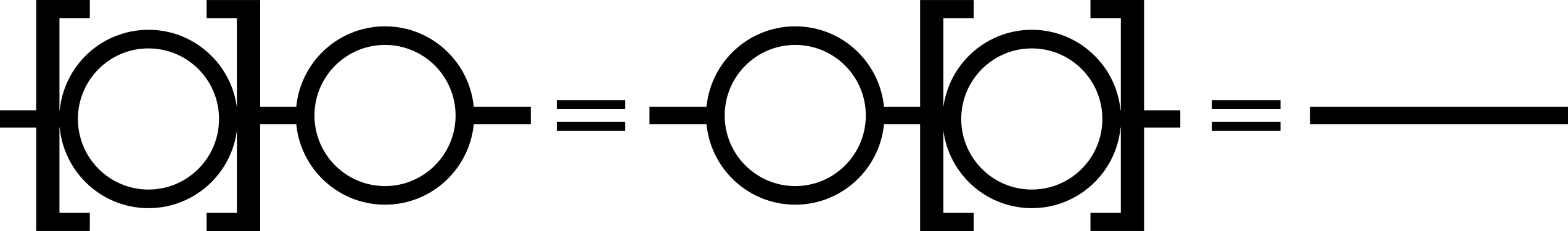}}\equiv \mathbbm{1}$.
The canonical form is particularly useful because
the left and right orthogonal form of the MPS can be easily extracted using the relations
\begin{align}
  \raisebox{-0.4cm}{\includegraphics[width=0.4\columnwidth]{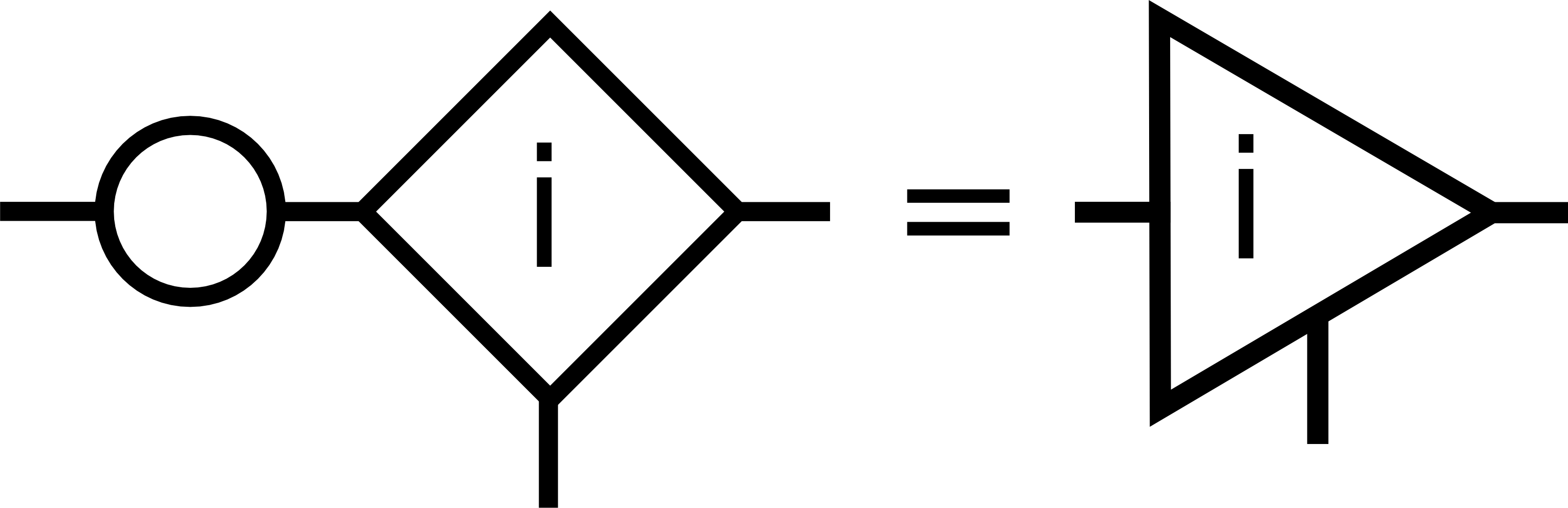}}\label{eq:merge_l}\\
  \raisebox{-0.4cm}{\includegraphics[width=0.4\columnwidth]{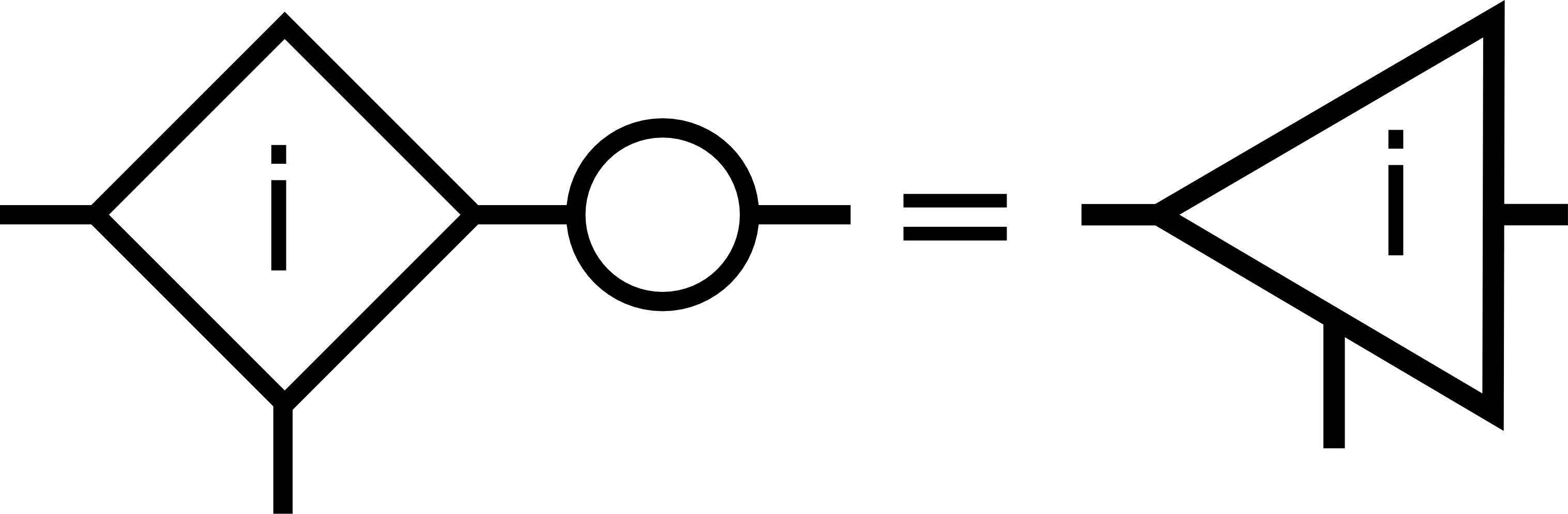}}.\label{eq:merge_r}
\end{align}
Finally, we introduce the parallel decomposition \cite{stoudenmire_real-space_2013}
\begin{align}\label{eq:parallel}
  &\ket{\Psi}=\\
  &\raisebox{-0.175cm}{\includegraphics[width=1.0\columnwidth]{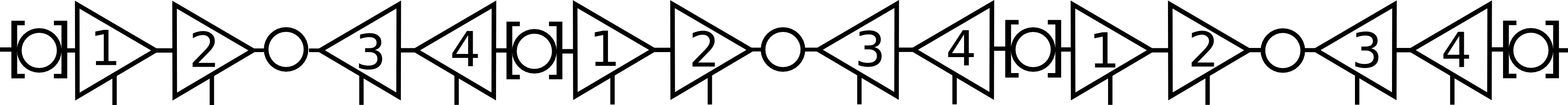}}\nonumber,
\end{align}
obtained e.g. from the canonical form by inserting identities 
\raisebox{-0.11cm}{\includegraphics[width=0.12\columnwidth]{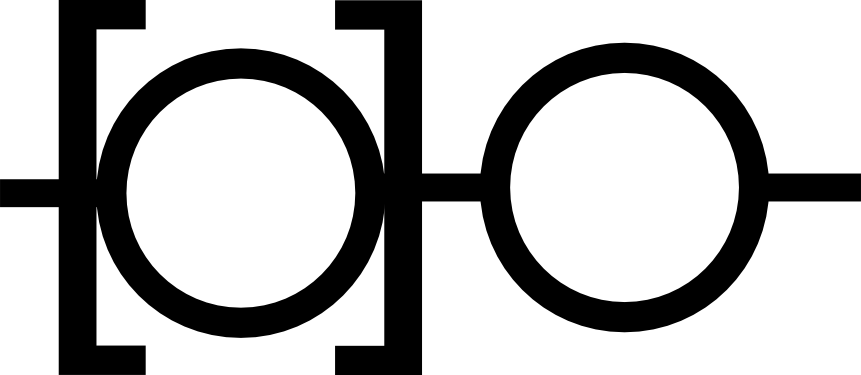}}
and using Eqs.(\ref{eq:merge_l}) and (\ref{eq:merge_r}).
Each unit-cell is here decomposed into a set of left and right orthogonal tensors, with 
matrices \raisebox{-0.11cm}{\includegraphics[width=0.08\columnwidth]{lambda_unlabeled.png}} 
and 
\raisebox{-0.11cm}{\includegraphics[width=0.08\columnwidth]{inverse.png}} between
the connection points. 
The positions of the matrices
\raisebox{-0.11cm}{\includegraphics[width=0.08\columnwidth]{lambda_unlabeled.png}} 
in \Eq{eq:parallel} are called the orthogonality centers. 
The matrices \raisebox{-0.11cm}{\includegraphics[width=0.08\columnwidth]{lambda_unlabeled.png}} 
can be shifted inside the unit-cell using a QR or SV decomposition,
\begin{align}\label{eq:shift}
  \raisebox{-0.4cm}{\includegraphics[width=0.2\columnwidth]{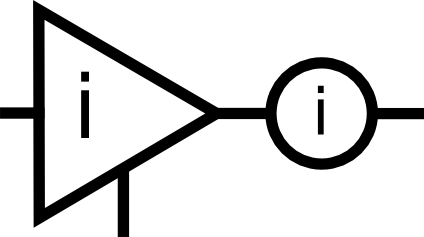}}=
\raisebox{-0.4cm}{\includegraphics[width=0.2\columnwidth]{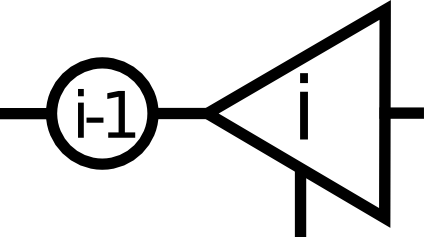}}
\end{align}
In particular, they can be shifted to the unit-cell boundaries, where they
can be absorbed into \raisebox{-0.11cm}{\includegraphics[width=0.08\columnwidth]{inverse.png}}.
A defining property of the parallel decomposition is that
this absorption results in 
$\raisebox{-0.19cm}{\includegraphics[width=0.17\columnwidth]{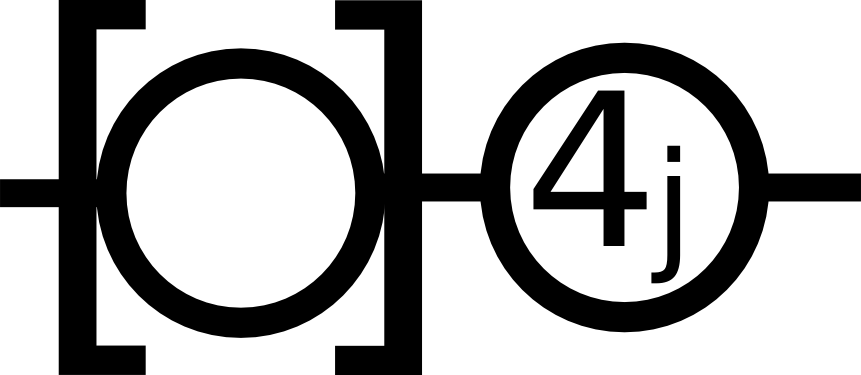}}=U_l$ and  
$\raisebox{-0.19cm}{\includegraphics[width=0.17\columnwidth]{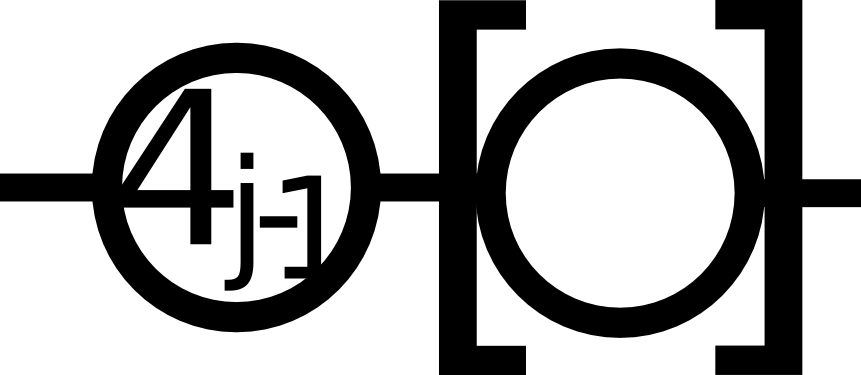}}=U_r$, 
with {\it unitary} matrices $U_l,U_r$. This implies that within each unit-cell local truncation
of the MPS bond-dimension can be done optimally.
The parallel decomposition is also central to the parallel version of DMRG on finite lattices 
\cite{stoudenmire_real-space_2013}. 

\begin{figure}
  \includegraphics[width=1.0\columnwidth]{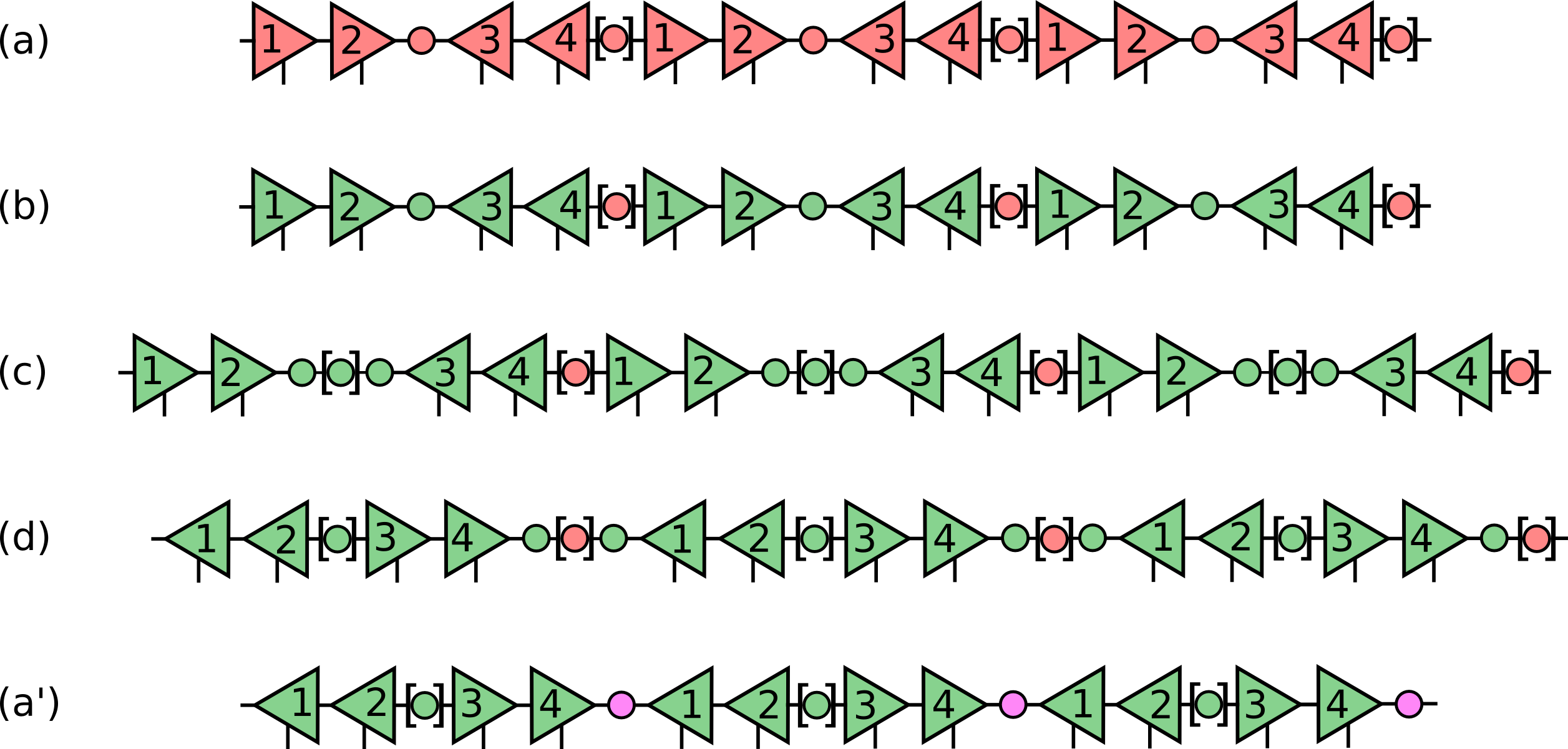}
  \caption{{\bf Outline of the first steps of an iDMRG calculation.} 
    (a) Initial state in parallel 
    decomposition. (b) State after all unit-cell tensors have been updated. (c) An identity 
    resolution is inserted at the center of each unit-cell. (d) The tensors 
    \protect\raisebox{-0.14cm}{\protect\includegraphics[width=0.1\columnwidth]{lambda_unlab_green.png}}
    are moved  to the left respectively right unit-cell boundaries. (a') The two matrices 
    \protect\raisebox{-0.18cm}{\protect\includegraphics[width=0.2\columnwidth]{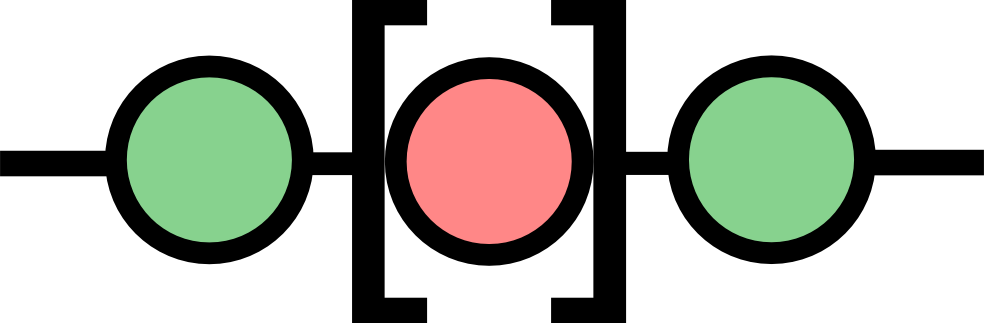}} are 
    contracted into 
    \protect\raisebox{-0.12cm}{\protect\includegraphics[width=0.1\columnwidth]{lambda_unlab_pink.png}}.
}
  \label{fig:idmrg}
\end{figure}

In the following we explain the different steps involved in the 
ground-state optimization of a periodic  Hamiltonian $H$ in the thermodynamic limit.
At each step of the optimization, the state is represented by $N$ unit-cell
tensors $\{A^{[1]},\dots,A^{[N]}\}$ which can be patched into an infinite MPS.
\Fig{fig:idmrg} gives an overview of the first five steps of the algorithm. 
Different colours of matrices and tensors indicate an update.
In \Fig{fig:idmrg} (a), 
the MPS is initialized with $N$ random tensors per unit-cell and brought into 
the parallel decomposition (red tensors and matrices in \Fig{fig:idmrg} (a)).
The initial matrices 
\raisebox{-0.14cm}{\includegraphics[width=0.1\columnwidth]{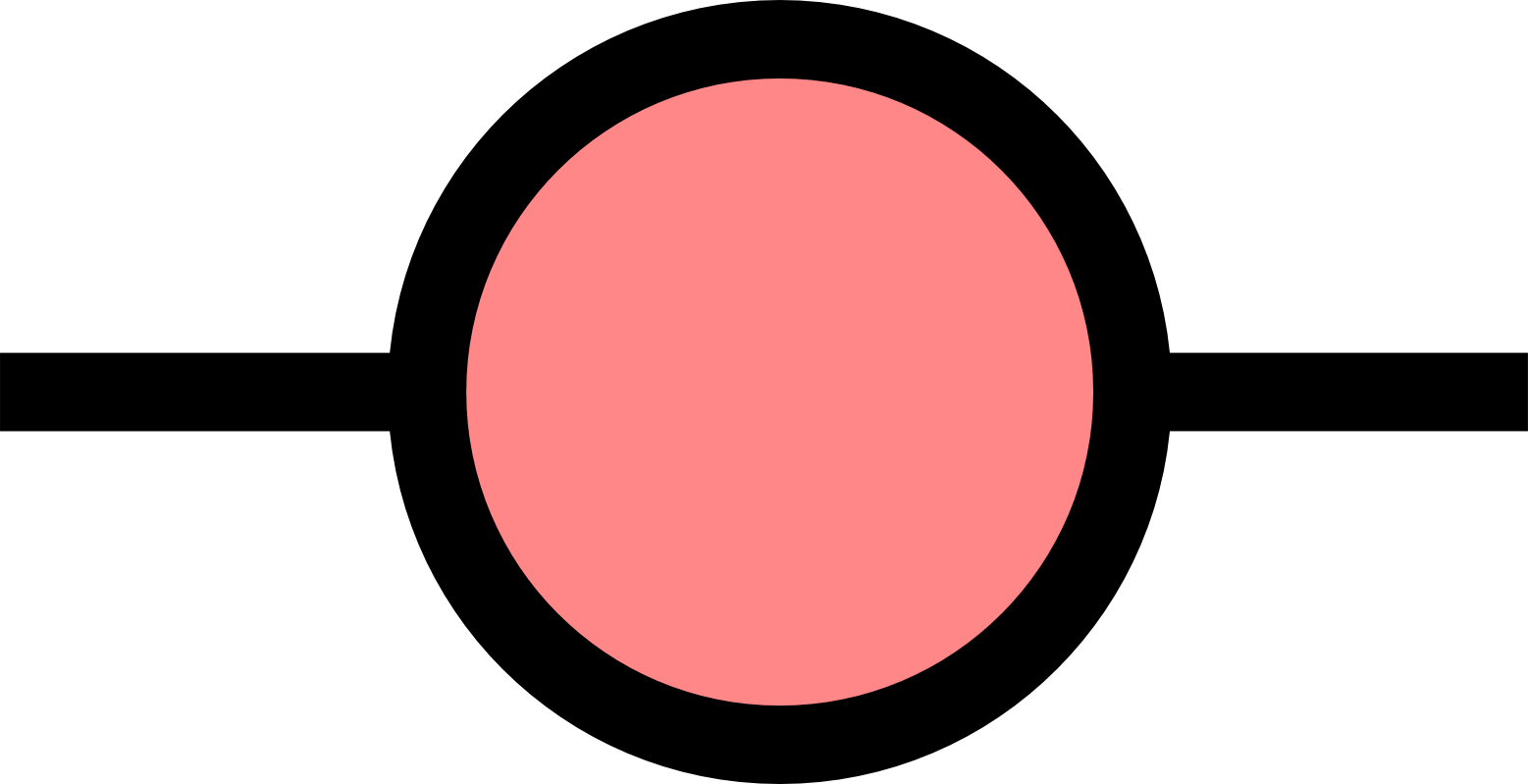}}
are diagonal, containing the Schmidt values. The tensors of each unit-cell are
then updated using standard variational optimization techniques for MPS (see below for
details),
see \Fig{fig:idmrg} (b). The updated objects are shown in green, and updates are 
identical between each unit-cell. 
Note that the matrices 
\raisebox{-0.14cm}{\includegraphics[width=0.1\columnwidth]{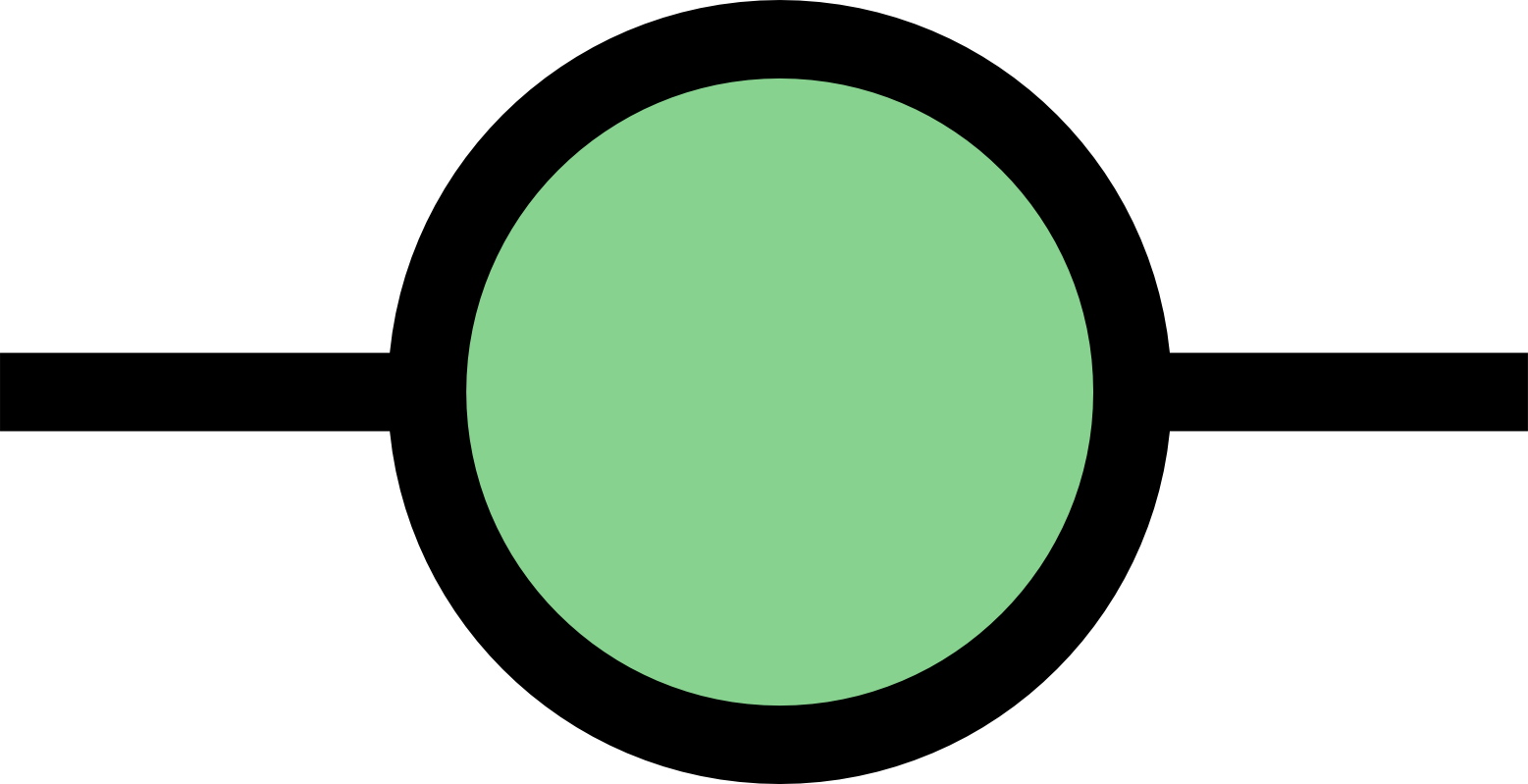}} 
are dense. Between each unit-cell resides a matrix
\raisebox{-0.26cm}{\includegraphics[width=0.1\columnwidth]{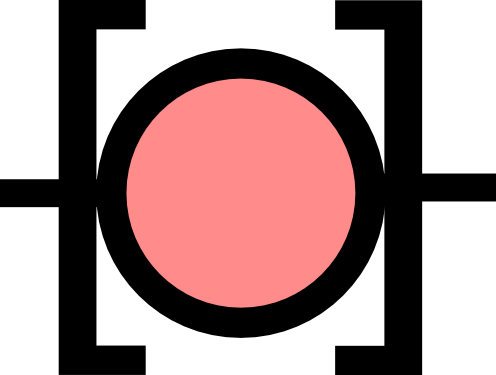}}
from the previous step.
(c) A resolution of identity is inserted. Using the QR or SV decomposition, 
the two matrices
\raisebox{-0.14cm}{\includegraphics[width=0.1\columnwidth]{lambda_unlab_green.png}}
at the orthogonality centers
are shifted left, respectively right (see \Eq{eq:shift}), until the boundary matrix
\raisebox{-0.26cm}{\includegraphics[width=0.1\columnwidth]{inverse_unlab_red.png}}
is reached (\Fig{fig:idmrg} (d)). The three matrices 
\raisebox{-0.18cm}{\includegraphics[width=0.26\columnwidth]{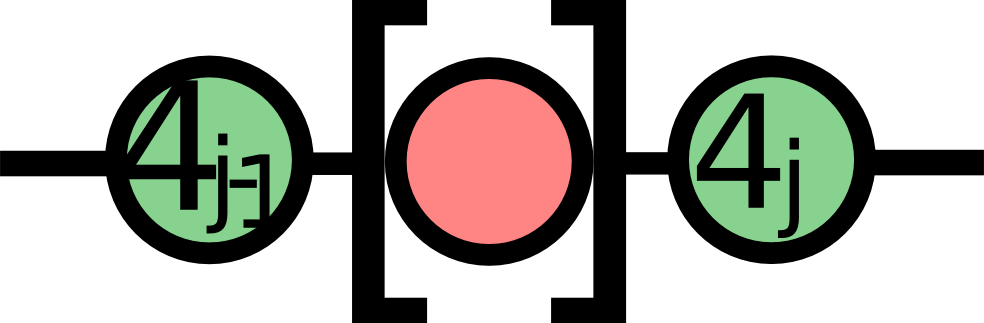}} are contracted
into a new matrix 
\raisebox{-0.12cm}{\includegraphics[width=0.1\columnwidth]{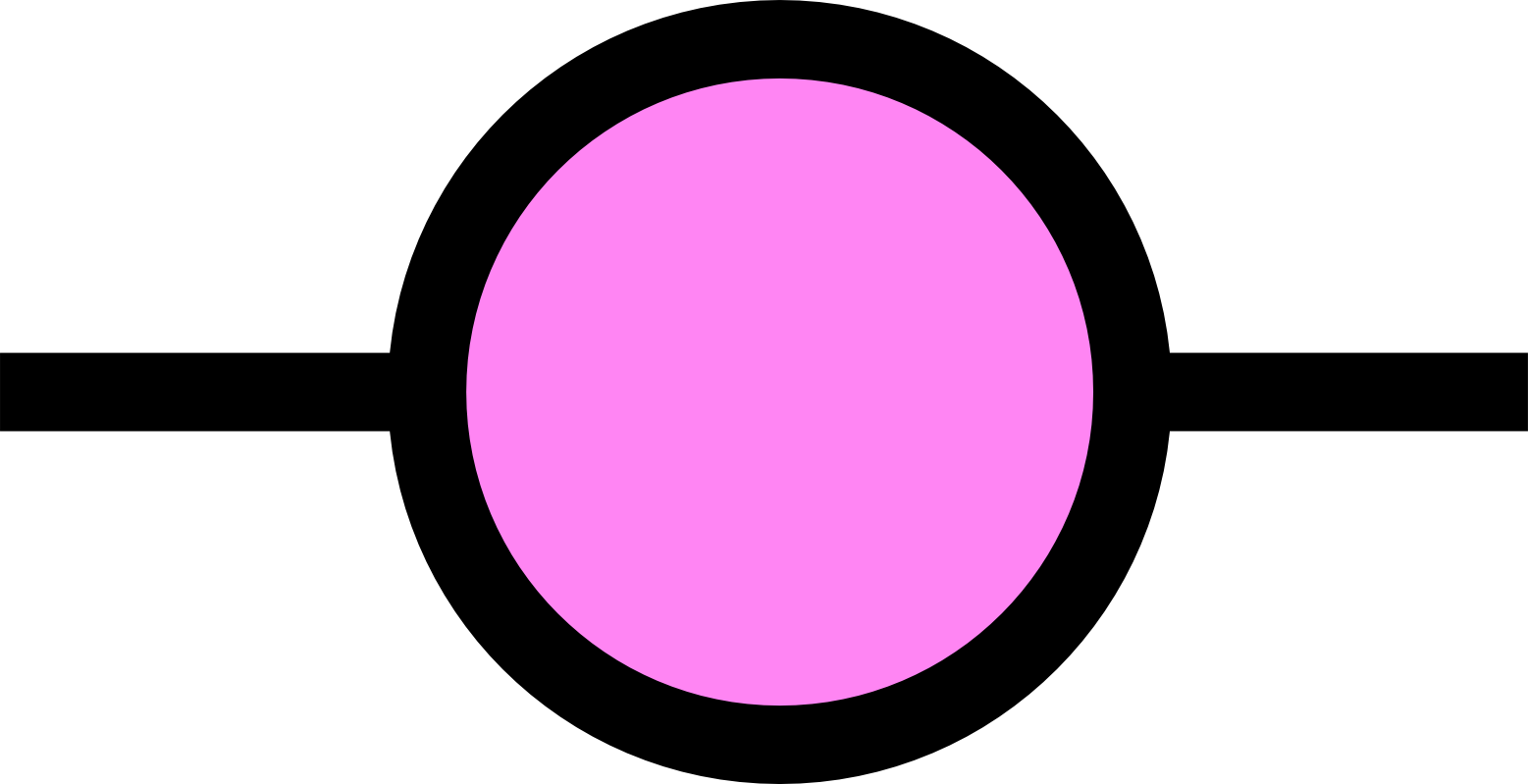}} (see 
\Fig{fig:idmrg} (a')). The 
state is now in what we call {\it quasi-parallel decomposition} with 
respect to a new unit-cell that
has been shifted by $N/2$. Quasi-parallel refers to the fact that
\raisebox{-0.16cm}{\includegraphics[width=0.14\columnwidth]{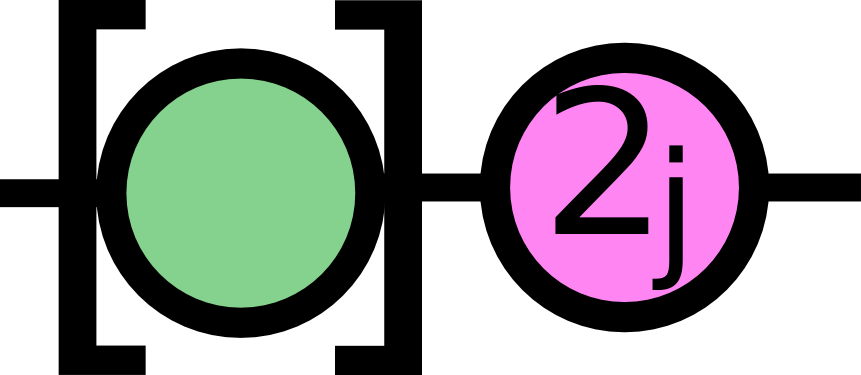}} and 
\raisebox{-0.18cm}{\includegraphics[width=0.14\columnwidth]{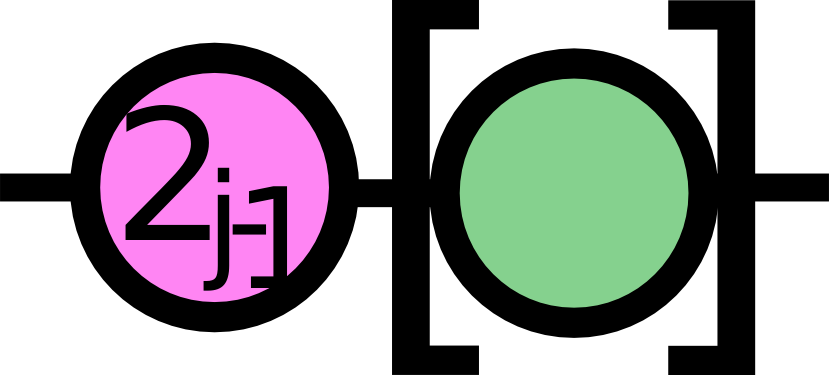}} 
are in general both non-unitary. Note that during the optimization, the matrices will
converge to unitary matrices. The degree of unitarity of the matrices 
\raisebox{-0.16cm}{\includegraphics[width=0.14\columnwidth]{inverse_1_lab.png}} and 
\raisebox{-0.18cm}{\includegraphics[width=0.14\columnwidth]{inverse_2_lab.png}} 
can be used as an alternative measure of convergence.
This completes the first iteration of the algorithm.
In the next iteration, the tensors of the new unit-cell
$\{N/2+1,\dots N, 1\dots N/2\}$ are then going to be updated. 
Over time, several slightly different variants of how to proceed 
at this point have been developed. We will below explain three different approaches, which
we find the most convenient one for our case.
Before, however, let us first explain in more detail the individual steps in \Fig{fig:idmrg}.

We start with the update from (a) to (b).
Within each unit-cell, we use standard variational MPS optimization 
\cite{white_density_1992,schollwock_density-matrix_2011} 
for a finite system to update the tensors. For a given, fixed unit-cell,
the updated tensors $\{\tilde A^{[1]},\dots, \tilde A^{[N]}\}$ are found
from minimizing the expectation value of the energy with respect to the unit-cell tensors, i.e.
\begin{align}
  \{\tilde A^{[1]},\dots, \tilde A^{[N]}\}=
  \underset{\{A^{[1]},\dots, A^{[N]}\}}{\text{argmin}}\frac{\braket{\Psi|H|\Psi}}{\braket{\Psi|\Psi}},
\end{align}
while keeping all tensors {\it outside} the unit-cell fixed.
Sweeping from left to right, i.e. $i=1\dots N$, the minimization proceeds by optimizing one 
tensor $A^{[i]}$ at a time. 
For each site $i$, this minimization leads to a sparse eigen-value equation with an effective
Hamiltonian $H_{eff}$. 
For example, for the initial optimization of site $i=1$, $H_{eff}$ has the form
\begin{align}\label{eq:Heff}
  H_{eff}=\cdots\raisebox{-0.78cm}{\includegraphics[width=0.4\columnwidth]{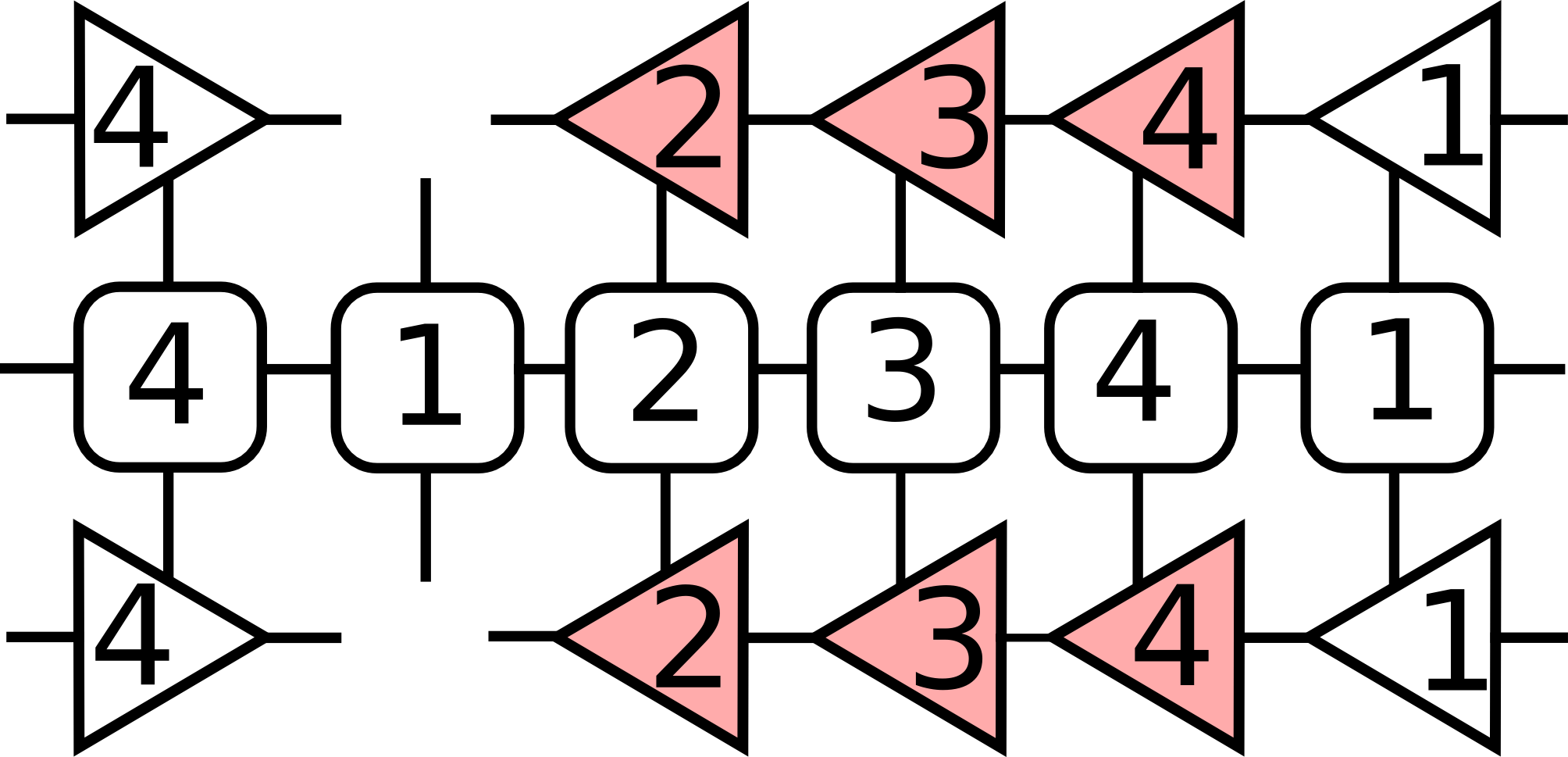}}\cdots\;.
\end{align}
The unit-cell tensors are highlighted in red,
and we have used a Matrix Product Operator representation 
\cite{schollwock_density-matrix_2011} (see Appendix \ref{app:contract} below)
of the Hamiltonian \Eq{eq:Ham_eps2},
\begin{align}
  H=\cdots\raisebox{-0.47cm}{\includegraphics[width=0.4\columnwidth]{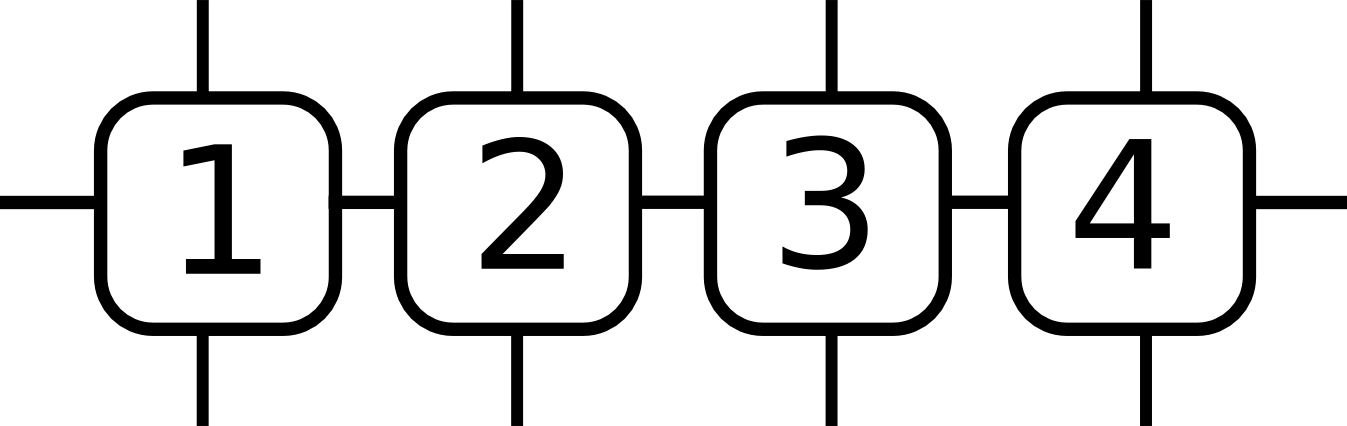}}\cdots\;.
\end{align}
$C_{\alpha\beta}^{nn'}\equiv $
\raisebox{-0.45cm}{\includegraphics[width=0.12\columnwidth]{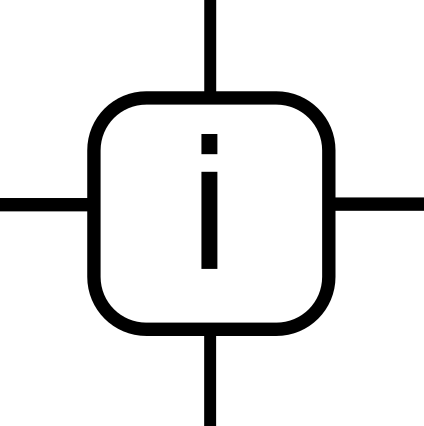}} is a sparse four-leg
tensor (see Appendix \ref{app:contract} below).
Using standard contraction techniques for MPS (Appendix \ref{app:contract})
the infinite tensor network in \Eq{eq:Heff} can be contracted,
resulting in the expression 
\begin{align}\label{eq:Heff_2}
  H_{eff}=\raisebox{-0.88cm}{\includegraphics[width=0.45\columnwidth]{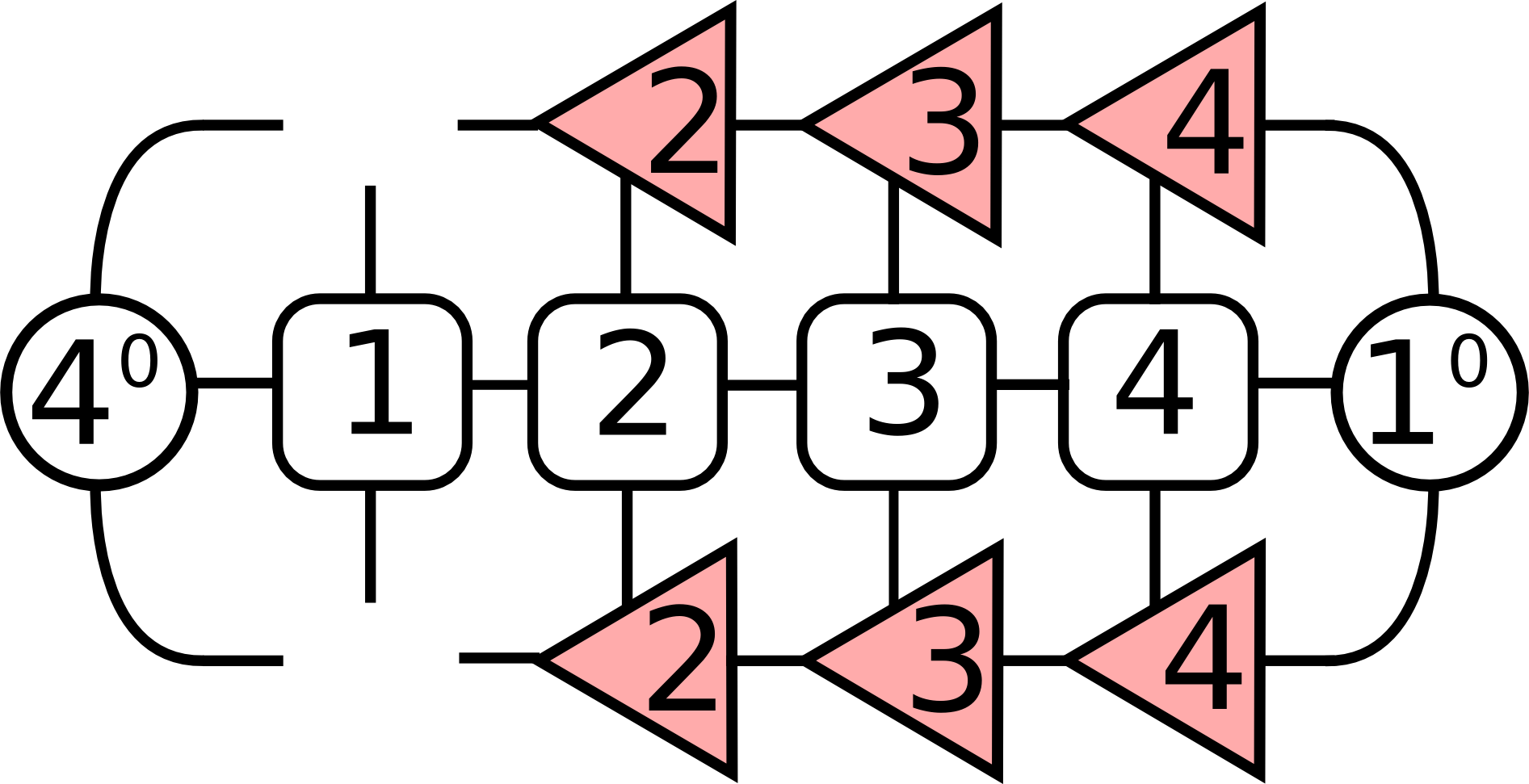}}\;.
\end{align}
We call \raisebox{-0.32cm}{\includegraphics[width=0.05\columnwidth]{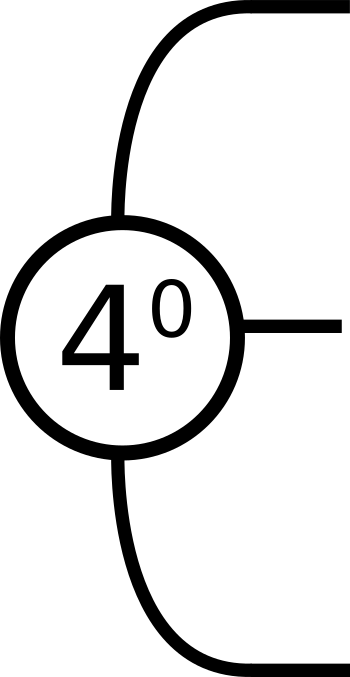}} and 
\raisebox{-0.32cm}{\includegraphics[width=0.05\columnwidth]{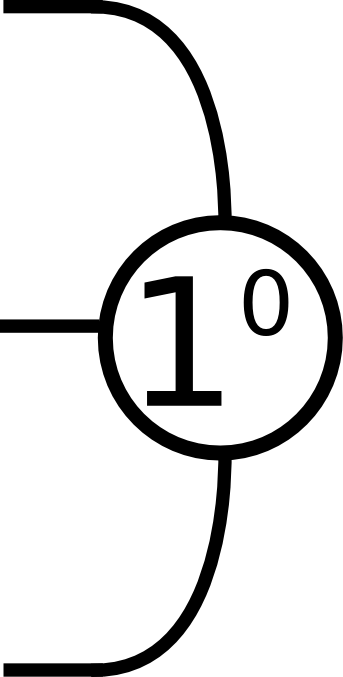}} the
effective left and right {\it unit-cell environments}.
The labels indicate the index of the last MPS tensor that has been contracted into 
\raisebox{-0.32cm}{\includegraphics[width=0.05\columnwidth]{L_0.png}} and
\raisebox{-0.32cm}{\includegraphics[width=0.05\columnwidth]{R_0.png}}. The superscripts
are the iDMRG iteration number.
\raisebox{-0.32cm}{\includegraphics[width=0.05\columnwidth]{L_0.png}}
is thus the left environment of tensor $i=1$ at iteration $q=0$.
(and likewise for \raisebox{-0.32cm}{\includegraphics[width=0.05\columnwidth]{R_0.png}}).
An updated $\tilde A^{[i]}\equiv\raisebox{-0.2cm}{\includegraphics[width=0.1\columnwidth]{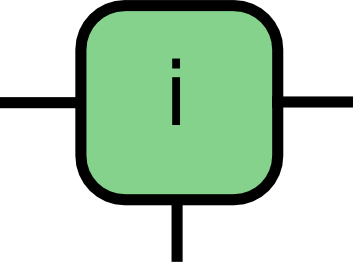}}$ 
is obtained from the (properly reshaped and normalized) 
lowest eigen-vector of $H_{eff}$. Using a QR or SV decomposition, it is decomposed into
\begin{align}
  \raisebox{-0.4cm}{\includegraphics[width=0.4\columnwidth]{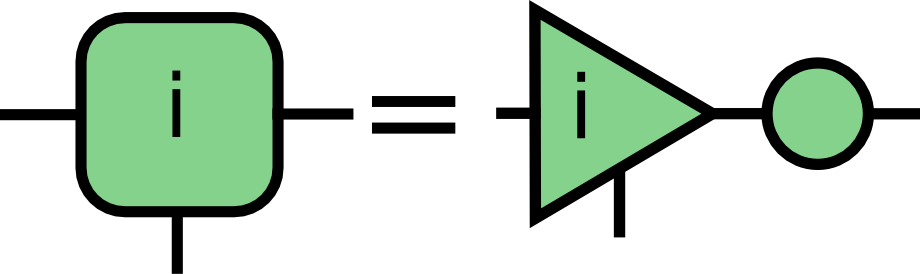}}
\end{align}
and inserted back into the MPS. One then moves one site to the right, where the above
procedure is repeated with
\begin{align}
  H_{eff}=\raisebox{-0.78cm}{\includegraphics[width=0.45\columnwidth]{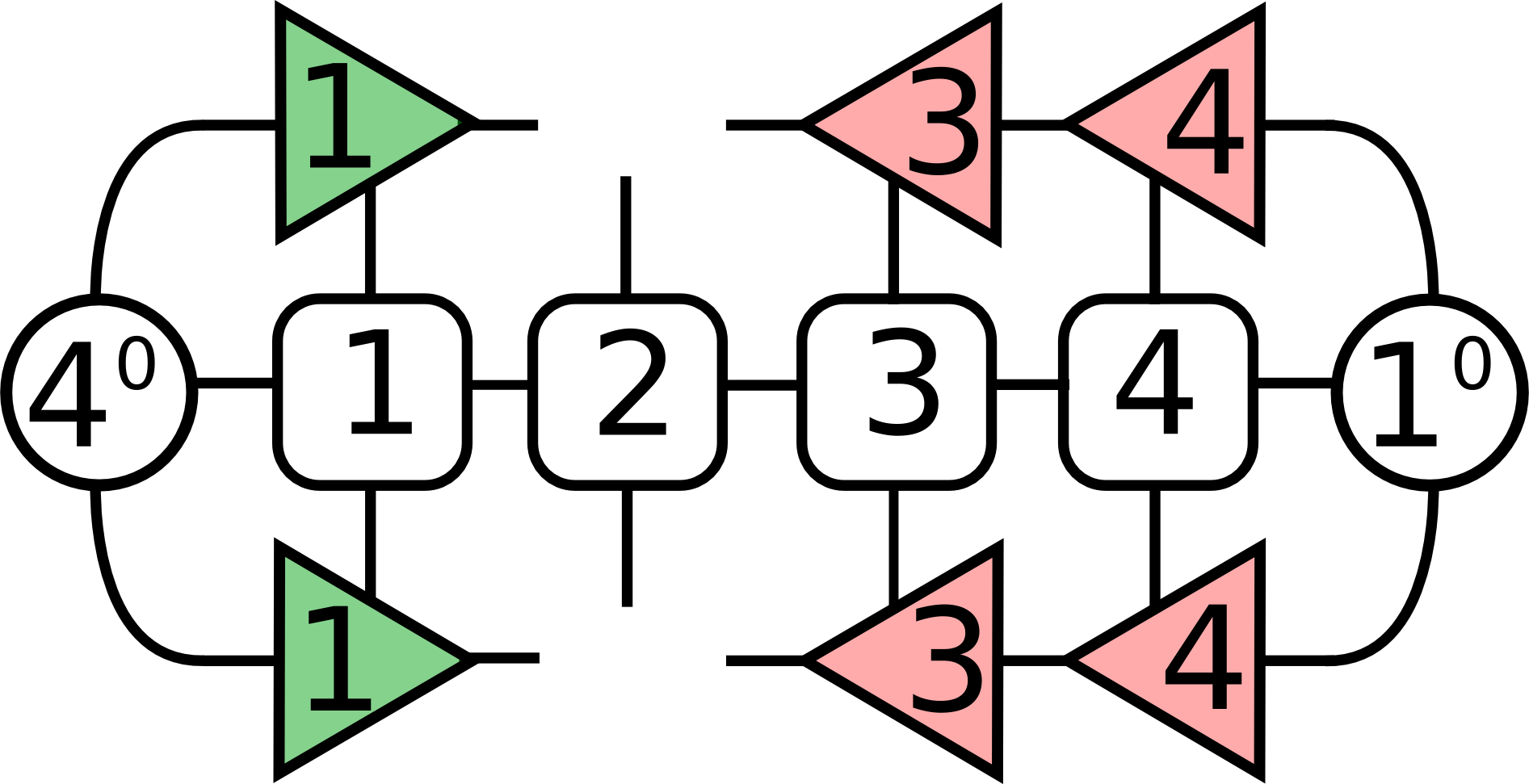}}.
\end{align}
The tensors \raisebox{-0.45cm}{\includegraphics[width=0.09\columnwidth]{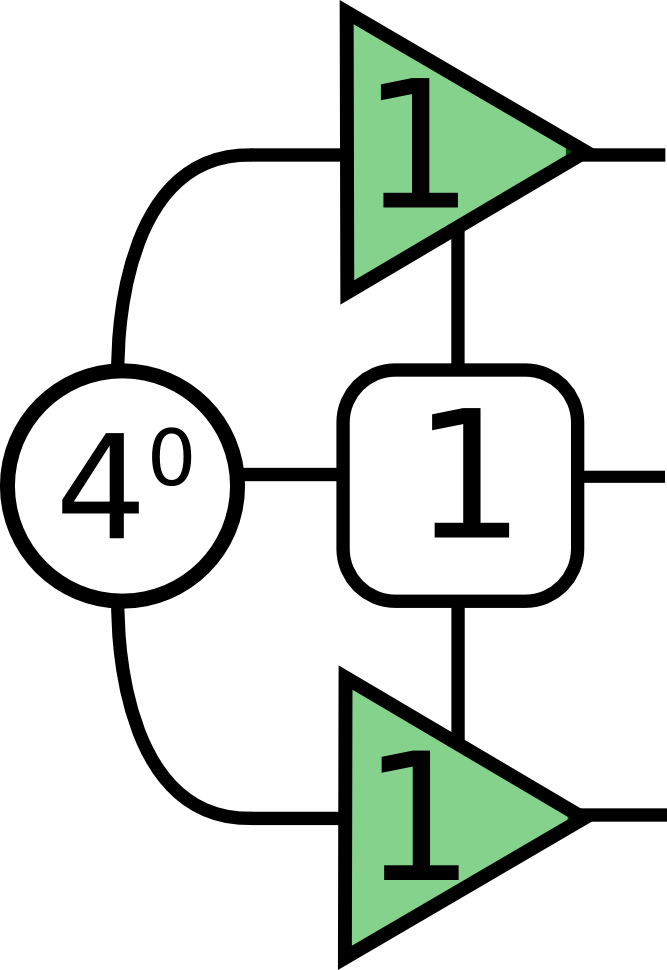}}
and 
\raisebox{-0.43cm}{\includegraphics[width=0.13\columnwidth]{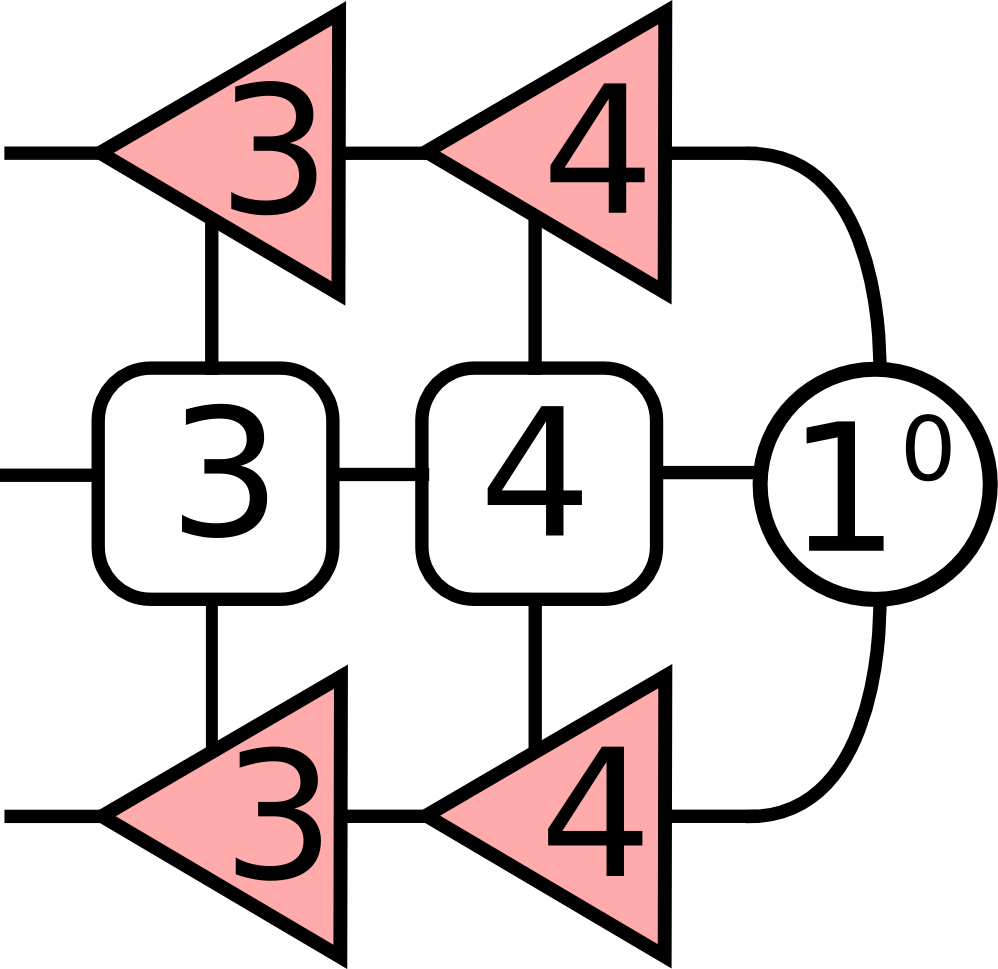}}
are called the effective left and right environments of site $i=2$.
The optimization keeps sweeping right and left (using a similar update method for the left sweep)
until sufficient convergence is reached (we only sweep once from left to right
for our applications).
The orthogonality center is then moved to the central bond, where a resolution of the identity
\raisebox{-0.17cm}{\includegraphics[width=0.14\columnwidth]{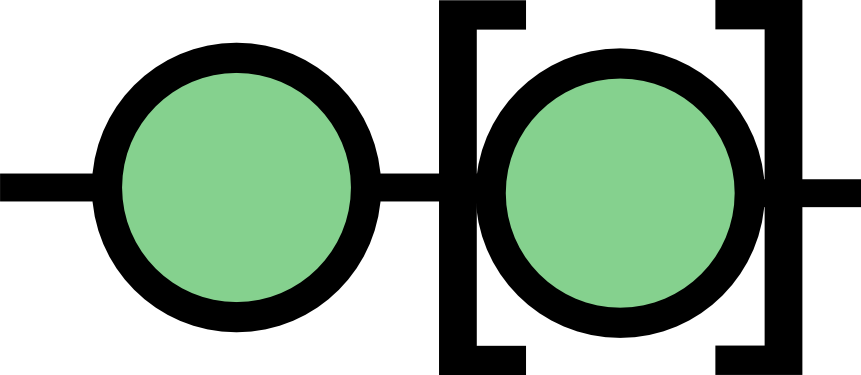}}
is inserted. The matrices
\raisebox{-0.17cm}{\includegraphics[width=0.1\columnwidth]{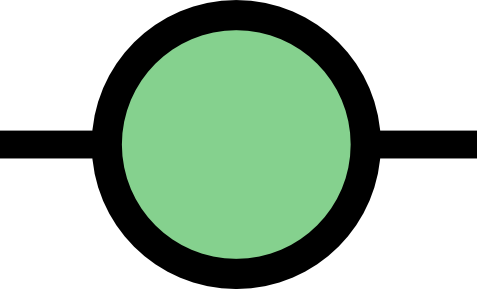}}
are shifted to the left and right boundary of the unit-cell (see \Fig{fig:idmrg} (c) and (d))
where they are contracted into 
$\raisebox{-0.12cm}{\includegraphics[width=0.1\columnwidth]{lambda_unlab_pink.png}}
=
\raisebox{-0.24cm}{\includegraphics[width=0.25\columnwidth]{centernew_lab.png}}$
(see \Fig{fig:idmrg} (a')). The state is now in a quasi-parallel decomposition
with respect to a unit-cell that has been shifted by $N/2$. Note, however, that
it is no longer in its canonical form. 
The next iteration consists of updating the tensors
$\{\tilde A^{[N/2+1]},\dots,\tilde A^{[N]},\tilde A^{[1]},\dots \tilde A^{[N/2]}\}$
of this shifted unit-cell.
To this end, one needs to calculate an effective Hamiltonian for the shifted unit-cell.
Several slightly different methods can be used to proceed at this point. In the following, we
will explain two of them in more detail.

In the iDMRG algorithm, the calculation of a new
effective Hamiltonian is achieved by recycling the left and right unit-cell environments
calculated during the previous iteration. The left and right unit-cell 
environments at iteration $q=1$,
\raisebox{-0.32cm}{\includegraphics[width=0.05\columnwidth]{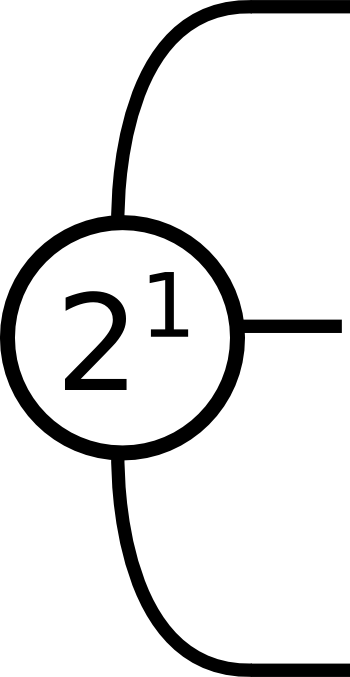}} and
\raisebox{-0.32cm}{\includegraphics[width=0.05\columnwidth]{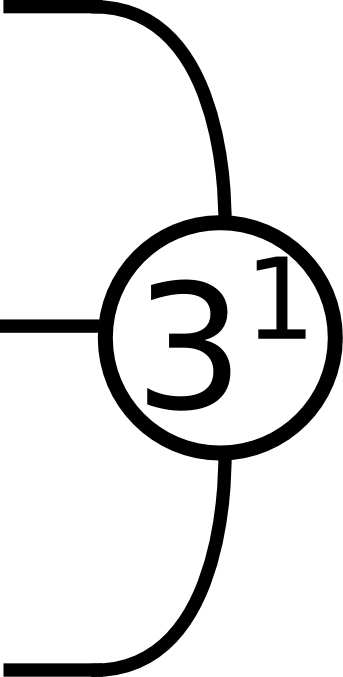}},
in this case are given by
\begin{align}
  \raisebox{-0.6cm}{\includegraphics[width=0.08\columnwidth]{L_1.png}}=\label{eq:idmrg_envs_l}
  \raisebox{-1.0cm}{\includegraphics[width=0.25\columnwidth]{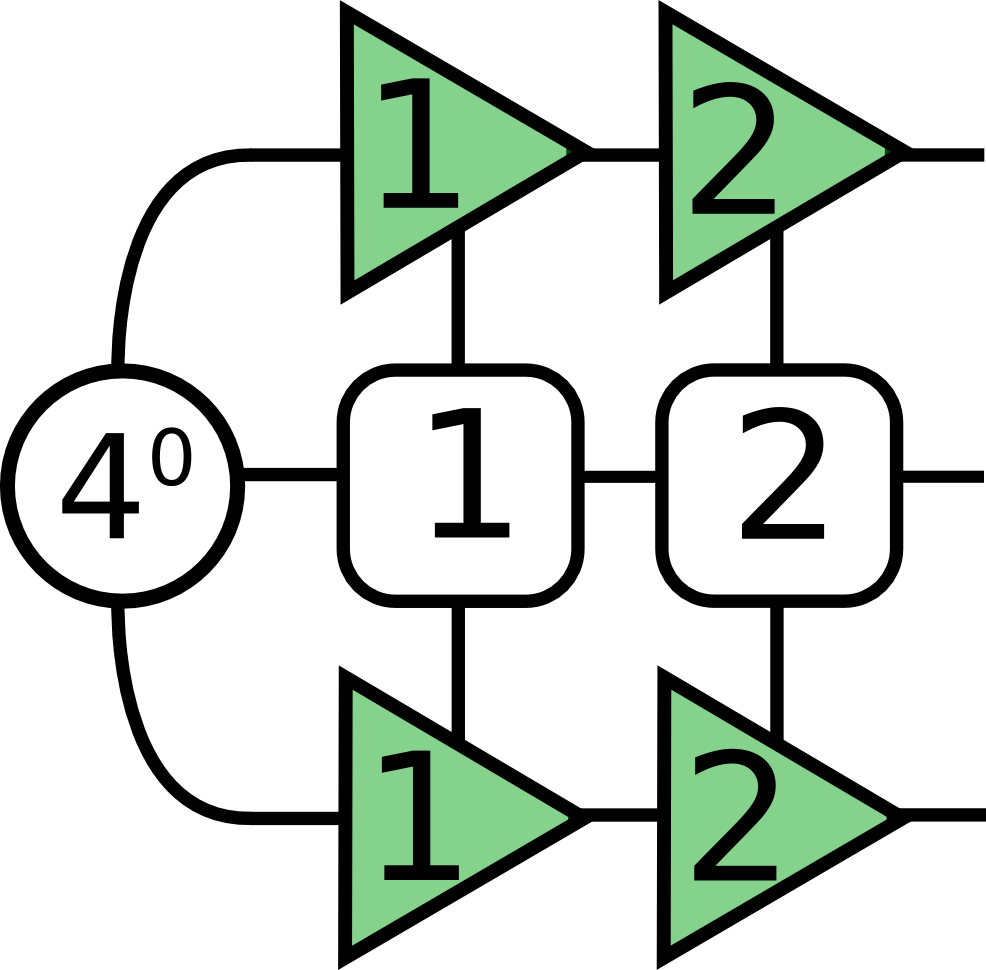}}\\
  \raisebox{-0.6cm}{\includegraphics[width=0.08\columnwidth]{R_1.png}}=
  \raisebox{-1.0cm}{\includegraphics[width=0.25\columnwidth]{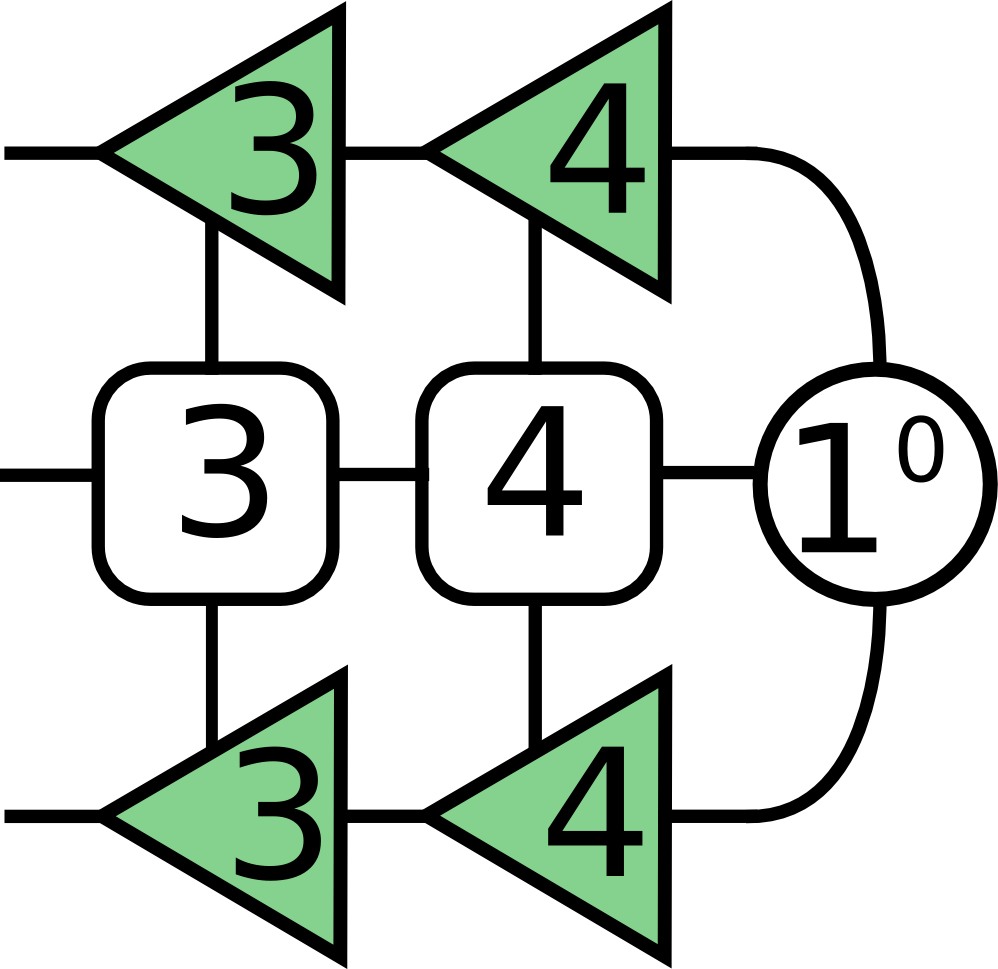}},
\label{eq:idmrg_envs_r}
\end{align}
The effective Hamiltonian for e.g. site $i=3$ during the second iteration $q=1$ assumes
the form
\begin{align}\label{eq:Heff_3}
  H_{eff}=\raisebox{-0.9cm}{\includegraphics[width=0.45\columnwidth]{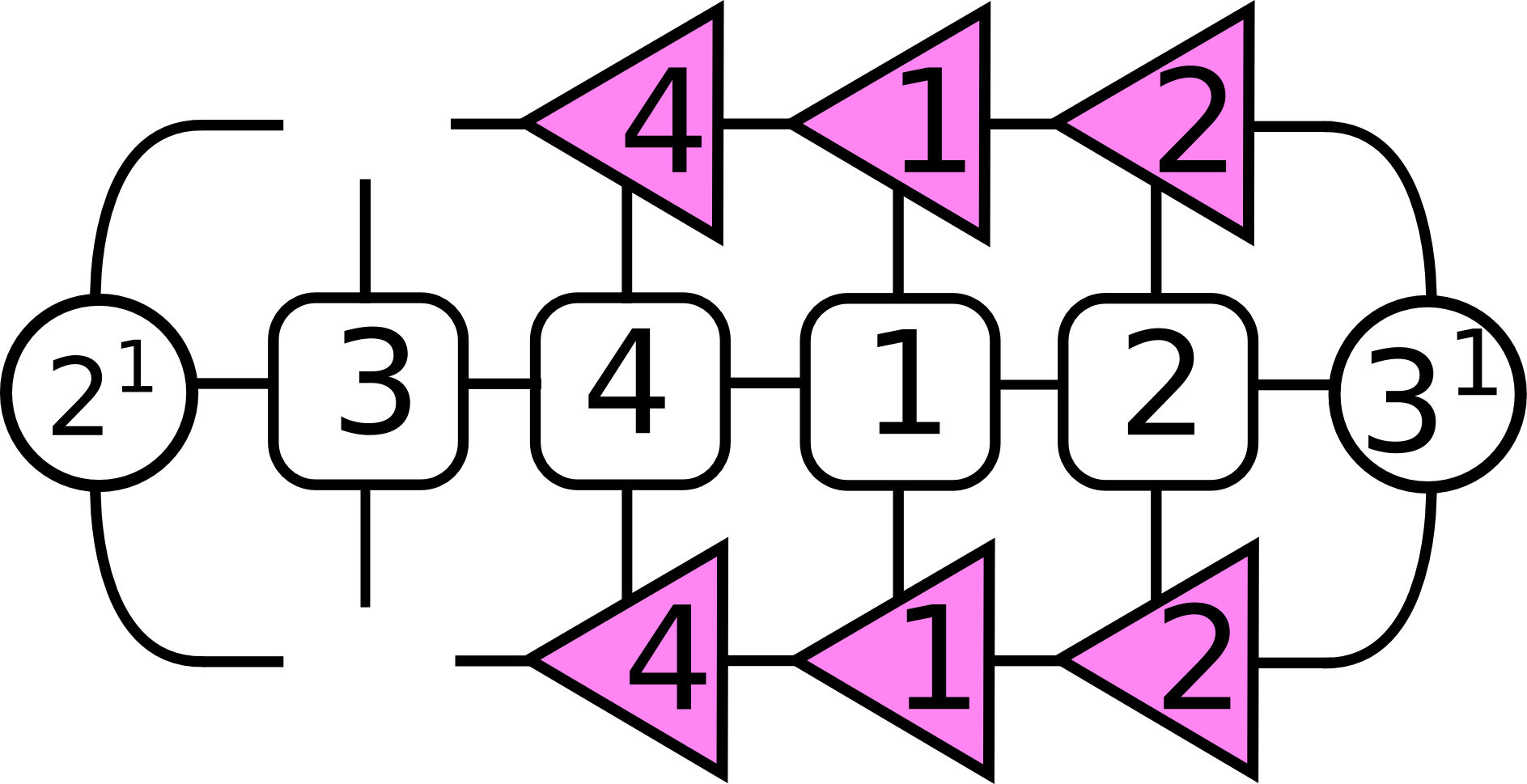}}.
\end{align}
We have here highlighted the currently optimized unit-cell tensors in pink.
From here on the method proceeds by repeatedly cycling through the steps (b)-(a')
of \Fig{fig:idmrg}, and updating the left and right unit-cell environments as described above.

While this method works well in general, an inconsistency arises after the first iteration:
the unit-cell environments in Eqs. (\ref{eq:idmrg_envs_l}) and (\ref{eq:idmrg_envs_r})
are calculated by adding updated tensors to 
\raisebox{-0.32cm}{\includegraphics[width=0.05\columnwidth]{L_0.png}} and 
\raisebox{-0.32cm}{\includegraphics[width=0.05\columnwidth]{R_0.png}}. 
These two, however, have both been obtained from contracting an infinite tensor network 
containing the tensors of the {\it initial} MPS. They do not
contain any information of the update and are thus different 
from the ones that would be obtained using the {\it current} MPS
(which has been updated within each unit-cell).
Thus, instead of recycling the environments after each iteration,
one could also regauge the MPS into its canonical form and recalculate 
the effective left and right unit-cell environments. This approach is seen
to result in slightly lower energies of the optimized MPS.

For the case of $N=1$, a very similar approach has been proposed for the setting of homogeneous cMPS
\cite{ganahl_continuous_2017}, termed gradient optimization
for cMPS. 
The gradient optimization for homogeneous lattices with $N=1$
can be obtained by slightly tweaking the above method, as we will now describe in the following. 
Consider a state with $N=1$
in its parallel decomposition, i.e. 
\begin{align}
  \ket{\Psi}=\raisebox{-0.25cm}{\includegraphics[width=0.45\columnwidth]{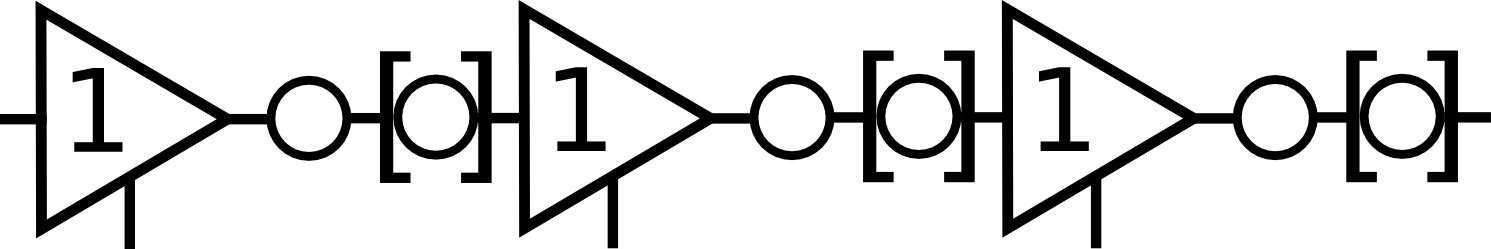}}\;.
\end{align}
The effective unit-cell Hamiltonian for this state is given by
\begin{align}\label{eq:Heff_homogeneous}
  H_{eff}=\cdots
  \raisebox{-1.05cm}{\includegraphics[width=0.28\columnwidth]{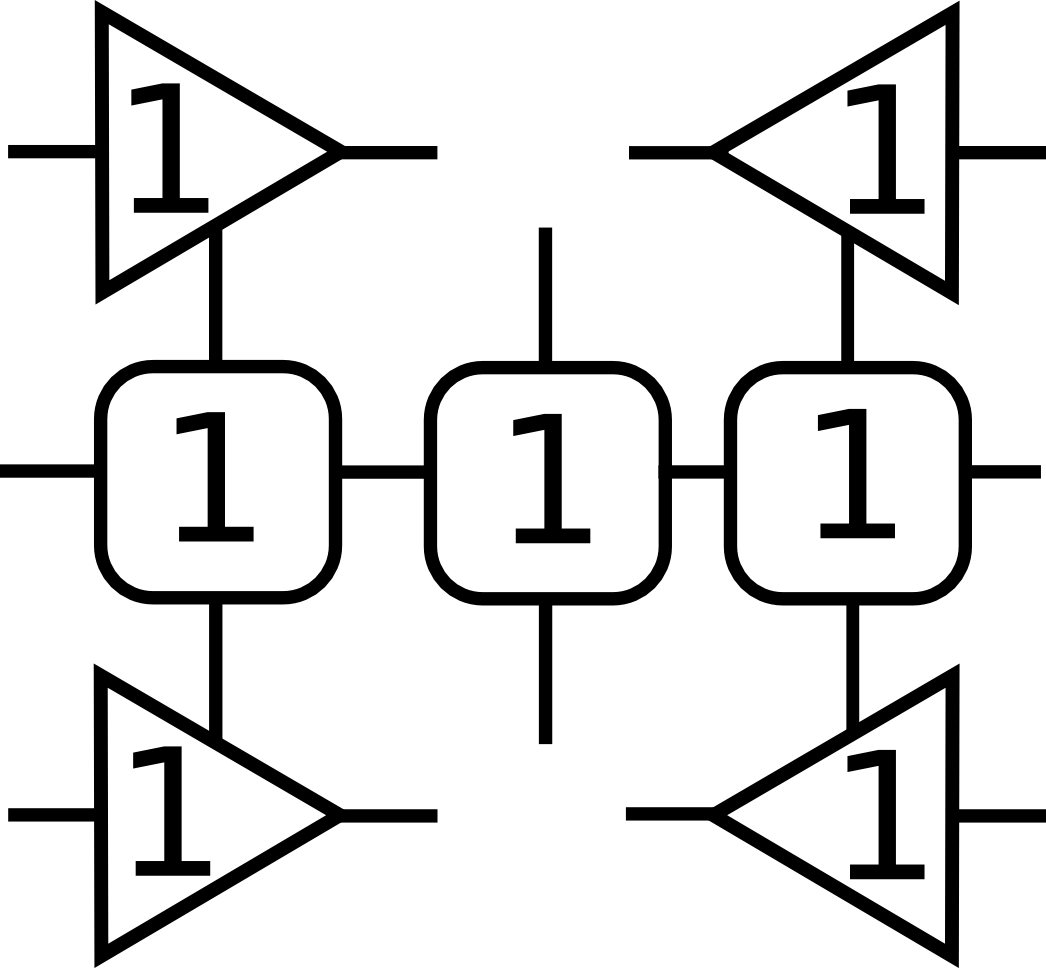}}\cdots=
  \raisebox{-0.7cm}{\includegraphics[width=0.23\columnwidth]{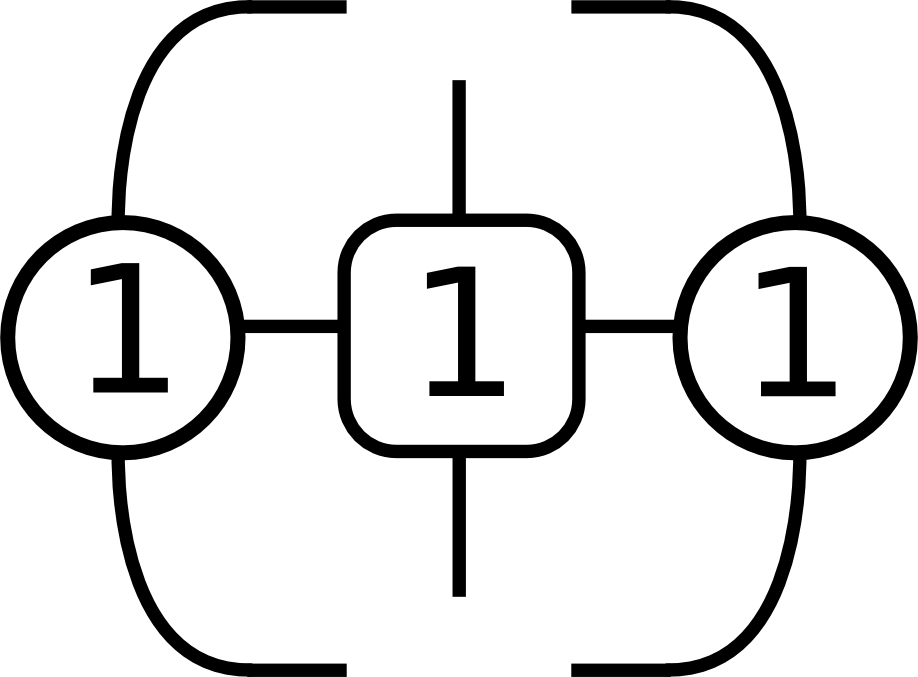}}\;.
\end{align}
The update of the tensor 
\raisebox{-0.25cm}{\includegraphics[width=0.15\columnwidth]{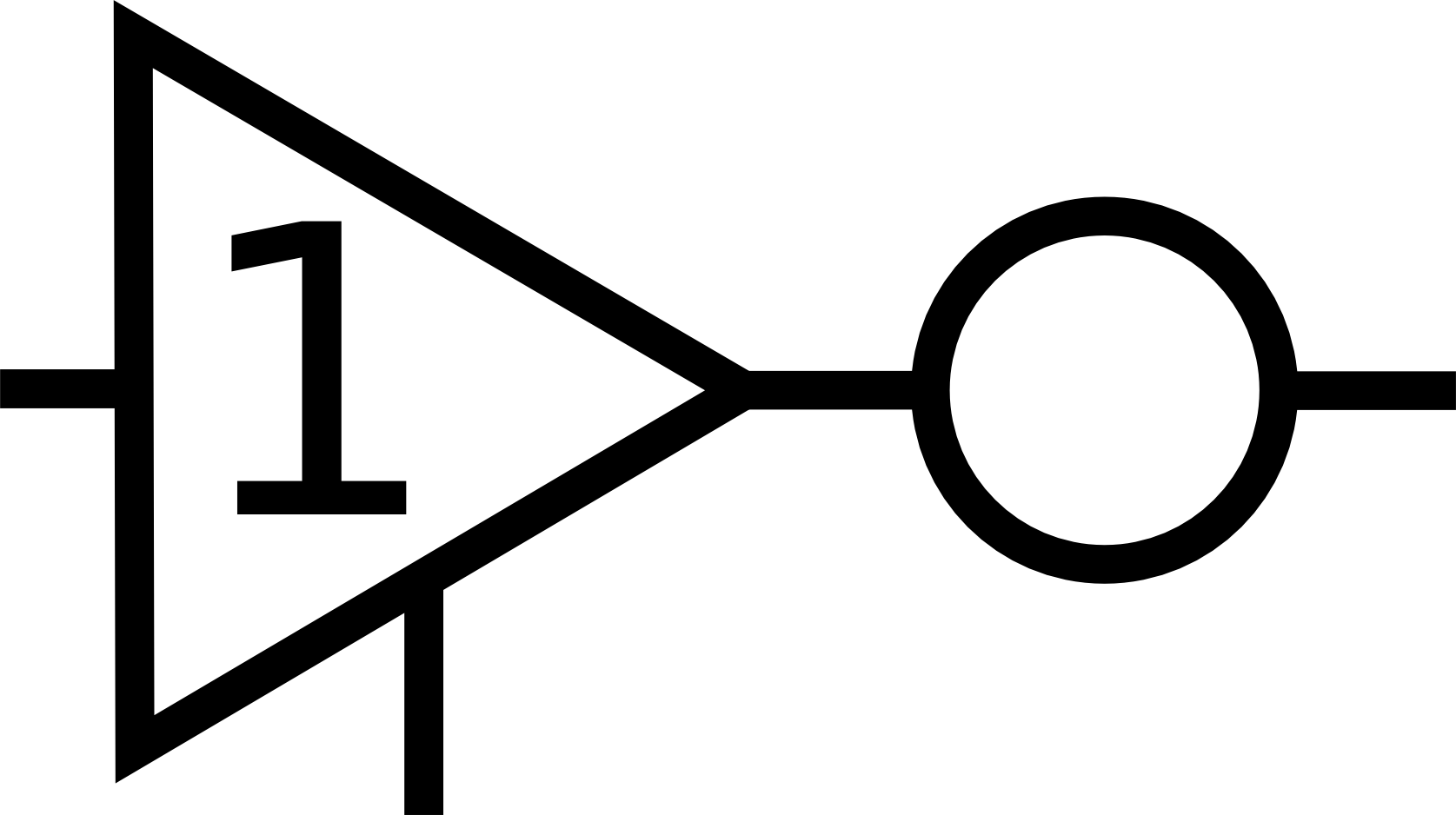}}
proceeds now in a different manner than described above.
Instead of finding an approximate lowest eigen-vector of \Eq{eq:Heff_homogeneous}
we calculate a local gradient 
\raisebox{-0.35cm}{\includegraphics[width=0.15\columnwidth]{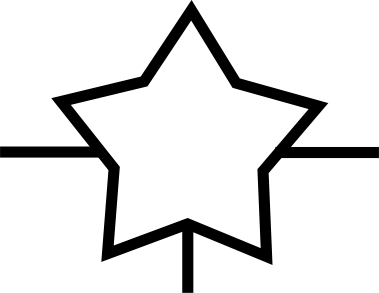}},
with
\begin{align}
  \raisebox{-0.55cm}{\includegraphics[width=0.2\columnwidth]{gradientTensor.png}}\equiv
  \raisebox{-0.9cm}{\includegraphics[width=0.3\columnwidth]{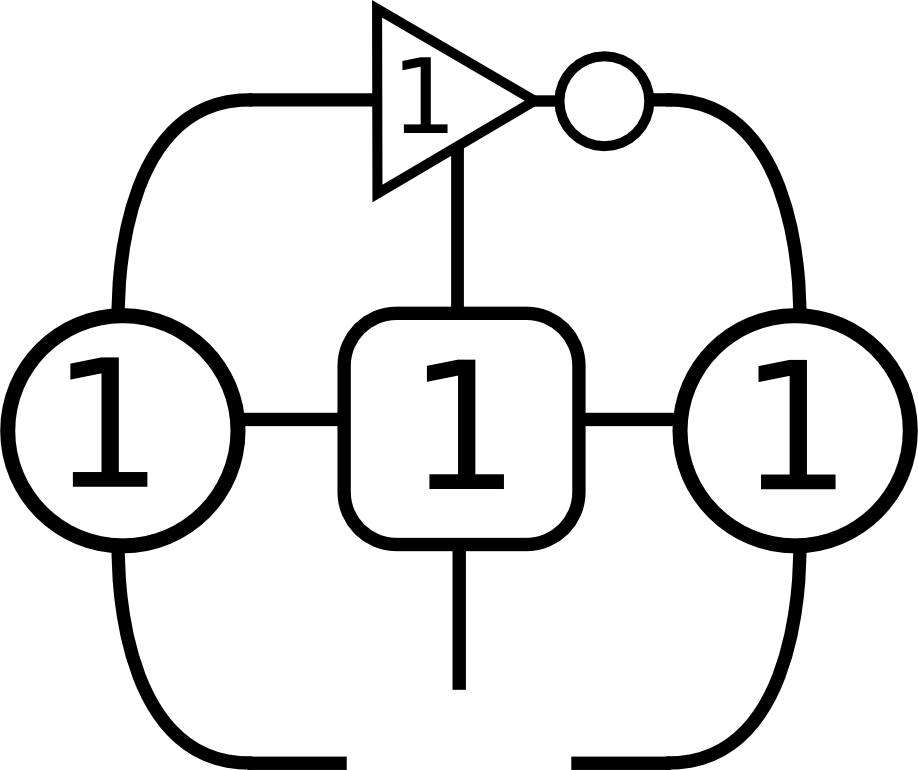}}\;,
\end{align}
and use it to update the tensor:
\begin{align}
  \raisebox{-0.43cm}{\includegraphics[width=0.13\columnwidth]{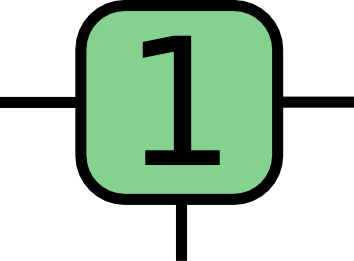}}\equiv
  \raisebox{-0.43cm}{\includegraphics[width=0.2\columnwidth]{centerTensor.png}}-\alpha\;
  \raisebox{-0.4cm}{\includegraphics[width=0.15\columnwidth]{gradientTensor.png}}.
\end{align}
$\alpha$ is a small positive number. 
If $\alpha$ is chosen such as to minimize the energy expectation value
\begin{align}
  \raisebox{-0.9cm}{\includegraphics[width=0.3\columnwidth]{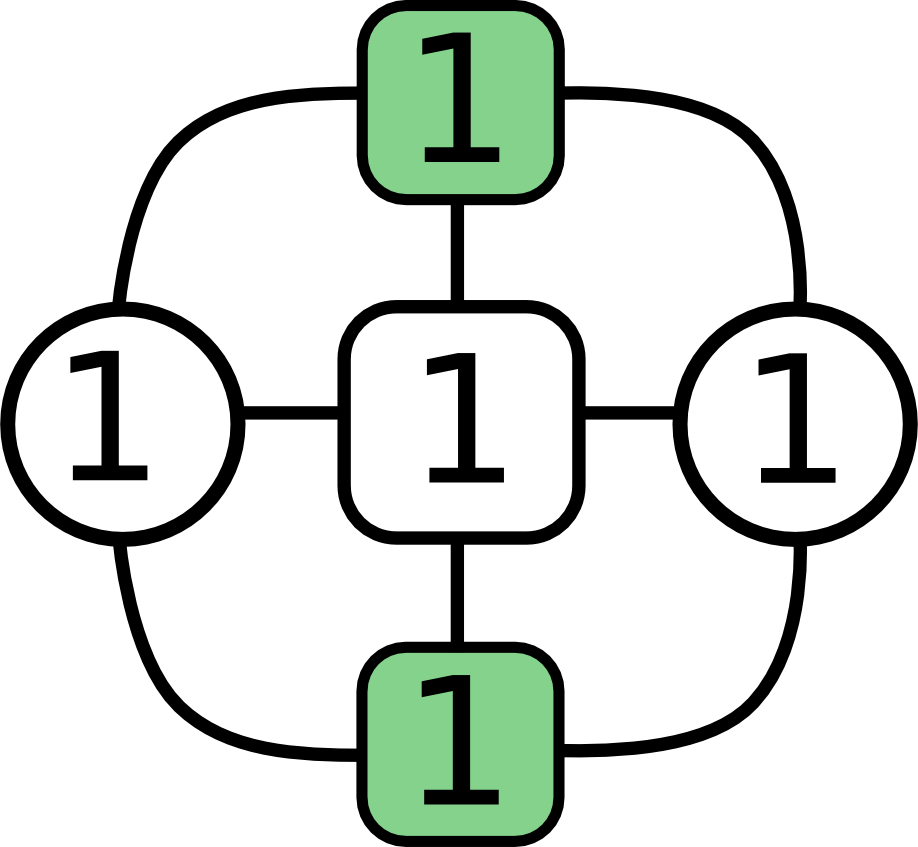}},
\end{align}
the update is equivalent to finding the lowest energy
state in the Krylov space spanned by
$\{\ket{\raisebox{-0.16cm}{\includegraphics[width=0.1\columnwidth]{centerTensor.png}}}, 
\ket{\raisebox{-0.2cm}{\includegraphics[width=0.1\columnwidth]{gradientTensor.png}}}\}$.
After this update, the state is in the form
\begin{align}
  \ket{\Psi}=\raisebox{-0.25cm}{\includegraphics[width=0.35\columnwidth]{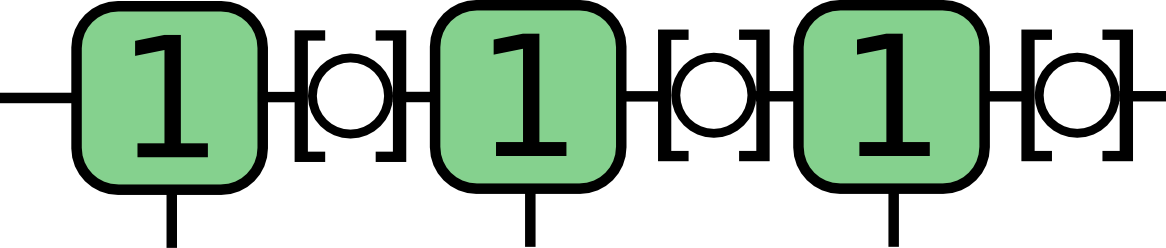}}.
\end{align}
The matrices \raisebox{-0.11cm}{\includegraphics[width=0.08\columnwidth]{inverse.png}} are
now absorbed back into 
\raisebox{-0.25cm}{\includegraphics[width=0.12\columnwidth]{updatedTensor.png}}
from the right hand side, and the iteration is restarted.
A related approach for 1d lattice models which avoids
the possibly ill-condition inversion 
\raisebox{-0.11cm}{\includegraphics[width=0.08\columnwidth]{inverse.png}}
has recently been proposed by Zauner-Stauber et al.\cite{zauner-stauber_variational_2017}.
For the case of cMPS, the gradient optimization greatly outperforms other optimization approaches.

As we have found numerically, recalculating the full environments gives more accurate 
results for ground-state energies than recycling the environments. However, for large $N\gg 1$,
it is slower than standard iDMRG due to the necessary recalculation
of the environments. For the model considered in this manuscript, 
the difference in accuracy between the two approaches is small enough
that in present manuscript we use the simple recycling method for all calculations.
In order to facilitate the extraction the cMPS content of the
optimized lattice MPS, we use slightly modified techniques for shifting the orthogonality 
center and regauging the state, as described in detail in Appendix \ref{app:shift}.

\section{Contracting \Eq{eq:Heff}}\label{app:contract} 
In this Appendix we discuss in detail how to contract the infinite tensor network 
appearing in \Eq{eq:Heff}. In the following we discuss the case of a nearest neighbor 
Hamiltonian. The extension to more general Hamiltonians follows straightforwardly.
Consider the case of a Hamiltonian of the form
\be\label{eq:HamApp}
H=\sum_{i\in\mathbbm{Z}}h_{i,i+1}=\sum_{i\in\mathbbm{Z}}\;\raisebox{-0.22cm}{\includegraphics[width=0.08\columnwidth]{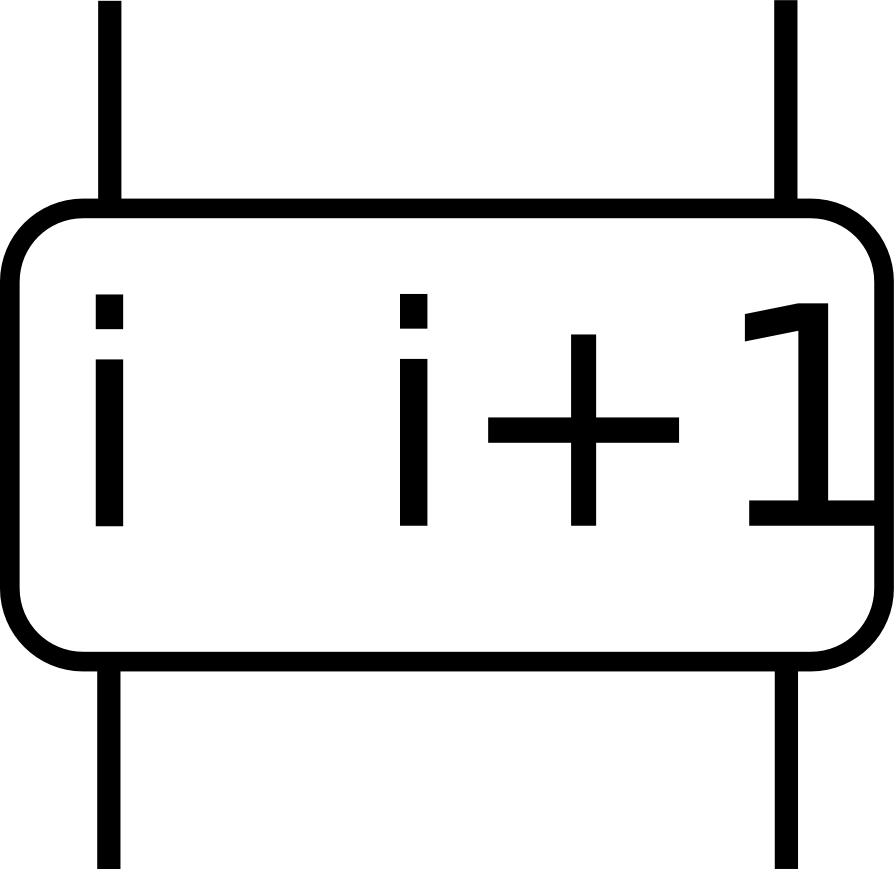}},
\ee
where we have introduced a diagrammatic notation for the two-site operators 
$h_{i,i+1}$.
We assume further a periodicity of the Hamiltonian over $N=4$ sites (see main text), 
such that we can decompose \Eq{eq:HamApp} into
\begin{align}
H&=\sum_{j\in\mathbbm{Z}}H_j\\
H_j&=\sum_{i=1}^{4}h_{i,i\oplus1}=\sum_{i=1}^4\;\raisebox{-0.22cm}{\includegraphics[width=0.08\columnwidth]{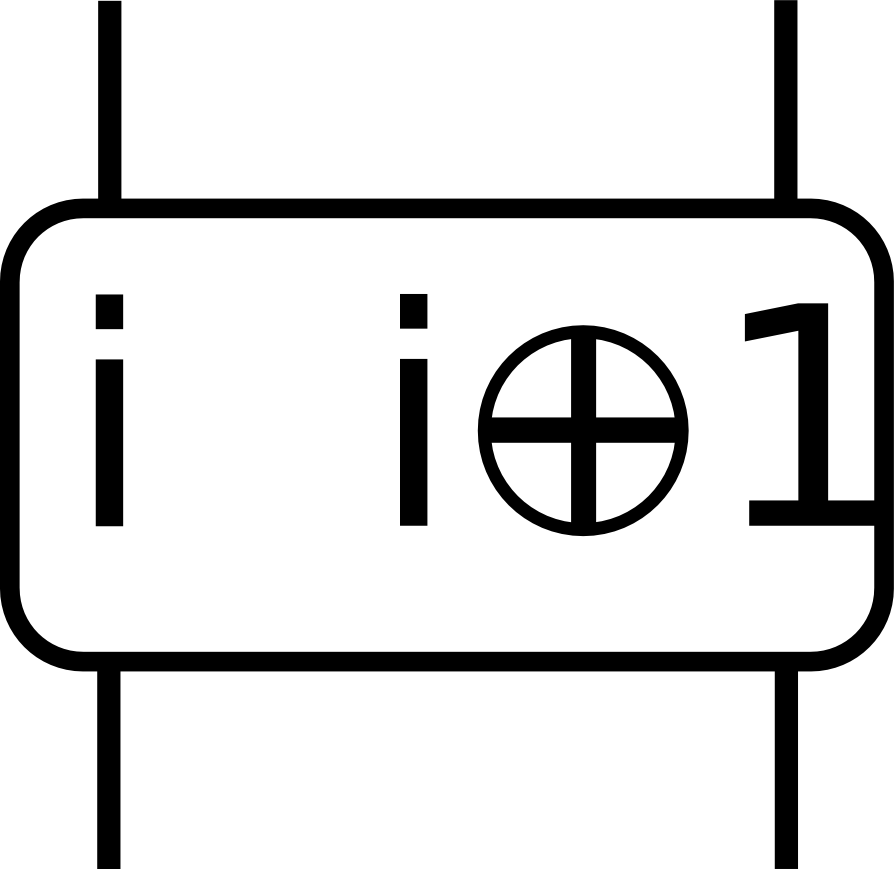}},
\end{align}
with $j$ running over the unit-cells, and $i\oplus 1\equiv i\,\textrm{mod}\,N+1$.
Using standard techniques \cite{schollwock_density-matrix_2011}, $H$ can be written in an
MPO decomposition 
\begin{align}
  H=\cdots\raisebox{-0.45cm}{\includegraphics[width=0.4\columnwidth]{MPO.png}}\cdots,
\end{align}
where numbers label sites $i$ inside the unit-cell. Horizontal lines denote auxiliary
indices (with a bond-dimension $M$), and vertical lines are physical indices of the MPO.
For example, for the Hamiltonian \Eq{eq:Ham_eps2}, a possible, non-unique MPO
decomposition is given by
\begin{align}\label{eq:MPO}
  &\raisebox{-0.6cm}{\includegraphics[width=0.15\columnwidth]{MPOTensor.png}}=\\
  &\left(
    \begin{array}{ccccccc}
      \mathbbm{1}&&&&&&\\
      c_i^{\dagger}&&&&&&\\
      c_i&&&&&&\\
      c_i^{\dagger}c_i&&&&&&\\
      \mathcal{P}^0_i&&&&&&\\
      c_i^{\dagger}c_i&&&&&&\\
      \mu_i c_i^{\dagger}c_i&\frac{-1}{2m\eps^2}c_i&\frac{-1}{2m\eps^2}c_i^{\dagger}&\frac{g}{\eps}c_i^{\dagger}c_i&\frac{1}{2m\eps^2}c_i^{\dagger}c_i&\frac{1}{2m\eps^2}\mathcal{P}^0_i&\mathbbm{1}
    \end{array}
  \right)\nonumber,
\end{align}
where we have only written entries different from 0. 
The MPO bond dimension is in this case $M=7$.
For later reference we introduce the boundary MPO tensors
\begin{align}
  \raisebox{-0.6cm}{\includegraphics[width=0.125\columnwidth]{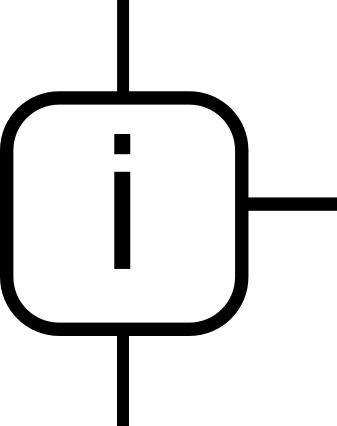}}\qquad,\qquad
  \raisebox{-0.6cm}{\includegraphics[width=0.125\columnwidth]{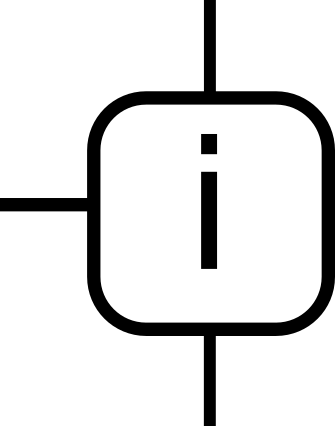}}\;.
\end{align}
In our MPO convention,
\raisebox{-0.4cm}{\includegraphics[width=0.1\columnwidth]{MPOTensorleft.png}} is
given by the last column of \Eq{eq:MPO}, and
\raisebox{-0.4cm}{\includegraphics[width=0.1\columnwidth]{MPOTensorright.png}}
is given by its first row.
We refer the reader to the literature \cite{schollwock_density-matrix_2011} for
a detailed introduction to the MPO formalism.

Written out explicitly for the case $N=4$ we have
\be
\sum_{i=1}^{4}h_{i,i\oplus1}=\quad
\raisebox{-2.cm}{\includegraphics[width=0.3\columnwidth]{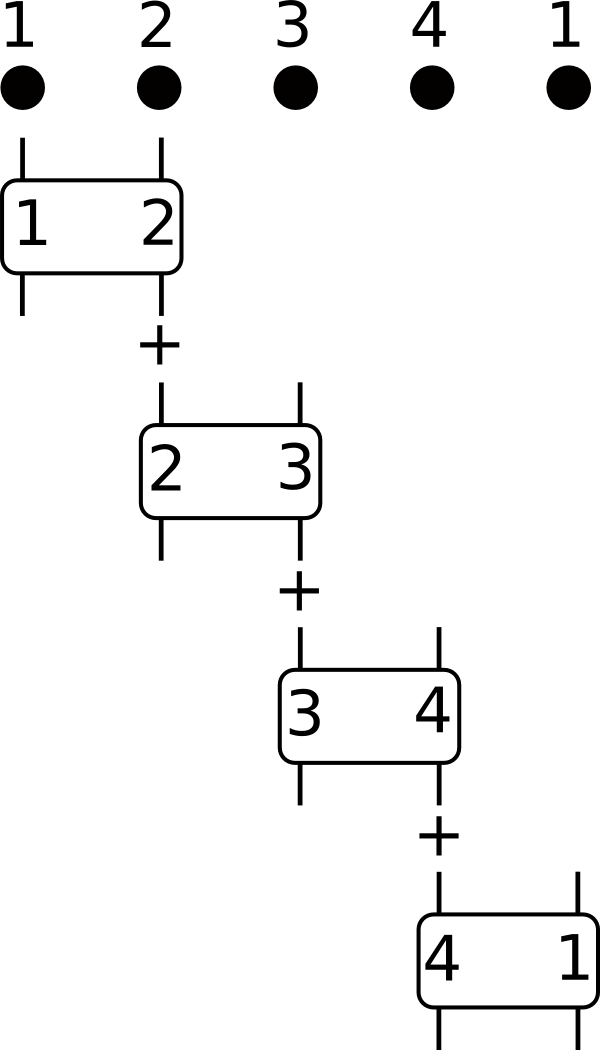}}.
\ee
For a fixed unit-cell we will now calculate the right unit-cell environment (see main text)
\be
\raisebox{-0.6cm}{\includegraphics[width=0.08\columnwidth]{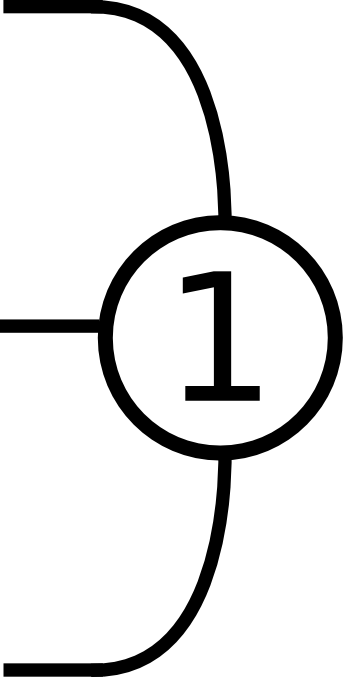}}\equiv\raisebox{-0.85cm}{\includegraphics[width=0.3\columnwidth]{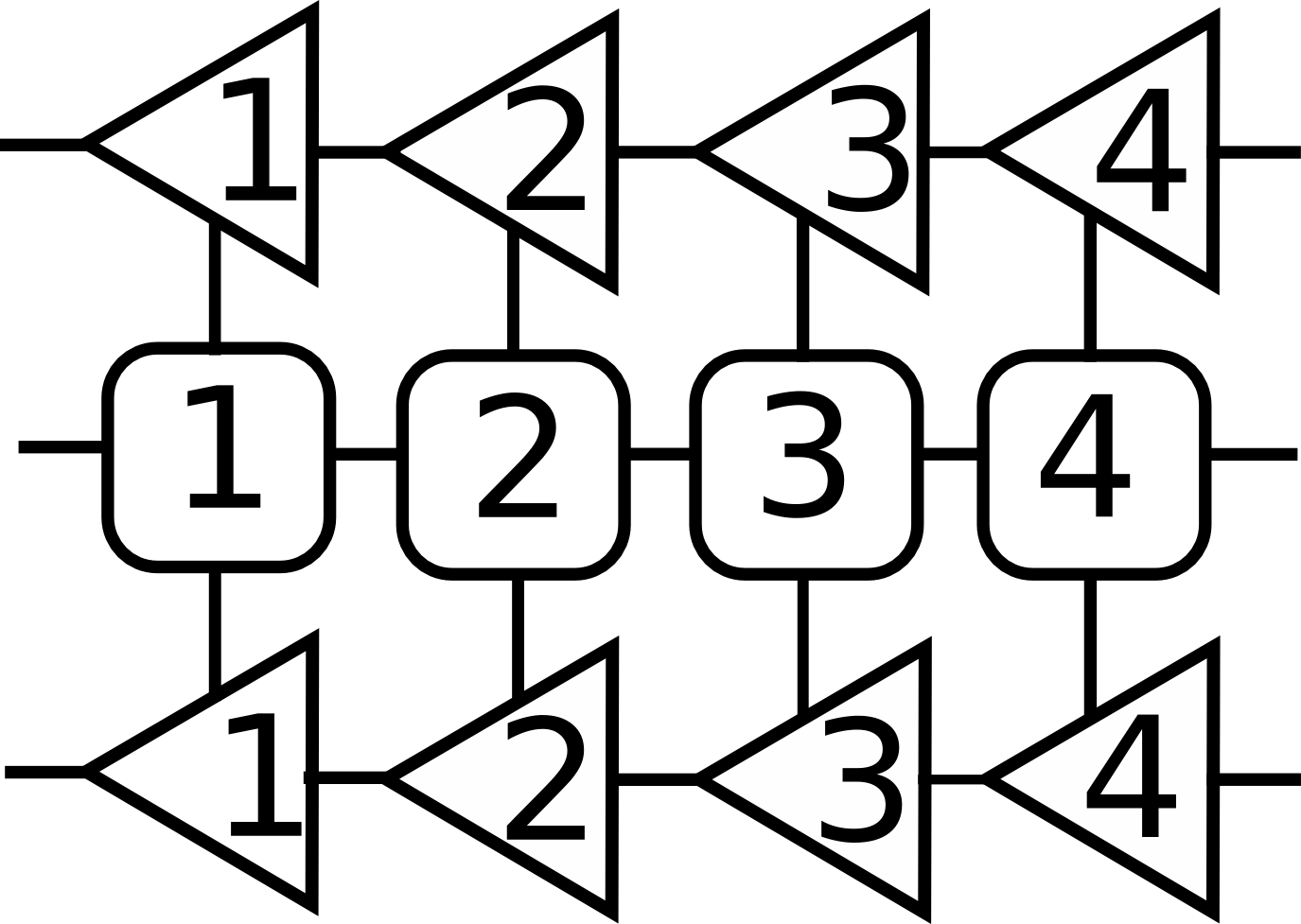}}\dots\;.
\ee
For an MPO with e.g. bond dimension $M$, this would be a vector of $M$ $D\times D$ matrices.
In vector notation, it is
\begin{align}\label{eq:Hr}
  \left(
    \begin{array}{c}
      \raisebox{-0.0cm}{\includegraphics[width=0.08\columnwidth]{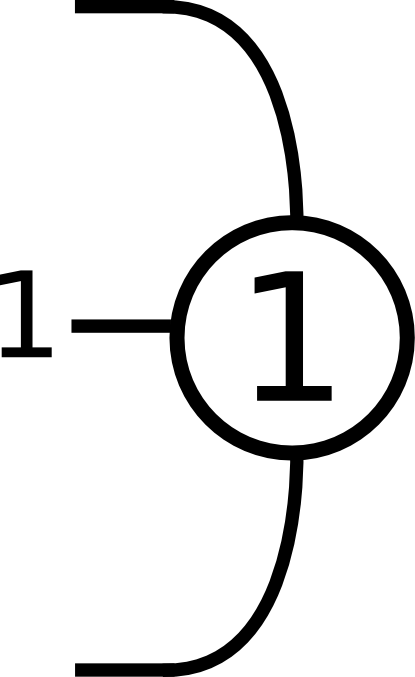}}\\
      \raisebox{0.1cm}{\vdots}\\
      \\
      \raisebox{-0.0cm}{\includegraphics[width=0.08\columnwidth]{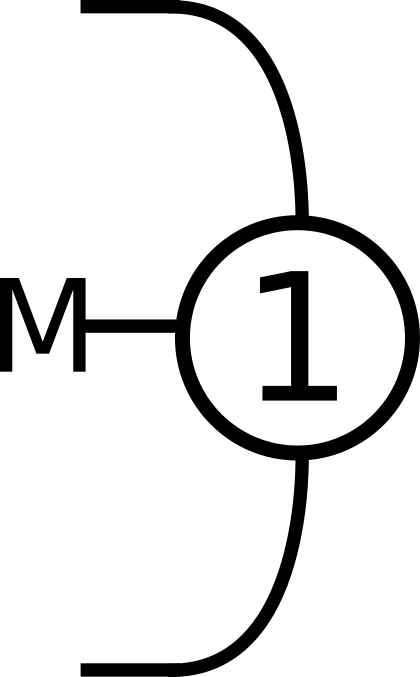}}\\
    \end{array}
  \right)=
  \left(
    \begin{array}{c}
      \raisebox{-0.7cm}{\includegraphics[width=0.1\columnwidth]{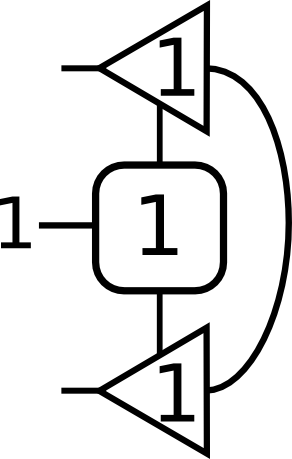}}\\
      \raisebox{0.3cm}{\vdots}\\
      \raisebox{-0.7cm}{\includegraphics[width=0.12\columnwidth]{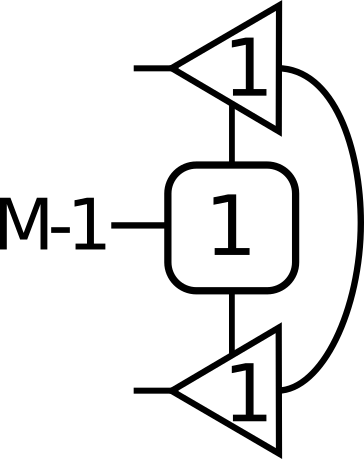}}\\
      \\
      \raisebox{-2.0cm}{\includegraphics[width=0.7\columnwidth]{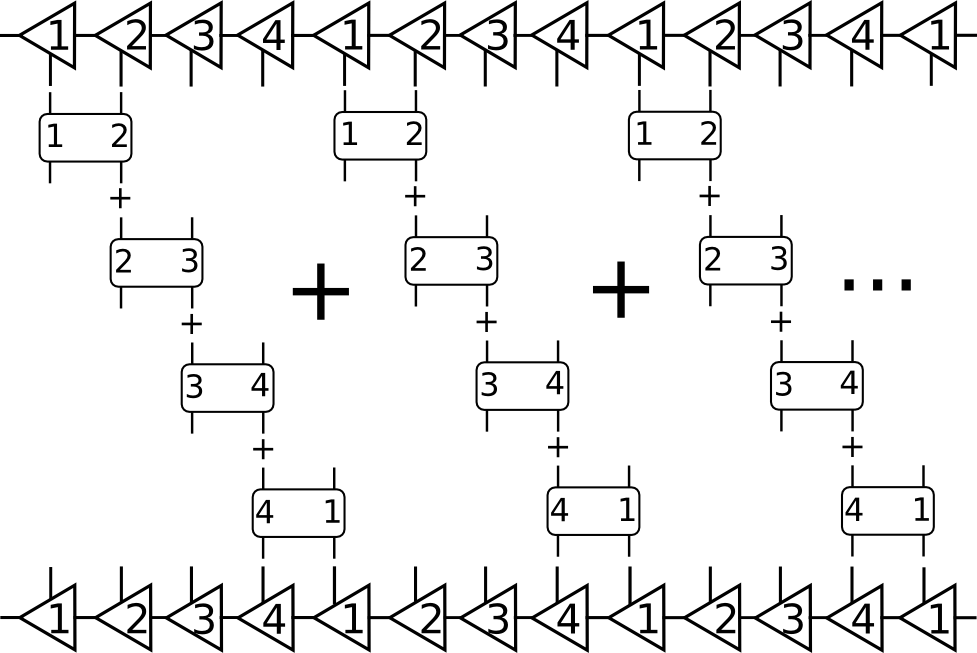}}
    \end{array}
  \right),
\end{align}
where we have explicitly written out the vector index.
Using an MPO decomposition of the unit-cell Hamiltonian,
\be
\raisebox{-0.3cm}{\includegraphics[width=0.5\columnwidth]{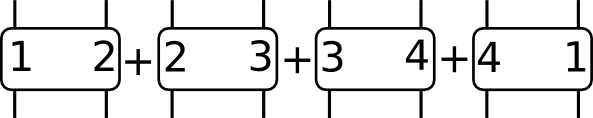}}=
\raisebox{-0.3cm}{\includegraphics[width=0.35\columnwidth]{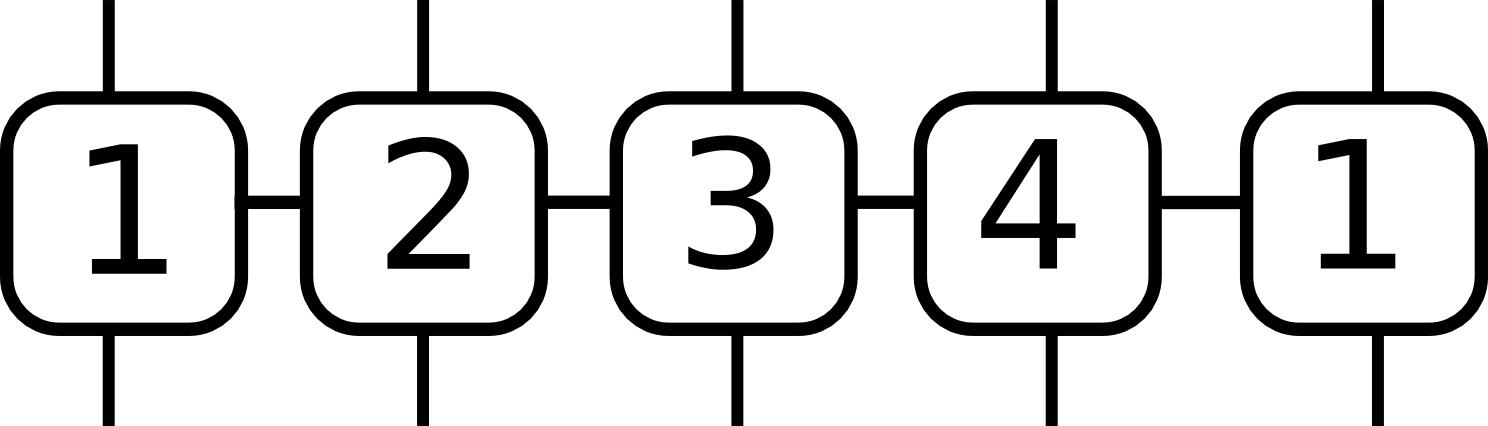}},
\ee
we can rewrite the last component of \Eq{eq:Hr} as 
\be\label{eq:geomsum_1}
\Bigg(\sum_{n=0}^{\infty}\Bigg[\;\;\raisebox{-0.5cm}{\includegraphics[width=0.2\columnwidth]{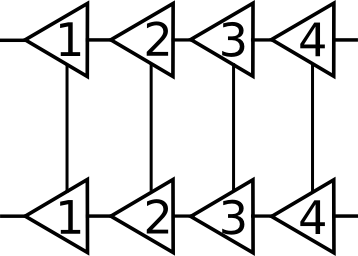}}\;\;\Bigg]^n\Bigg)
\raisebox{-0.5cm}{\includegraphics[width=0.25\columnwidth]{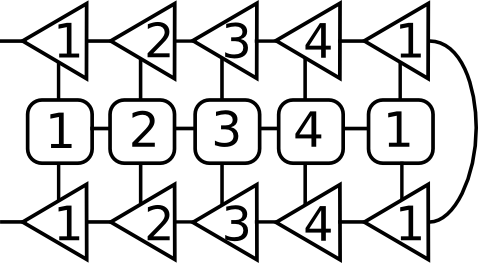}}
\ee
with the right-normalized unit-cell transfer operator 
\be
\mathbbm{E}_r\equiv\raisebox{-0.55cm}{\includegraphics[width=0.2\columnwidth]{UCTr.png}}.
\ee
and the renormalized unit-cell Hamiltonian 
\be
\ket{h_r}\equiv\raisebox{-0.5cm}{\includegraphics[width=0.25\columnwidth]{UCR.png}}.
\ee
The operator $\mathbbm{E}_r$ has a left and right dominant eigen-vector $\bra{l},\ket{\mathbbm{1}}$
to eigenvalue $\eta=1$. Thus, the geometric series in \Eq{eq:geomsum_1} diverges. 
This can be cured by restricting the operator $\mathbbm{E}_r$
to the orthogonal complement of the subspace $\ket{\mathbbm{1}}\bra{l}$, i.e. by 
the replacements 
\begin{align}
\mathbbm{E}_r&\ra\mathbbm{E}_{r\perp}=\mathbbm{E}_r-\ket{\mathbbm{1}}\bra{l}\\
\ket{h_r}&\ra \ket{h_r}_{\perp}=\ket{h_r}-\braket{l|h_r}\ket{\mathbbm{1}}
\end{align}
The geometric series \Eq{eq:geomsum_1} transforms into
\begin{align}
  \ket{H_r}_{\perp}\equiv\Big(\sum_{n=0}^{\infty}\left[\mathbbm{E}_{r\perp}\right]^n\Big)\ket{h_r}_{\perp}=
  \frac{1}{\mathbbm{1}-\mathbbm{E}_{r\perp}}\ket{h_r}_{\perp}
\end{align}
which can be solved iteratively for $\ket{H_r}_{\perp}$ using sparse solvers like 
e.g. the {\it lgmres} method from the {\it scipy} package.

The approach for calculating the left unit-cell environment 
starts from the expression
\be
\raisebox{-2.0cm}{\includegraphics[width=0.7\columnwidth]{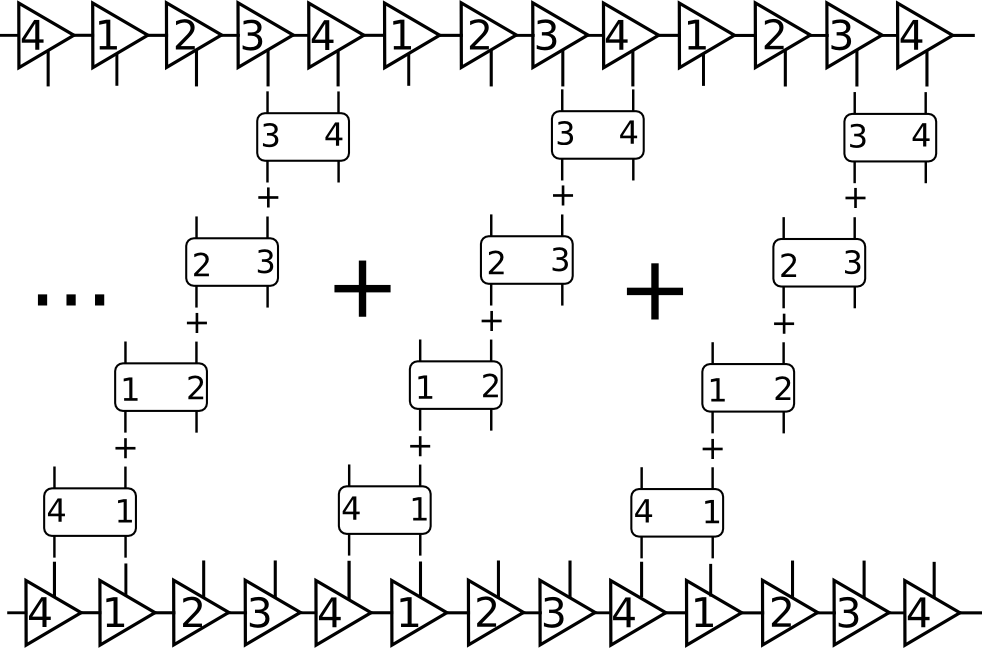}}
\ee
and then follows the same steps, using an MPO decomposition for the unit-cell
Hamiltonian of the form
\be
\raisebox{-0.3cm}{\includegraphics[width=0.5\columnwidth]{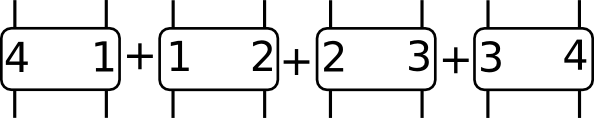}}=
\raisebox{-0.3cm}{\includegraphics[width=0.35\columnwidth]{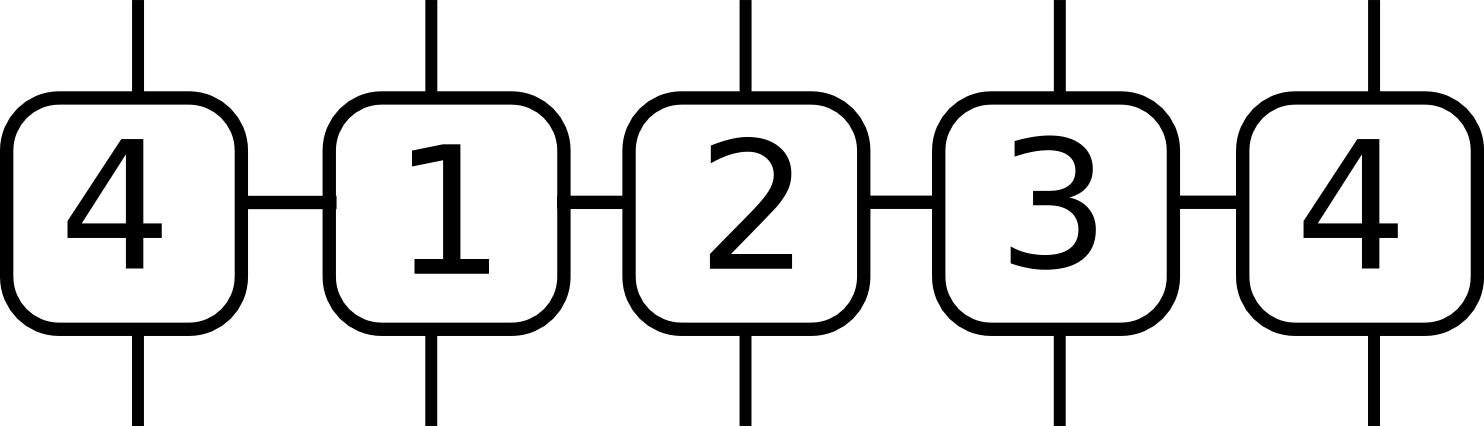}}.
\ee
With the definitions
\begin{align}
\mathbbm{E}_l&\equiv\raisebox{-0.55cm}{\includegraphics[width=0.2\columnwidth]{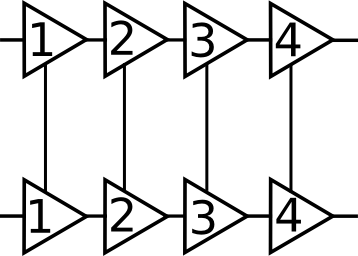}}\\
\bra{h_l}&\equiv\raisebox{-0.5cm}{\includegraphics[width=0.25\columnwidth]{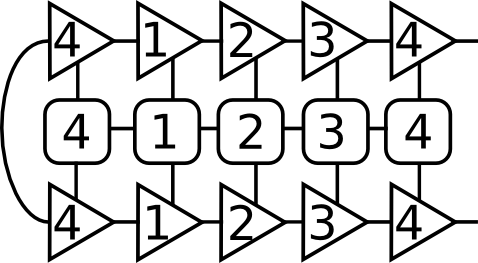}}
\end{align}
and the replacements
\begin{align}
\mathbbm{E}_l&\ra\mathbbm{E}_{l\perp}=\mathbbm{E}_l-\ket{r}\bra{\mathbbm{1}}\\
\bra{h_l}&\ra \bra{h_l}_{\perp}=\bra{h_l}-\braket{h_l|r}\bra{\mathbbm{1}}
\end{align}
one obtains 
\begin{align}
  \bra{H_l}_{\perp}\equiv\bra{h_l}_{\perp}\frac{1}{\mathbbm{1}-\mathbbm{E}_{l\perp}}.
\end{align}
Here, $\bra{\mathbbm{1}}$ and $\ket{r}$ are the dominant 
left and right eigen-vectors of $\mathbbm{E}_l$.
The final result is
\begin{align}\label{eq:Hr}
  \raisebox{-0.6cm}{\includegraphics[width=0.08\columnwidth]{R_unlab.png}}&=
  \left(
    \begin{array}{c}
      \raisebox{-0.7cm}{\includegraphics[width=0.1\columnwidth]{R1.png}}\\
      \raisebox{0.3cm}{\vdots}\\
      \raisebox{-0.7cm}{\includegraphics[width=0.12\columnwidth]{RM-1.png}}\\
      \\
      \ket{H_r}_{\perp}
    \end{array}
    \right)\\
  \raisebox{-0.6cm}{\includegraphics[width=0.08\columnwidth]{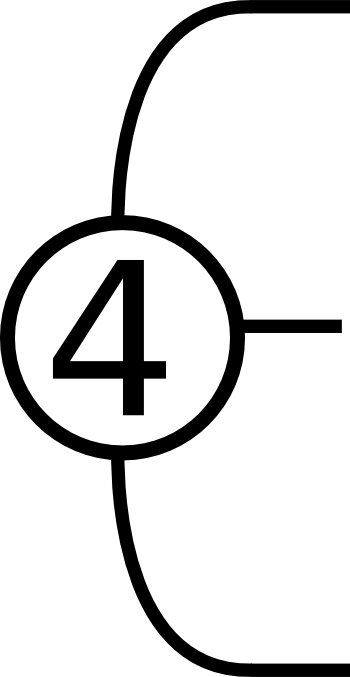}}&=
  \left(
    \begin{array}{cccc}
      \bra{H_l}_{\perp},&
      \raisebox{-0.58cm}{\includegraphics[width=0.1\columnwidth]{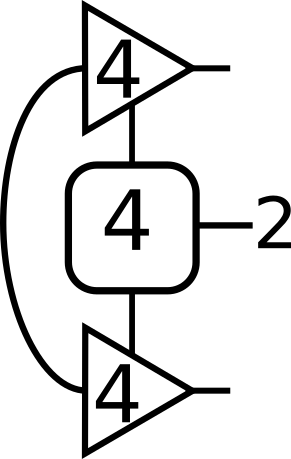}}\;,&
      \cdots,&
      \raisebox{-0.58cm}{\includegraphics[width=0.1\columnwidth]{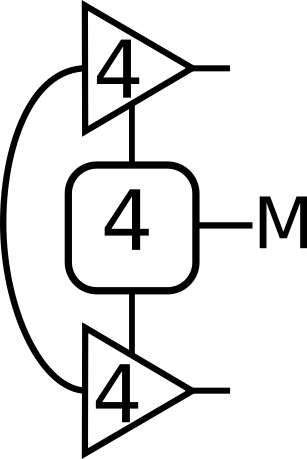}}
    \end{array}
  \right)
\end{align}

\end{document}